\documentclass{emulateapj}

\usepackage{graphicx}
\usepackage{epstopdf}
\usepackage{natbib}
\usepackage{enumerate}
\usepackage{color}

\newcommand{\be}{\begin{equation}}
\newcommand{\ee}{\end{equation}}

\newcommand{\ifm}[1]{\relax\ifmmode#1\else$\mathsurround=0pt #1$\fi}
\newcommand{\kms}{\ifmmode\,{\rm km}\,{\rm s}^{-1}\else km$\,$s$^{-1}$\fi}
\newcommand{\kpc}{\ifmmode\,{\rm kpc}\else kpc\fi}
\newcommand{\Mpc}{\ifmmode\,{\rm Mpc}\else kpc\fi}

\newcommand{\ltsima}{$\; \buildrel < \over \sim \;$}
\newcommand{\lsim}{\lower.5ex\hbox{\ltsima}}
\newcommand{\gtsima}{$\; \buildrel > \over \sim \;$}
\newcommand{\gsim}{\lower.5ex\hbox{\gtsima}}

\long\def\symbolfootnote[#1]#2{\begingroup\def\thefootnote{\fnsymbol{footnote}}\footnote[#1]{#2}\endgroup}

\def\apjl{\textit{Astrophys. J., Lett.}}        % Astrophysical Journal, Letters
\def\apjs{\textit{Astrophys. J., Suppl. Ser.}}  % Astrophysical Journal, Supplement
\def\aap{\textit{Astron. Astrophys.}}          % Astronomy and Astrophysics
\shorttitle{The Evolution of Central Group Galaxies}
\shortauthors{Feldmann et al.}
\begin{document}

\title{The Evolution of Central Group Galaxies in Hydrodynamical Simulations}


\author{R. Feldmann\altaffilmark{1}, C. M. Carollo\altaffilmark{1}, L. Mayer\altaffilmark{2}, A. Renzini\altaffilmark{3}, G. Lake\altaffilmark{2}, T. Quinn\altaffilmark{4}, G. S. Stinson\altaffilmark{4,5} and G. Yepes\altaffilmark{6}}


\altaffiltext{1}{Institute for Astronomy, ETH Z\"urich, 
		 Wolfgang-Pauli-Strasse 16, 8093 Z\"urich, Switzerland}
\altaffiltext{2}{Institute for Theoretical Physics, University of Zurich,
                 Winterthurerstrasse 190, 8057 Z\"urich, Switzerland}	 
\altaffiltext{3}{INAF Osservatorio Astronomico di Padova, Vicolo Dell'Osservatorio, 5, 35122 Padova, Italy} 
\altaffiltext{4}{Astronomy Department, University of Washington, Seattle, WA, 98195-1580, USA}
\altaffiltext{5}{Department of Physics and Astronomy, McMaster University, Hamilton, Ontario, L85 4M1, Canada}
\altaffiltext{6}{Grupo de AstroÞsica, Universidad Autonoma de Madrid, 28049 Madrid, Spain}


\begin{abstract}

We trace the evolution of central galaxies in three $\sim{}10^{13}$ $M_\odot$ galaxy groups simulated at high resolution in cosmological hydrodynamical simulations. 

In all three cases, the evolution in the group potential leads, at $z=0$, to central galaxies that are massive, gas-poor early-type systems supported by stellar velocity dispersion and which resemble either elliptical or S0 galaxies. Their $z\sim{}2-2.5$ main progenitors are massive ($M_*\sim{}3-10\times{}10^{10}$ $M_\odot$), star forming ($20-60 M_\odot$/yr) galaxies which host substantial reservoirs of cold gas ($\sim{}5\times{}10^{9}$ $M_\odot$) in extended gas disks. Our simulations thus show that star forming galaxies observed at $z\sim{}2$ are likely the main progenitors of central galaxies in galaxy groups at $z=0$.

At $z\sim{}2$ the stellar component of all galaxies is compact, with a half-mass radius $< 1$ kpc. The central stellar density stays approximatively constant from such early epochs down to $z=0$. Instead, the galaxies grow inside-out, by acquiring a stellar envelope outside the innermost $\sim{}2$ kpc. Consequently the density \emph{within the effective radius} decreases by up to two orders of magnitude. Both major and minor mergers contribute to most ($70^{+20}_{-15}\%$) of the mass accreted outside the effective radius and thus drive an episodical evolution of the half-mass radii, particularly below $z=1$. In situ star formation and secular evolution processes contribute to $14^{+18}_{-9}\%$ and $16^{+6}_{-11}\%$, respectively. Overall, the simulated galaxies grow by a factor $\sim{}4-5$ in mass \emph{and} size since redshift $z\sim{}2$. 

The short cooling time in the center of groups can foster a ``hot accretion'' mode. In one of the three simulated groups this leads to a dramatic rejuvenation of the central group galaxy at $z<1$, affecting its morphology, kinematics and colors. This episode is eventually terminated by a group-group merger. Mergers also appear to be responsible for the suppression of cooling flows in the other two groups. Passive stellar evolution and minor galaxy mergers gradually restore the early-type character of the central galaxy in the cooling flow group on a timescale of $\sim{}1-2$ Gyr. Although the average properties of central galaxies may be set by their halo masses, our simulations demonstrate that the interplay between halo mass assembly, galaxy merging and gas accretion has a substantial influence on the star formation histories and $z=0$ morphologies of central galaxies in galaxy groups.

\end{abstract}


\section{Introduction}
\label{sect:Introduction}

The study of galaxy groups is motivated by the fact that they host many galaxies in the local universe \citep{2004MNRAS.348..866E} and provide an environment in which galaxy interactions and merging are preferentially able to drive the morphological evolution of galaxies (e.g. \citealt{1998ApJ...496...39Z}). In addition, cluster galaxies likely experienced at some point in their history preprocessing in groups (e.g. \citealt{2002ASPC..257..123Z}, \citealt{2008ApJ...688L...5K}, \citealt{2009ApJ...694.1349P}, \citealt{2009arXiv0904.2851P}). Observations also indicate that bound, relaxed groups in the local Universe often host massive ($>10^{11} M_\odot$) early-type galaxies at their centers (\citealt{1998ApJ...496...39Z}, \citealt{2009ApJ...695..900Y}). The exploration of galaxy evolution in groups therefore provides the link between the study of isolated galaxies that preferentially populate lower mass halos ($\lesssim$ a few $10^{12}$ $M_\odot$) and the analysis of member galaxies of the more massive galaxy clusters ($\gtrsim10^{14}$ $M_\odot$).

With the advent of large surveys it has now become feasible to identify and study the evolution in galaxy groups both at low (e.g. \citealt{2006MNRAS.366....2W}, \citealt{2006MNRAS.372.1161W}, \citealt{2009ApJ...695..900Y}, \citealt{2009arXiv0901.1150G}, \citealt{Carolloetal2009}, \citealt{Cibineletal2009a, Cibineletal2009b, Cibineletal2009c}) and high redshifts (e.g. \citealt{2009arXiv0903.3411K}) with an exquisite statistics. The study of early type galaxies in galactic halos has gained considerable attention \citep{2003ApJ...590..619M, 2007ApJ...658..710N, 2009arXiv0903.1636N}, but the high computational demands necessary to resolve reliably the structural properties of individual galaxies pose a challenge for works that address the galaxy group scale \citep{2003MNRAS.346..135K, 2005MNRAS.361.1216K, 2006MNRAS.373..503O, 2007MNRAS.376...39O, 2008ApJ...672L.103K}. Typically a spatial resolution of $\gtrsim{}$ 1 kpc is reached \citep{2003MNRAS.346..135K, 2006MNRAS.373..503O, 2007MNRAS.376...39O, 2008ApJ...672L.103K} or the simulations are not fully cosmological \citep{2005MNRAS.361.1216K}. In this work we present a set of high-resolution simulations of galaxy groups that allow not only a study of the evolution of their central galaxies, but also enable us to investigate a rich satellite population within the virial radius of the group. Here, we put particular focus on the structural and kinematic properties of the central group galaxies and their most massive progenitors over time, relate their morphological properties to their assembly and gas accretion histories, and analyze the evolution of masses, sizes and densities of the stellar component since $z\sim{}1.5$. In particular, we address the following questions: 

\begin{enumerate}
\item \emph{Is the  $\Lambda$CDM concordance model together with our current understanding of the baryonic physics able to reproduce massive galaxies that resemble those observed at the centers of $z=0$ galaxy groups?} We perform our numerical experiments with the same code (Gasoline; \citealt{2004NewA....9..137W}), with the same set of parameters, and at a resolution comparable to the one that has been used to study the formation and evolution of disk and dwarf galaxies (\citealt{2007MNRAS.374.1479G}, \citealt{2008ASL.....1....7M}, \citealt{2008arXiv0812.0379G}, \citealt{Governato2009}). Therefore our simulations are not specifically tuned towards modeling massive galaxies in groups, but rather we apply a successful physical model to a yet unexplored mass regime.

\item \emph{What are the specific properties of the $z\sim{}2$ progenitors of $z=0$ central group galaxies and how do they compare with the galaxies observed at those redshifts?} One of the biggest challenges in observational cosmology is to connect and relate galaxy populations at different times. Here the simple question is whether some of galaxies that we can observe at high redshifts, such as 
distant red galaxies \citep{2003ApJ...587L..79F}, 
restframe UV/optical selected galaxies (BM/BX \citealt{2004ApJ...607..226A}, \citealt{2004ApJ...604..534S}),
or BzK selected galaxies \citep{2004ApJ...617..746D}, contribute to the precursor population of local central group galaxies, and, if so, what physical mechanisms play a role in the evolution of such systems from high redshift to the current epoch.

\item \emph{Which physical processes contribute to the mass and size evolution of massive galaxies over the $z\sim{}2$ to $z=0$ time span?}
Our simulations show that the progenitors of central group galaxies grow by a factor of 4-5 in mass and size between $z\sim{}2$ and $z=0$. Natural questions are therefore (i) which physical mechanism, such as major and minor merging or star formation, dominates the mass and size evolution and (ii) how does the mass growth relate to the size evolution; in particular, how steep is the size growth per unit mass? For example, minor merging has been suggested as the main driver in the evolution of \emph{passive}, compact galaxies (e.g. \citealt{2007ApJ...658..710N,2009arXiv0903.1636N, 2009ApJ...697.1290B}) as it allows for a fast size growth per unit mass. Since we find that the $z\gtrsim{}2$ progenitors of the $z=0$ central group galaxies are star forming, their mass and size evolution may proceed differently.

\item \emph{What is the impact of different assembly and gas accretion histories in groups at a fixed $z=0$ halo mass scale of $\sim{}10^{13}$ on the detailed physical properties of their massive central galaxies?} The assembly of the stellar component of the central galaxies is not simply a reflection of the assembly of the dark matter halo, but depends on the properties of the accreted baryons. As their halo mass grows above $10^{12} M_{\odot}$ it is expected that  these galaxies will shift from cold mode to hot mode accretion (\citealt{2005MNRAS.363....2K}; \citealt{2006MNRAS.368....2D}; \citealt{2009ApJ...694..396B}). However, the accretion mode may also be influenced by details of the environment, such as by the properties of the surrounding network of intergalactic gas filaments (\citealt{2005MNRAS.363....2K}). By $z\sim{}0$ all three groups have similar virial masses. They are therefore ideally suited for studying the impact of the rich variety of stochastic physical processes on the properties of their $z=0$ central galaxies.
\end{enumerate}

% Organization
The paper is organized as follows. In sections \ref{sect:Simulations} and \ref{sect:Methodology} we describe the simulation set-up and data analysis, respectively. In section \ref{sect:Results}  we present the properties of the simulated central galaxies over time and analyze the evolution of their masses, sizes and densities. We summarize our findings in section \ref{sect:summaryCentralGalaxies} and point out open questions in section \ref{sect:openq}. In the appendix we describe how we deal with artificially enhanced central star formation (appendix \ref{sect:SFCorr}), present our resolution test strategy (appendix \ref{sec:ResolutionTest}) and provide supplementary material (appendix \ref{sect:SupplementaryFigures}).

 %%%%%%%%%%%%%%%%%%%%%%%%%%%%%%%%%%%%%%%%%%%%%%%%%%%%%%%%%%%%%%

\section{Simulations}
\label{sect:Simulations}

The galaxy groups are selected from a DM-only simulation \citep{2007MNRAS.375..489H} based on their Friends-of-Friends (FoF) masses \citep{1983ApJS...52...61G, 1985ApJ...292..371D} at redshift $z=1$ using a standard linking length of 0.2. The employed FoF mass range is 0.8 - 1.2 $\times{}10^{13}$ M$_\odot$. We note that while the original DM run uses WMAP1 cosmological parameters \citep{2003ApJS..148..175S} our re-simulations are performed with a WMAP3 cosmology (see Table \ref{tab:cosmology}, \citealt{2007ApJS..170..377S}) and therefore the mass selection corresponds to a selection at redshift $z\sim{}0.7$ in our cosmology. 
$G1$ is an isolated group without any halo above $1.5\times{}10^{12}M_\odot{}h^{-1}$ within 5 Mpc/$h$ at $z=0$ in WMAP3. The matter overdensity $\delta=\rho/\bar{\rho}-1$ within this radius is close to 0 ($\delta=0.2$). $G3$ on the other hand comprises a cluster of $1.6\times{}10^{14}M_\odot{}h^{-1}$ and two massive groups of $3.4$ and $5.8\times{}10^{13}M_\odot{}h^{-1}$ within this radius ($\delta=15.0$). Group $G2$ lies in-between the two extremes ($\delta=1.4$).

\begin{table}
\begin{center}
\caption{Cosmological parameters used in this work.\label{tab:cosmology}}
\renewcommand{\arraystretch}{1.25}
\begin{tabular}{cccccc}
\tableline
\tableline
$\Omega_m$ & $\Omega_\Lambda$ & $\Omega_b$ & $h$ & $\sigma_8$ & $n$ \\ \tableline
& & & & &\\[-0.32cm]
0.24 & 0.76 & 0.04185 & 0.73  & 0.77 & 0.96 \\
\end{tabular}
\end{center}
\end{table}

The initial power spectrum has been generated with \texttt{linger} \citep{1995astro.ph..6070B}. Regions enclosing each galaxy group are 
refined with several layers of resolution using \texttt{grafic-2} \citep{2001ApJS..137....1B}. The highest-resolution region is defined as the initial Lagrangian patch that contains the set of particles that enter at any point a sphere of radius $R(z)$ around the particular group center. The group center is defined as the position of the particle with the highest density within the group halo (or its main progenitor). We use $R(z)=2\times{}R_\mathrm{vir}(z=0)$ fixed in comoving coordinates. The highest-resolution region of each group is embedded into spherical Lagrangian patches of increasing radius and decreasing resolution. Gas particles are added only in the highest-resolution regions to reduce the computational costs.

Our simulation strategy is as follows: We simulate three galaxy groups ($G1$, $G2$, $G3$) with SPH at ``intermediate resolution'' (see Table \ref{tab:simResolution}) down to $z=0$. In addition, we simulate our fiducial group $G2$ at ``high resolution'', i.e. 8 times higher mass and 2 times higher force resolution, down to $z\sim{}0.1$ and denote this simulation as $G2-HR$. We have resimulated this group also at a varying number of coarser mass and/or force resolution in order to study the impact on our results. We keep the softening length fixed in physical coordinates from redshift 0 to a high redshift, here 10, and fixed in comoving coordinates for higher redshifts, as found beneficial in simulations of disk galaxies (e.g. \citealt{2004ApJ...607..688G, 2007MNRAS.374.1479G}) or galaxy clusters (e.g. \citealt{2006MNRAS.367.1641B}). We discuss the results of our resolution tests in appendix \ref{sec:ResolutionTest}.

\begin{table*}
\begin{center}
\caption[Mass and force resolutions of the simulations]{Particle masses and spline softenings of the different resolutions.\label{tab:simResolution}}
\renewcommand{\arraystretch}{1.25}
\footnotesize
\begin{tabular}{lrrrrrr}
\tableline
\tableline
           & m$_\mathrm{DM}$ & m$_\mathrm{gas}$ & m$_\mathrm{star}$ & $\epsilon_\mathrm{DM}$ & $\epsilon_\mathrm{bar}$ \\
Resolution & ($M_\odot$ $h^{-1}$)                      & ($M_\odot$ $h^{-1}$)      & ($M_\odot$ $h^{-1}$)       & (kpc $h^{-1}$)                                   & (kpc $h^{-1}$) \\
\tableline
intermediate & $3.7\times{}10^7$ & $7.9\times{}10^6$ & $2.3 \times{}10^6$ & 0.73\tablenotemark{a} & 0.44\tablenotemark{b} \\
high & $4.7\times{}10^6$ & $9.9\times{}10^5$ &         $2.9 \times{}10^5$ & 0.36                  & 0.22 \\
\tableline
\tableline
\end{tabular}
\normalsize
\tablecomments{The second column denotes the mass of the dark matter particles in the zoom-in region. The following columns indicate the initial mass of gas particles, the initial mass of spawned star particles, the gravitational softening of dark matter particles and the gravitational softening of baryonic (gas and star) particles, respectively. }
\end{center}
\tablenotetext{a}{For $G2$ a slightly larger gravitational softening of 0.88 kpc $h^{-1}$ has been used.}
\tablenotetext{b}{For $G2$ a slightly larger gravitational softening of 0.53 kpc $h^{-1}$ has been used.}
\end{table*}

The simulations are performed with the TreeSPH code Gasoline \citep{2004NewA....9..137W} which is based on the parallel, multiple time stepping N-body code PKDGrav \citep{2001PhDT........21S}. The gravity tree opening angle is 0.525 above redshift 2 and 0.7 below it.

We use a standard radiative cooling scheme for primordial (metal-free) gas \citep{2004NewA....9..137W, 2006MNRAS.373.1074S}. We discuss changes expected due to metal cooling in appendix \ref{sect:MetalCooling}. The simulations include a spatially uniform UV background field \citep{1996ApJ...461...20H}. The UV background reduces the formation of galaxies with low masses ($\lesssim{}10^9 M_\odot$) and also implies a lower limit on the halo mass of the order of $\sim{}10^{11} M_\odot$ above which accreted gas is shock heated \citep{2009ApJ...694..396B}.

We model the star formation and feedback following \cite{2006MNRAS.373.1074S}. Stars are formed in a probabilistic fashion if the gas density is larger than 0.1 cm$^{-3}$, the gas temperature is lower than 15'000 K, the gas is in an overdense ($\delta{}>55$) region and part of a convergent flow. The star formation efficiency is set to 0.05. The simulations incorporate supernova type Ia and type II feedback and mass loss of star particles due to stellar winds. Each star particle is treated as a single stellar population with Scalo IMF \citep{1979ApJS...41..513M} and the parametrization of \cite{1996A&A...315..105R} of the stellar tracks.
The modelling of the type II supernovae makes use of the analytic blastwave scenario of \cite{1977ApJ...218..148M}. In particular, a thermal feedback of $4\times{}10^{50}$ erg per supernova is injected into neighbouring gas particles which have their cooling shut-off for the time corresponding to the end of the snowplow phase of the blastwave. The cooling is not disabled for supernovae of type Ia. Over its lifetime subgrid stellar winds return a substantial fraction of the mass of a star particle ($\sim{}40\%$) to its surrounding gas particles.

\section{Methodology}
\label{sect:Methodology}

% masses & luminosities
 We identify halos and galaxies and determine virial radii and masses with the help of the AMIGA halo finder (AHF, \citealt{2004MNRAS.351..399G, 2009ApJS..182..608K}). Halos extend out to the virial radius, which we define as the radius enclosing a mean matter density of $200/\Omega_m$ times the mean density of the universe, or (for subhalos) out to the tidal radius, whichever is smaller. Galaxy masses are then estimated both in real space and in projection. We identify and extract the central galaxy of our group using the star particle with the lowest potential in the halo as center. We assign the central galaxy a stellar mass by measuring the mass of each stellar particle in a sphere of 20 physical kpc around the center. The choice of this radius is to some extent arbitrary. However, the so determined stellar masses change typically by less than 10\% when the radius is varied within a factor of 2. We show in Appendix \ref{sec:ResolutionTest} that these masses are well converged (up to a $\sim{}20$\%)  below $z=2$. We observe some residual, artifical, star formation even at $z=0$ and correct our masses for this effect (see appendix \ref{sect:SFCorr}). This lowers the stellar masses at $z=0$ by $\sim{}20\%$ and by less at higher redshifts.
  
% Mock images
 Much of our analysis is done on mock images. The image projections are chosen to be along the axes of the moment-of-inertia tensor of the stellar mass within 10 kpc. The (reduced) moment-of-inertia tensor $\mathcal{T}$ is defined as
\[
\mathcal{T}_{ij}=\sum_k\frac{m_kx_k^ix_k^j}{r_k^2},
\]
where $x^i_k$ is the $i-$th component of the vector pointing from the halo center to the $k$-th particle position, $r_k$ is the distance from the halo center and $m_k$ denotes the mass of the $k$-th particle. The mock images have a pixelscale of 0.25 physical kpc per pixel, i.e. comparable to the gravitational softening of 0.3 kpc in our high resolution simulations $G2-HR$. The mass (or light) of a given stellar particle is assigned to its nearest grid point (NGP) in the projected image. We use the NGP approach for all mock images except Fig.\ref{fig:GroupOverview}, where we employ instead an adaptive cloud-in-cell method. In this case the mass (or light) of each particle is spread into a cube of a size that is proportional to the local interparticle distance. In either way, contours lines of constant surface mass density are created with SAOImage \texttt{DS9}.

% radii
We define as effective radius the radius that includes half the stellar mass or flux within 20 kpc around the center of the respective galaxy. The estimate may include a correction according to the minimal star formation correction scheme, see section \ref{sect:SFCorr}. For the $z=0$ central galaxies we additionally derive effective radii from Sersic or deVaucouleurs fits to the surface mass-density or surface brightness using the intensity values determined by running the \texttt{IRAF} task \texttt{ellipse} on mock images.
The images are created by projecting along the minor axis of the moment-of-inertia tensor of the stellar component. In the employed weighted least square fit we only use intensity values for circularized radii between $3\times{}\epsilon_\mathrm{bar}$ and $20$ kpc and consider integer Sersic indices in the range 1-8. We adopt the Sersic index with the lowest $\chi^2$ value as our best fit solution. We find similar results whether we use mass- or $I$-band flux-weighted effective radii and refer to the former unless mentioned otherwise.

% SFR
The star formation rates (SFR) that we quote refer to the stellar mass formed within 20 kpc around the main progenitor of the central group galaxy (excluding contribution from substructure) within a time-span of $\sim{}100$ Myr. Note that an ``archaeological`` measurement at $z=0$ would give much larger SFRs at earlier times because it measures the formation of all stars that are within the galaxy by $z=0$, rather than the SFR of the most massive progenitor that existed at a given time.

% color
We calculate the luminosity of each stellar particle from its age and metallicity assuming a single stellar population with Chabrier IMF \citep{2003MNRAS.344.1000B}. 
We measure the $^{0.1}(g-r)$ and $^{0.1}(r-i)$ colors and the $^{0.1}M_i$ absolute magnitude of the central galaxies at $z=0$ in the $z=0.1$ redshifted SDSS filter system of \cite{2003ApJ...592..819B} and the rest-frame SDSS r-band magnitude $M_r$. 
We follow \cite{2003ApJ...592..819B} in determining the colors and magnitudes from within a circular aperture with a diameter of 4 Petrosian radii. To avoid biases towards bluer colors from artificial star formation in the unresolved, central region of a $z=0$ central galaxy we fix the mass-to-light ratio within $2\times{}\epsilon_\mathrm{bar}$ to its value \emph{at} $2\times{}\epsilon_\mathrm{bar}$. In addition, we estimate the amount of mass deposition due to artifical star formation in the central region and subtract a corresponding, excessive flux from our mock images within a circle of 2 $\epsilon_\mathrm{bar}$ around the center. We note that the colors of the $z=0$ central galaxies do not include a correction for dust absorption.
In a similar way we determine the colors and magnitudes of the main progenitors at higher redshifts (see Table \ref{tab:PropertiesZ25Phot} for the set of employed filter transmission curves). We classify the most massive progenitors with the help of the fluxes measured within a projected radius of 8 physical kpc ($\sim{}$1\arcsec),  excluding the light from resolved satellites. We do not apply a mass-to-light ratio fix or a subtraction of central flux, mainly because we expect central star formation to be a physical reality at high redshifts and the impact of additional artificial star formation is consequently small. Nonetheless, the quoted color and magnitude errors include variations arising from different viewing angles and from the difference between correcting or not-correcting for artificial star formation. Absorption by dust may affect the colors and magnitudes of the gas-rich central galaxies at high redshifts and we therefore explore extinction corrections \citep{2000ApJ...533..682C} with $A_V$ ranging from 0 to $>1$. At high redshifts ($z>2$) we use as default value $A_V=0.8$ which is the mean value found in the (almost) mass-limited sample of \citep{2008ApJ...682..896K, 2008ApJ...677..219K} and comparable to the median value in the SINS sample (\citealt{2009arXiv0903.1872F}, see also \citealt{2006Natur.442..786G}). Colors and magnitudes are given in the AB system, unless explicitly noted otherwise.

% velocity & dispersion
In order to measure stellar line-of-sight rotation velocity and velocity dispersion we first project the galaxies along the intermediate axis of their moment-of-inertia tensor. Then, we put a slit of extent and width of 20 kpc and 2 kpc, respectively, along the major axis of the projected image and measure velocity and dispersion in 24 bins along the slit. We proceed in an analogous way for gas velocities and velocity dispersions. The velocity dispersion within the effective radius is then estimated from the mass-weighted average of all bins within
$2\times{}\epsilon_\mathrm{bar}$ and 3.5 kpc ($\sim{}R_\mathrm{eff}$) of the velocity dispersion profile. Statistical errors are estimated from a bootstrapping analysis. The statistical errors are typically much smaller than systematic effects that arise from using either velocity moments or Gauss-Hermite polynomials to fit rotation velocities and dispersion. Due to the latter we estimate that our velocities and dispersions are accurate to $\sim{}10$\%.

% Bulge/disk ratio
Bulge-to-disk ratios of the central galaxies are measured on mock images with GALFIT \citep{2002AJ....124..266P}. We compute noise (``sigma'') maps based on the particle-per-pixel number in mock images assuming a Poisson statistics. We then run GALFIT on each central galaxy 16 times covering the following cases: (i) edge-on vs. face-on projection, (ii) projected mass density vs. I-band flux density, (iii) a single component (Sersic profile) vs. a double component (Sersic profile + exponential disk profile) model, (iv) including or excluding the central 2 softening lengths from the fit. While GALFIT very often produces good ($\chi^2{}\sim{}1$) and sensible fits, in some cases it can be stuck in a non-optimal solution (either not a good fit ($\chi^2{}>2$) or a clearly unphysical solution). In these cases we either changed our initial parameters, guided by visual inspection, or introduced some constraints that cut away the unphysical solution space. 

% division of origin of the stars (accreted, from merger, in-situ)
We compute the contribution to the growth of the stellar envelope by monitoring the stellar mass flux into the region between the effective radius and 20 kpc (our defined size of the central galaxy) among pairs of successive snapshots. The merger flux is the net flux of stars into this region that have formed at an earlier time outside the central galaxy. The stellar transfer flux is defined as the net flux of stars that formed in situ earlier on and are now crossing the mean effective radius between the two snapshots. Stellar particles that form from on snapshot to the other in the shell enclosed by the effective radius and 20 kpc define the in situ star formation flux. These fluxes represent smoothed quantities averaged over the time interval between successive snapshots. We checked that our results are not affected by our sampling frequency of the snapshots.
An alternative methode to obtain fluxes is to apply the continuity equation using positions and velocities from a single snapshot. However, unless the time resolution is exquisitely high this approach is not well suited to measure the average flux from non-continuous, clumpy accretion events and galaxy mergers for a given, fixed radius.

% splitting between major/minor mergers and accretion events
In this paper we consider a merger between two galaxies as major (minor) if the stellar mass ratio $\mathcal{R}$ is larger (smaller) than 1:3.5. Sometimes stars form in places that are not identified as bound halos by our halo finder. In this case we say that these star particles originate from unresolved objects. In order to determine whether a particular star particle is added to the central galaxy by a major merger, minor merger or from an unresolved source we proceed as follows. First we identify the last satellite halo to which this star particle belonged before it finally merged with the central galaxy. We keep track of whether the star particle is stripped from this last satellite halo before it finally enters a 20 kpc boundary around the central galaxy. If the star particle first appears in a resolved satellite located within a 20 kpc radius around the central galaxy, we say that the star particle forms by a merger-induced star formation event in the satellite.
In case no such satellite halo can be identified we distinguish between star particles that form within 20 kpc around the central galaxy (in situ star formation) and star particles that form outside this radius (unresolved origin). 

\section{Results}
\label{sect:Results}

In this section we address the four major questions that we raise in the introduction, namely: (i) How do predictions from ab-initio, cosmological, hydrodynamical simulations compare with observations of nearby central group galaxies? (ii) How do early-type central group galaxies at $z=0$ relate to galaxies observed at $z\sim{}2$? (iii) How do the main progenitors of central group galaxies evolve in their most fundamental parameters, namely mass and size? and (iv) How does the assembly history affect the evolution of central group galaxies? 

\subsection{Overview of the groups and their central galaxies}

\begin{figure*}
\begin{center}
\begin{tabular}{c}                            
\includegraphics[width=160mm]{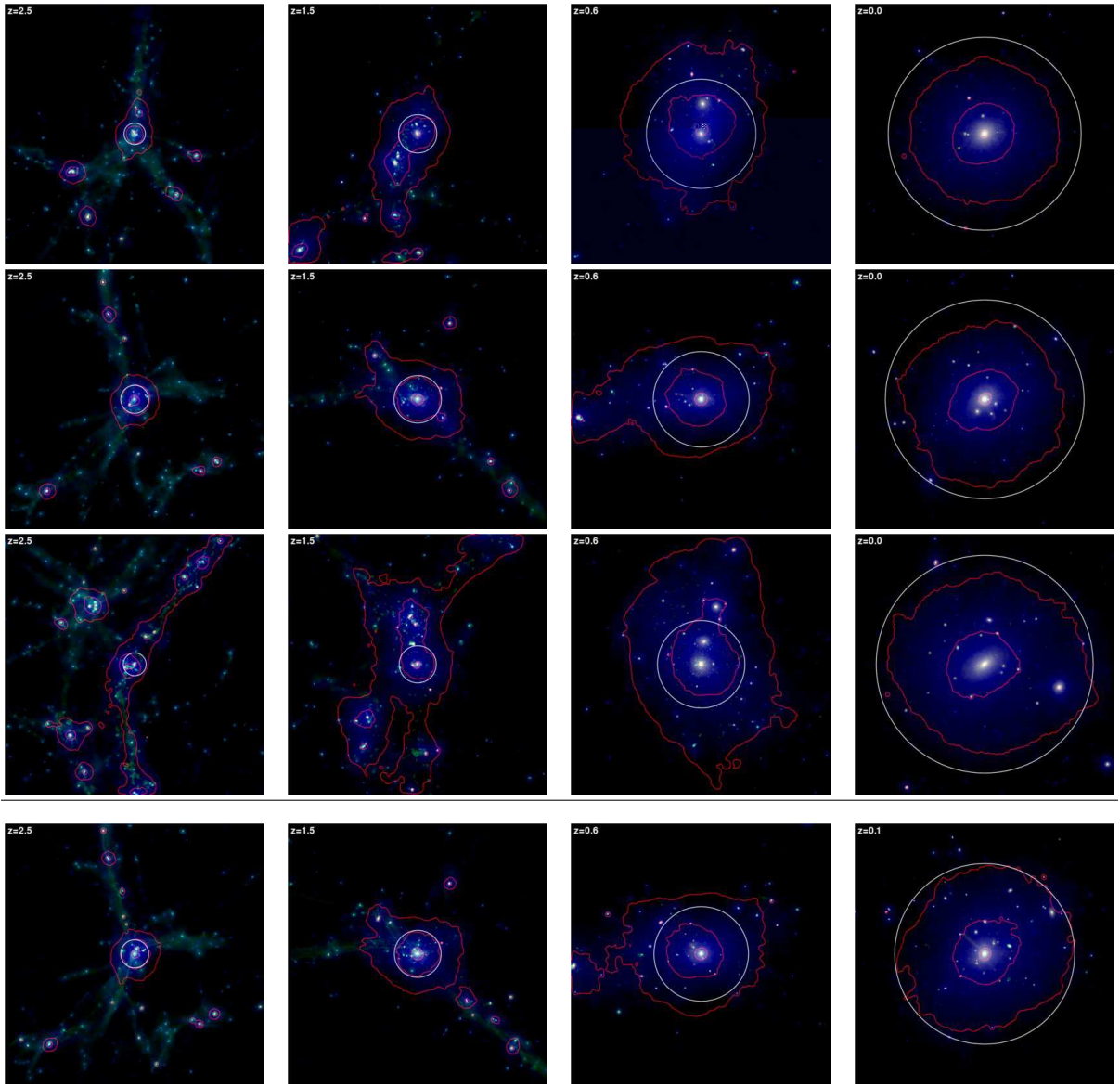}
\end{tabular}
\caption[The appearance simulated galaxy groups]{The appearance of the simulated galaxy groups as function of redshift. The size of each image is 1.2 Mpc $\times$ 1.2 Mpc. The four columns correspond to $z=2.5$, $z=1.5$, $z=0.6$ and $z=0$ ($z=0.1$ for $G2-HR$), while the rows correspond to the different groups $G1$, $G2$, $G3$, and $G2-HR$ (from top to bottom). Color coded are dark matter in blue (from 3.6 to 1460 $M_\odot$ pc$^{-2}$), cold gas (here the gas with $T<2.5\times{}10^5$ K) in green, stellar matter in yellow (both from 0.9 to 365 $M_\odot$ pc$^{-2}$), and hot gas (here the gas with $T\geq{}2.5\times{}10^5$ K) as red surface mass isocontours (3 contour lines at 1, 4.5 and 20 $M_\odot$ pc$^{-2}$). The white circle shows the virial radius of each group at the indicated redshift.\label{fig:GroupOverview}}
\end{center}
\end{figure*}

%% Overview of groups
In Fig. \ref{fig:GroupOverview} we show how the three groups $G1$, $G2$ and $G3$ and their progenitors evolve from $z\sim{}2.5$ to $z=0$. At $z>2$ the group progenitors are surrounded by filaments of cold gas in agreement with a cold accretion picture, e.g. \cite{2005MNRAS.363....2K}. These filaments disappear at $z\lesssim{}1.5$.  The (shock-heated) hot gas is confined to halos and filaments, i.e. regions of high dark matter density. The contours of constant surface mass density of the hot gas can have very irregular shapes at $z>0.6$, but at lower redshift they become more regular. Stars and galaxies form in the places of high density in the cosmic filaments. A ``central galaxy'' forms at the bottom of the potential well of each group progenitor. Other galaxies approach and fall into the forming galaxy groups. Some of these galaxies merge quickly with the central galaxy, while others continue to orbit for a long time. At $z\sim{}0$ the virial masses of all three groups are in the range $1-2\times{}10^{13} M_\odot$, although their amount of substructure varies substantially. The number of satellites with a stellar mass in excess of $10^{10}$ $M_\odot$ is 2 ($G1$), 8 ($G2$), 8 ($G3$, all at $z=0$) and 13 ($G2-HR$, at $z=0.13$).

%% Overview of central galaxies

\begin{figure*}
\begin{center}
\begin{tabular}{c}                            
\includegraphics[width=160mm]{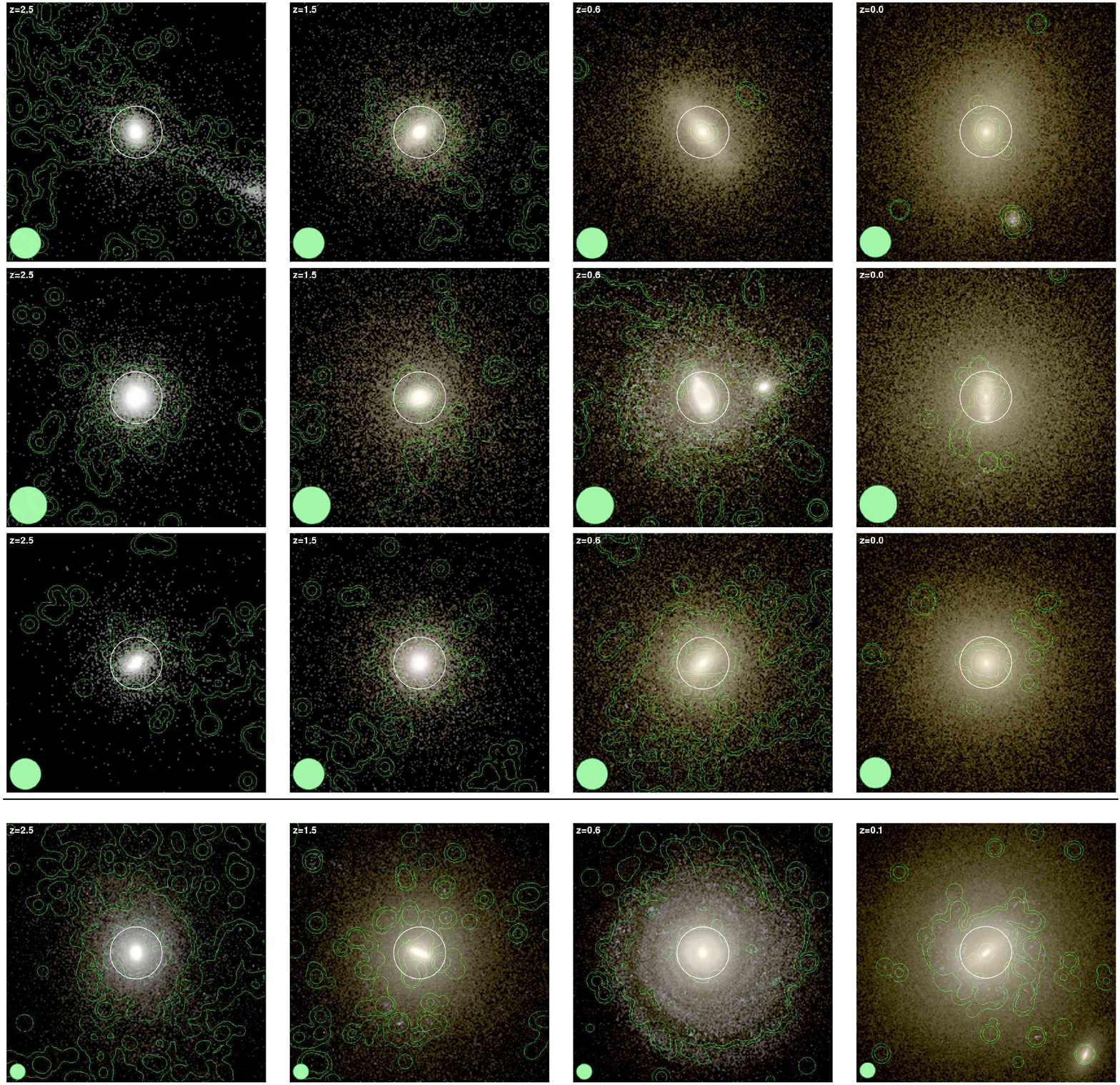}
\end{tabular}
\caption[Face-on images of the central group galaxies]{Mock images of the central group galaxies (and their main progenitors) as function of redshift. Each image is 20 physical kpc across and shows a face-on projection corresponding to the minor axis of the moment-of-inertia tensor of the stellar mass within a sphere of 10 kpc radius. The four columns correspond to $z=2.5$, $z=1.5$, $z=0.6$ and $z=0$ ($z=0.1$ for $G2-HR$). The first three rows correspond to $G1$, $G2$, and $G3$ (from top to bottom) while the last row shows the appearance of the central galaxy in the high resolution re-simulation $G2-HR$. When assessing the images note that  (i) the projections are along the line-of-sight of the virial radius of the galaxy group, (ii) the image orientations are arbitrary, hence the displayed disk orientations are not directly comparable, (iii) the addition of high frequency modes in the high resolution re-simulation and low-level numerical noise can lead to subtle timing and positioning differences at small scales and (iv) the central galaxy in group $G2$ ($G2-HR$) undergoes a substantial morphological evolution between $z=0.2$ and $z=0$. The RGB color channels of the images correspond to the surface brightness in the restframe Bessel B, R and I filterbands, respectively, and range from 13.5 mag arcsec$^{-2}$ to 22 mag arcsec$^{-2}$. Green contours indicate column densities of cold gas ($T<3.2\times{}10^4 K$) corresponding to 1, 10, 100, and 1000 $M_\odot$ pc$^{-2}$.  The green circle has a radius of 2 gravitational softenings $\epsilon_\mathrm{bar}$, indicating the resolution limit. The white circle encloses the central 2 kpc.\label{fig:EvolutionFaceOn}}
\end{center}
\end{figure*}

\begin{figure*}
\begin{center}
\begin{tabular}{c}                            
\includegraphics[width=160mm]{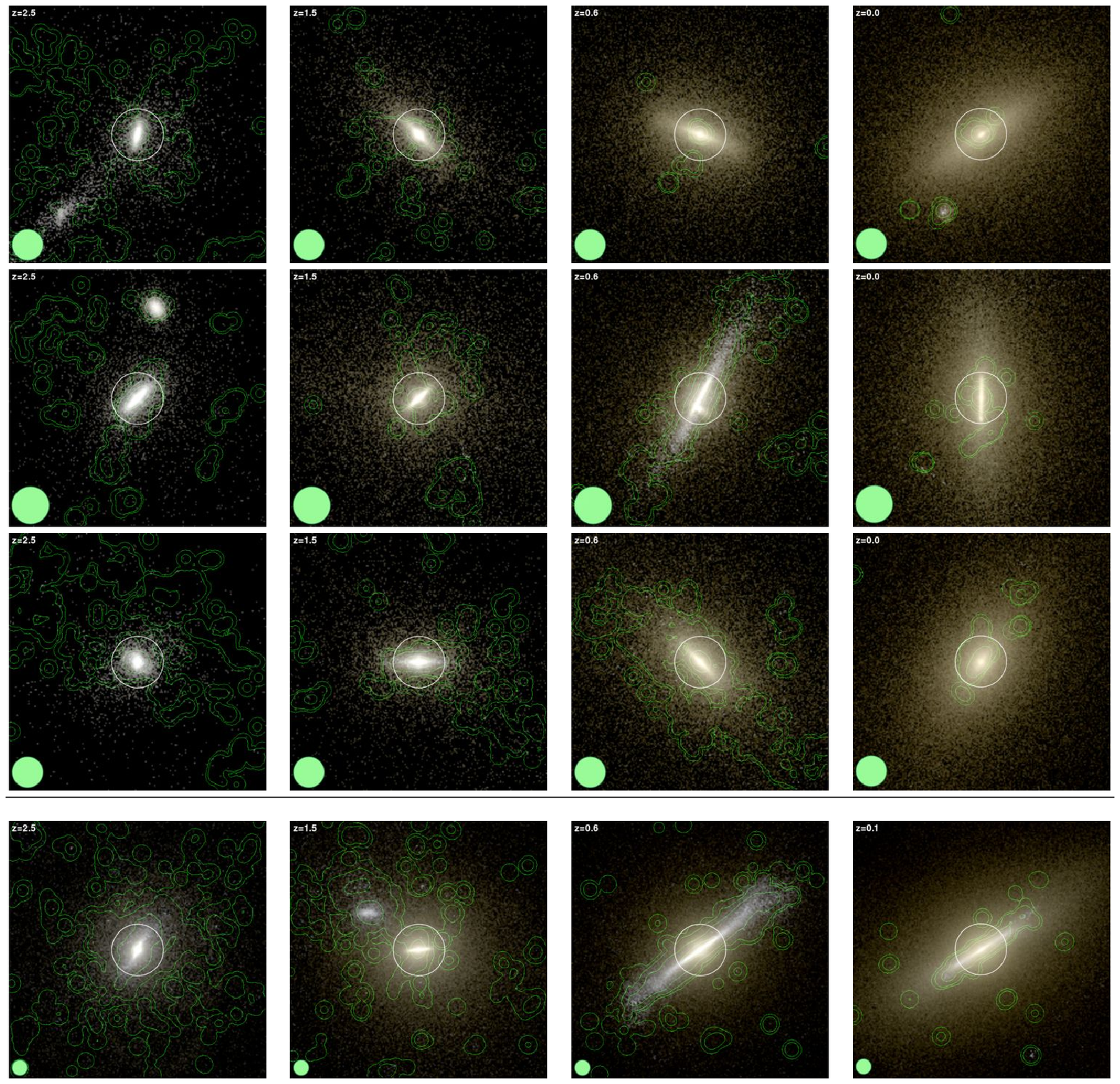}
\end{tabular}
\caption[Edge-on images of the central group galaxies]{As Fig. \ref{fig:EvolutionFaceOn} but showing instead an edge-on projection corresponding to the intermediate axis of the moment-of-inertia tensor of the stellar mass.\label{fig:EvolutionEdgeOn}}
\end{center}
\end{figure*}

Zooming in to galactic scale we show  in Fig.~\ref{fig:EvolutionFaceOn} and Fig.~\ref{fig:EvolutionEdgeOn} Bessel B, R and I band composite images of the central galaxies. Clearly, the gross evolution proceeds in all three cases in a similar fashion including a substantial change in color (from blue to red), morphology, size (from $<1$ kpc to several kpc) and cold gas content (from gas-rich at $z=2.5$ to gas-poor at $z=0$). The most notable differences between the three groups are: (i) the galaxies show a varying degree of disky-ness at $z=0$, (ii) at $z\sim{}0.6$ group $G2$ harbors a massive spiral galaxy and a substantial reservoir of cold gas, (iii) group $G2$ and its central galaxy appears more massive and evolved especially at $z\geq{}1.5$. We now proceed with a more quantitative analysis of the properties of the central galaxies. In the following we will, unless noted otherwise, include a minimal correction scheme for artificial, central star formation as described in appendix \ref{sect:SFCorr}. There we also discuss implications of more aggressive correction schemes.

%%%%%%%%%%%%%%%%%%%%%%%%%%%%%%%%%%%%%%%%%%%%%%%%%%%%%%
\subsection{The properties of the central galaxies at z=0}
\label{sect:PropZ0}

In this section we show that the central objects resemble early-type galaxies based on various criteria. We also compare the properties of the simulated central group galaxies with observations of central group galaxies or of massive early-type galaxies in general, and with results from previous simulations.

Galaxies can be split into early- and late-types according to different criteria, e.g. based on colors, Sersic index, Bulge-to-disk ratios, rotational support, or morphology. In addition, indirect indicators can be used - such as masses (the most massive galaxies are early-type galaxies) or gas fractions (early-type galaxies are often gas-poor). Using various criteria we now show the early-type nature of the simulated central group galaxies.

\begin{figure*}
\begin{center}
\begin{tabular}{ccc}                            
\includegraphics[width=55mm]{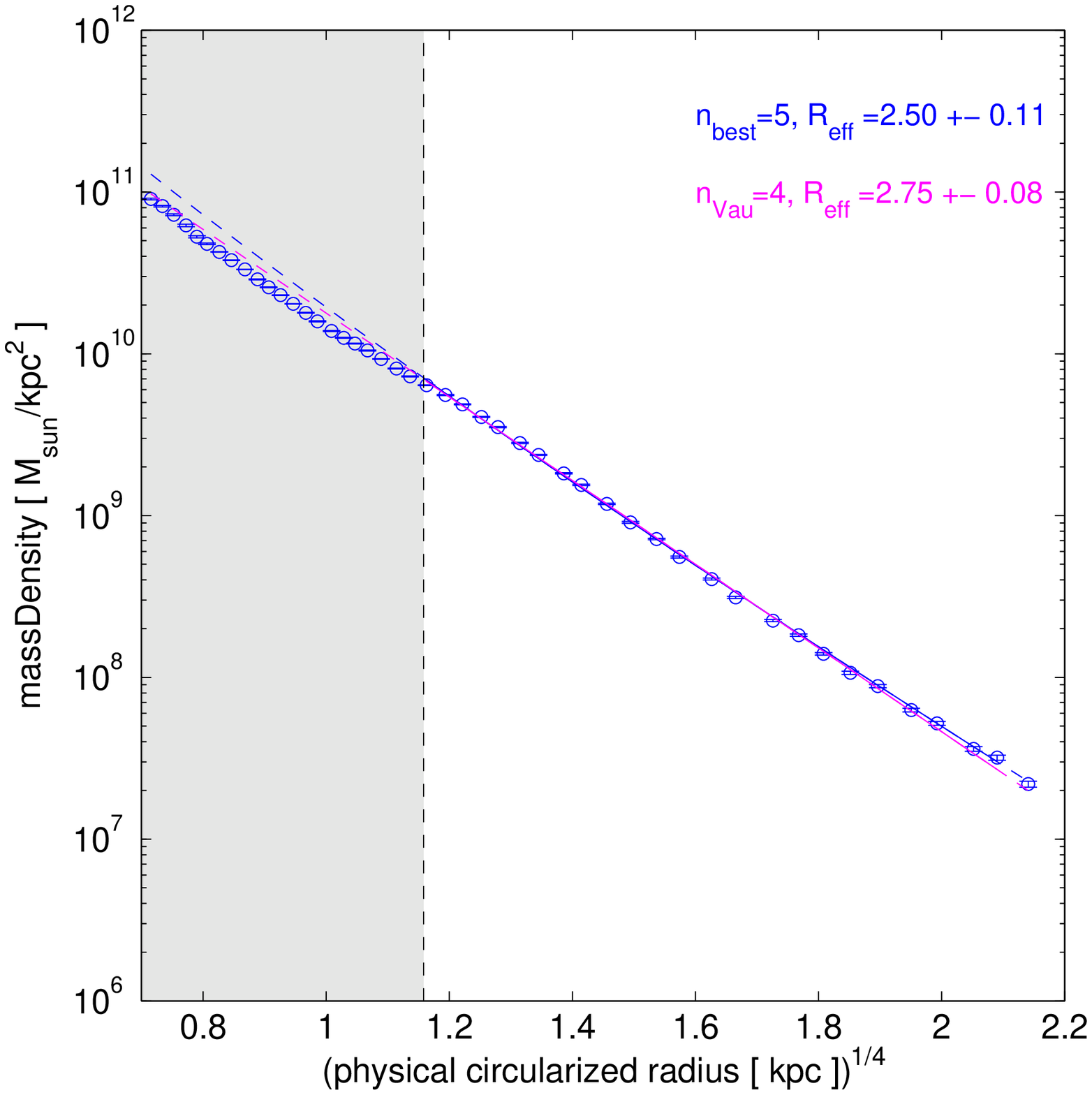} &
\includegraphics[width=55mm]{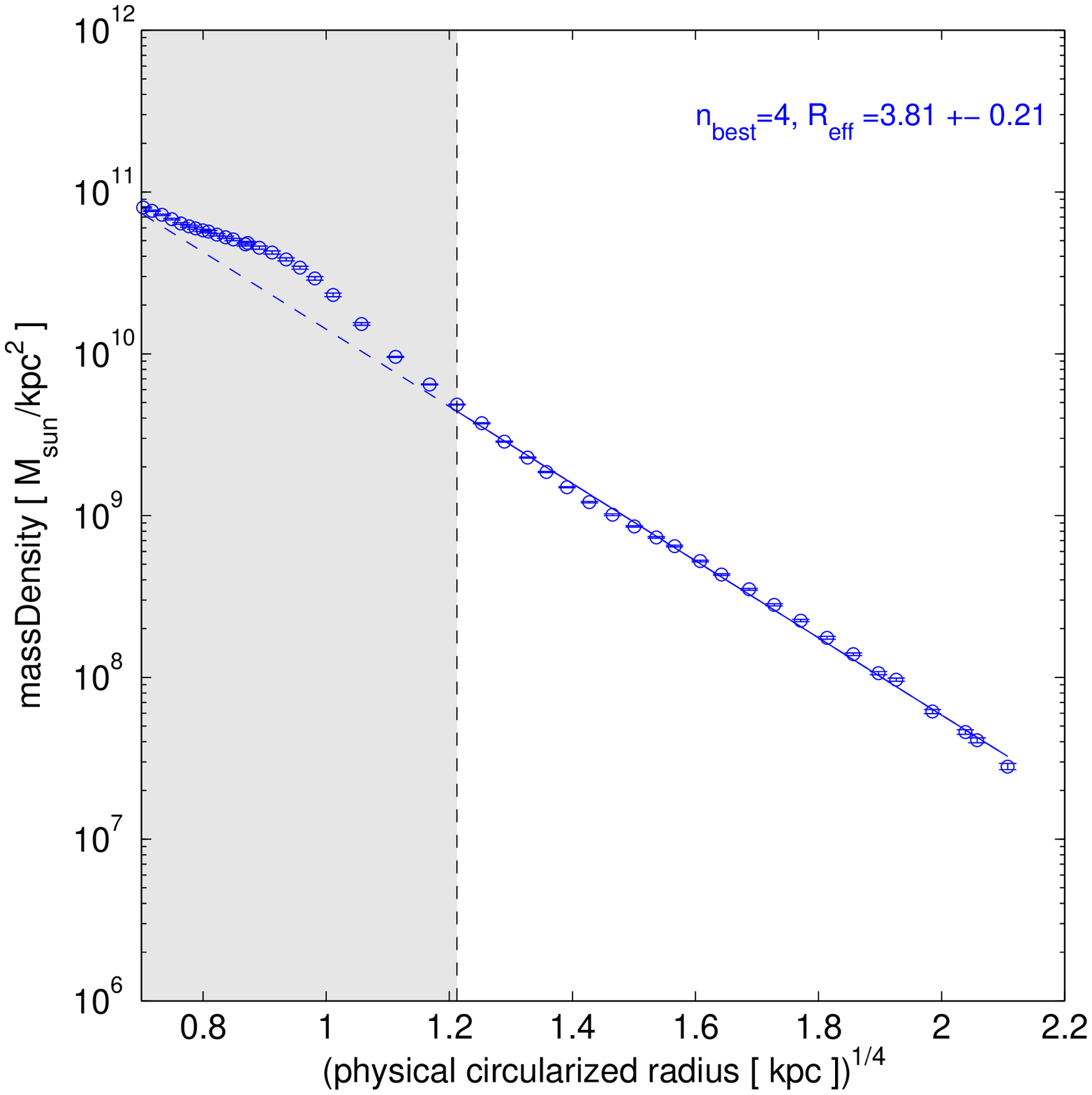} &
\includegraphics[width=55mm]{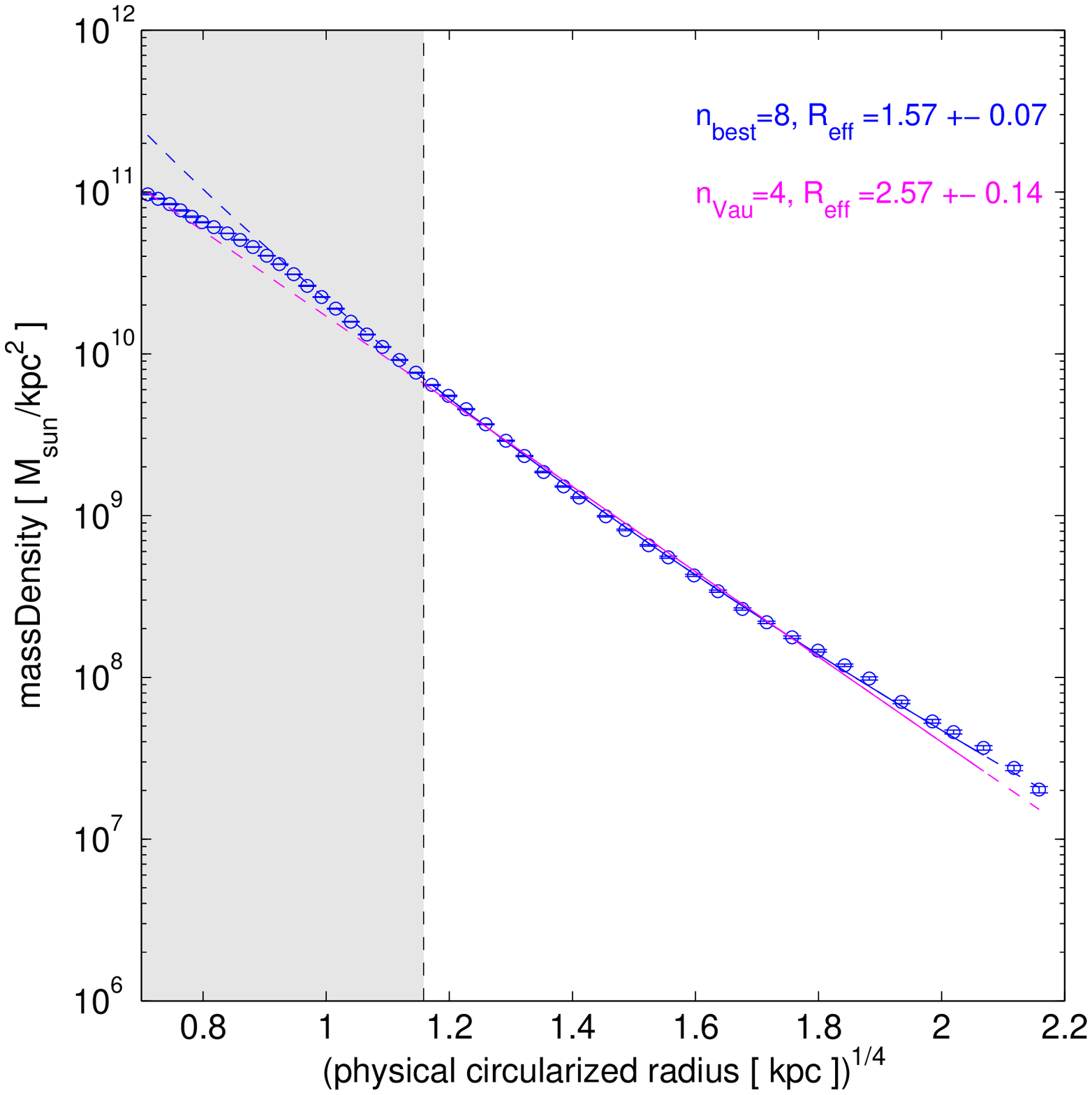}
\end{tabular}
\caption[The surface mass density of the central galaxies at $z=0$]{(From left to right) The surface mass density at $z=0$ of the central objects in the group $G1$, $G2$ and $G3$. The best-fit Sersic profile (blue line) with $n\in\{1,\ldots,8\}$, the best-fit deVaucouleurs profile (magenta line) and the derived effective radii are indicated. Fit solutions inside (outside) the fitting range are shown as solid (dashed) lines. The shaded region, which corresponds to the (projected) central region within 3 gravitational softening lengths $\epsilon_\mathrm{bar}$, has been excluded from the fit. \label{fig:SurfaceBrightness}}
\end{center}
\end{figure*}

The red  $^{0.1}(g-r)$  and $^{0.1}(r-i)$ colors of the central group galaxies, see Table \ref{tab:PropertiesZ0Phot}, are indicative of an aging stellar population.  Figure \ref{fig:SurfaceBrightness} shows that surface mass density profiles of the central galaxies are close to a deVaucouleurs profile. When fit with a Sersic law, the indices are  4, 5 and $\geq{}8$, respectively. Analyzing the mass density or I-band images with GALFIT \citep{2002AJ....124..266P} we find that $G1$ and $G3$ are very well fitted ($\chi^2\sim{}1-1.3$) by a single $n>3$ Sersic profile (independent of inclination and of whether or not we mask the central region). $G2$ can also be fitted with a single high-$n$ Sersic profile as long as the central two softening lengths are excluded. Alternatively, it can be well fitted with a two-component model in which either (a) a small disk (scale length $\sim{} 0.7$ kpc) is embedded in a more extended spheroid (effective radius of 3.4 - 5.5 kpc depending on whether or not the central region is masked), or (b) a central spheroid (1.6-1.8 kpc) is embedded in an extended, but faint disk (scale length $\sim{}6-7$ kpc). The first model is favored in edge-on projections, the second in face-on projection. The bulge-to-disk ratios of $G1$ and $G3$ are naturally very high and are typically $>3$. For $G2$ we obtain a bulge-to-disk ratio of 3 (edge-on) and 1.6 (face-on).
Also a visual inspection leads to the conclusion that the central galaxies are of early-type. The central galaxies in the groups $G1$ and $G3$ have an elliptical morphology while the central galaxy of $G2$ clearly shows some disk component. Spiral arms in the disk are revealed in its high resolution equivalent G2-HR at $z\sim{}0.6$, but disappear later on. At $z\sim{}0$ the central galaxy of group $G2$ resembles a S0 or E/S0 galaxy,  see Fig. \ref{fig:EvolutionFaceOn} and Fig. \ref{fig:EvolutionEdgeOn}.

\begin{figure*}
\begin{center}
\begin{tabular}{ccc}                            
\includegraphics[width=55mm]{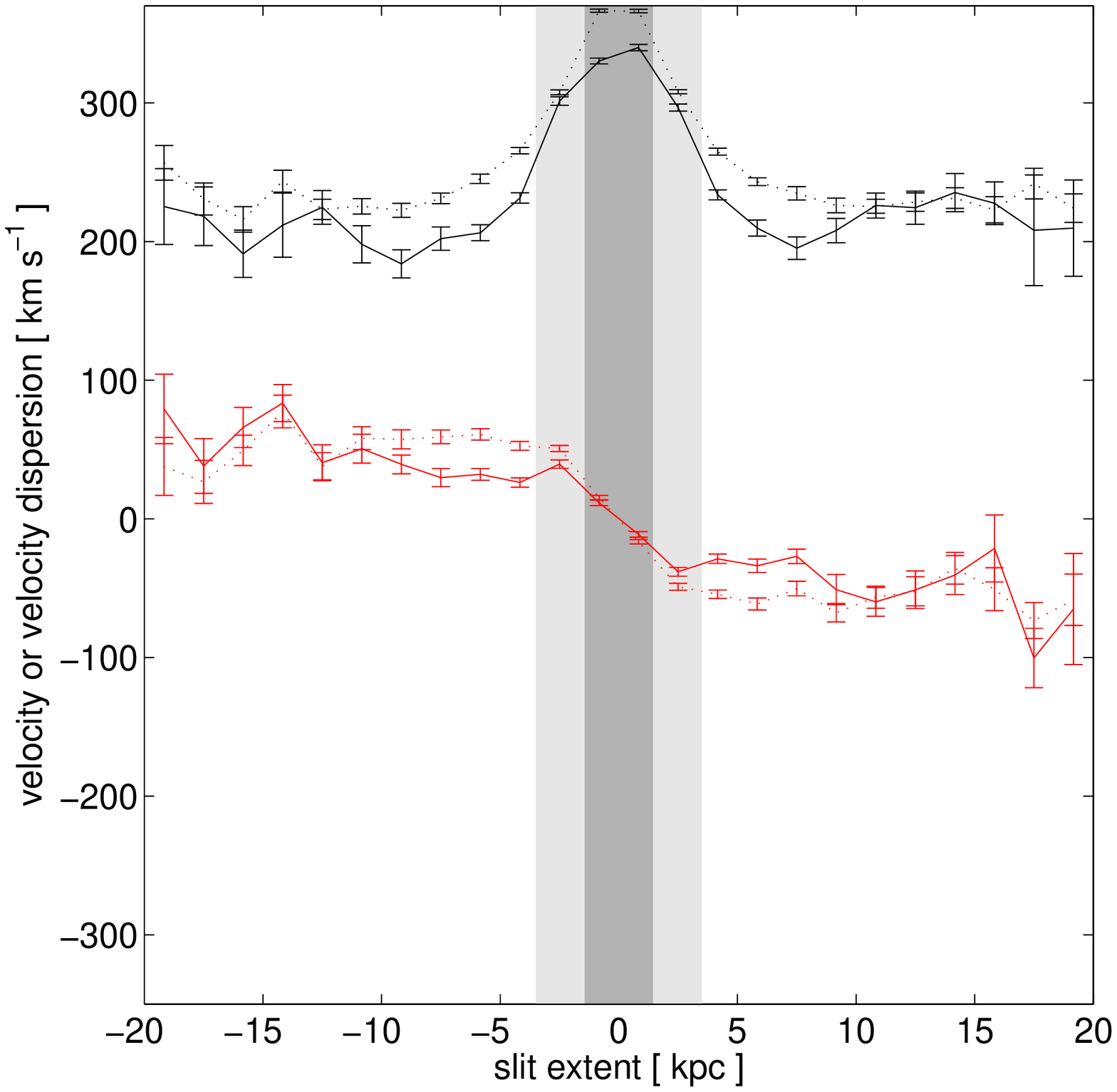} &
\includegraphics[width=55mm]{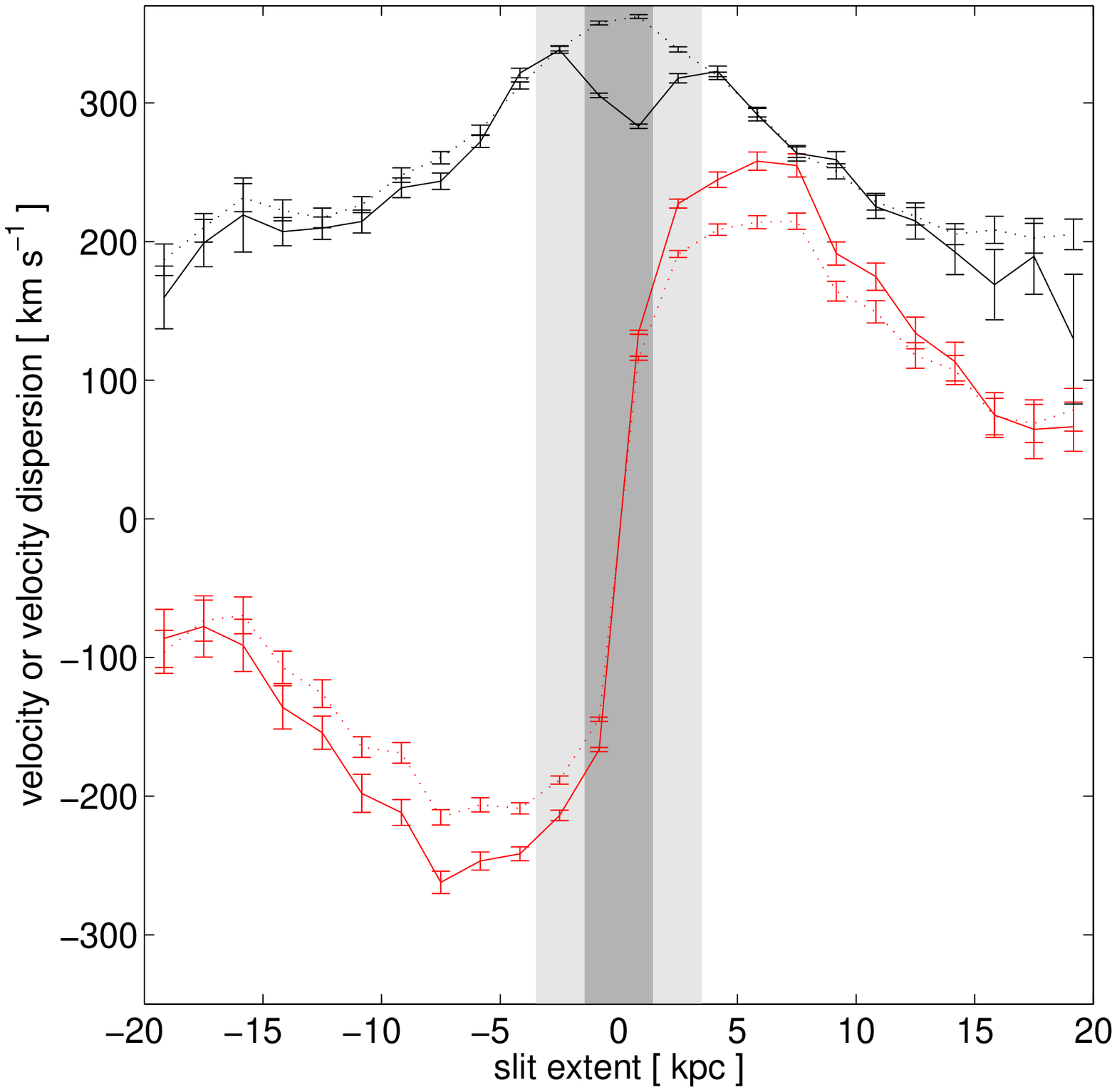} &
\includegraphics[width=55mm]{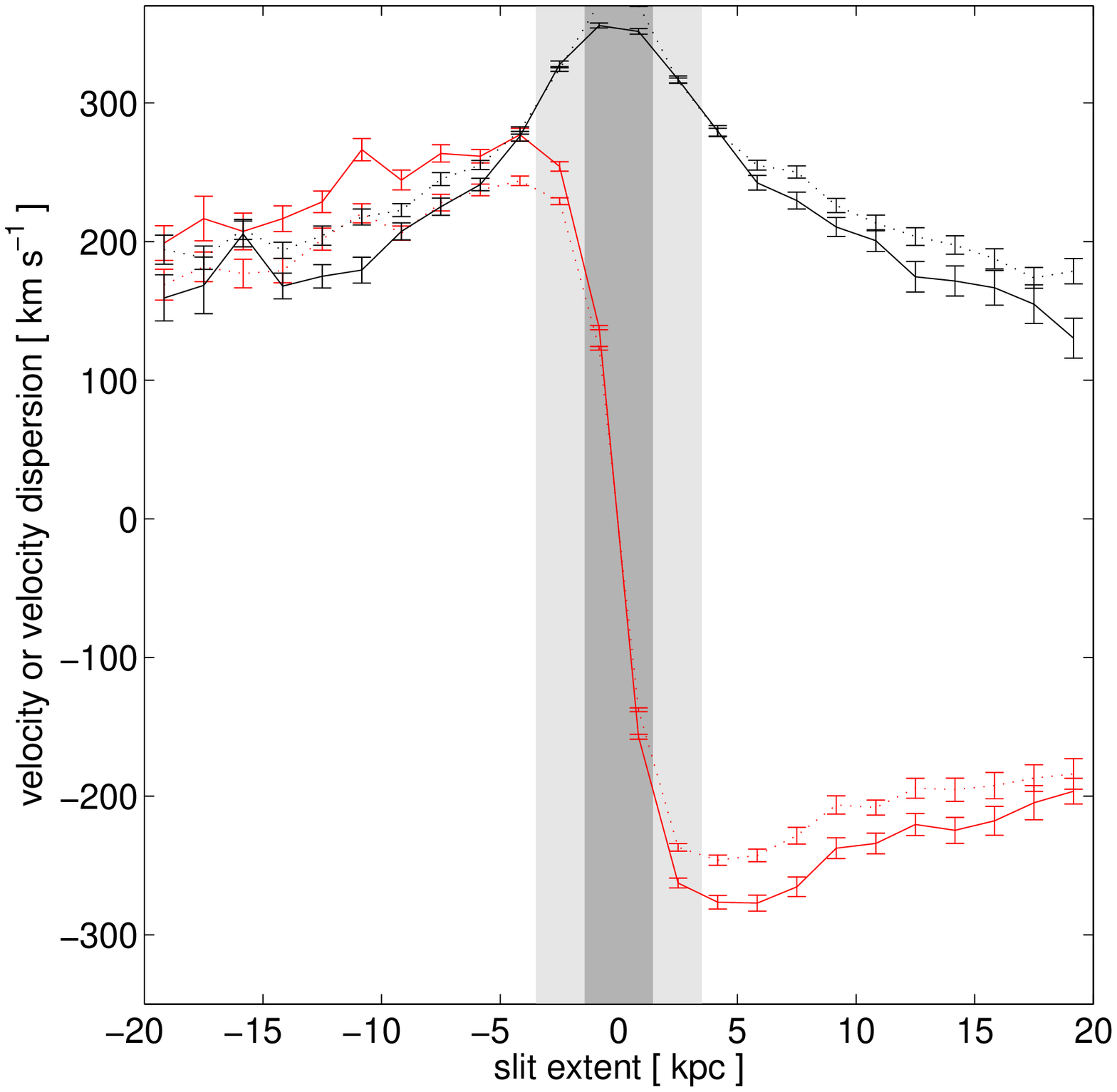}
\end{tabular}
\caption[Kinematic properties of the central galaxies at $z=0$]{The stellar rotation velocity (red) and velocity dispersion (black) along the axis which maximizes rotation velocity of the central objects in the three groups $G1$, $G2$ and $G3$ at $z=0$ (from left to right). The dashed curves show the velocities and velocities dispersions using moments of velocity, while the solid curves are calculated from fitting a 5th order Gauss-Hermite polynomial to the line-of-sight velocity distribution. The dark shaded area is affected by numerical resolution and excluded in the analysis. We quote in table \ref{tab:PropertiesZ0} the velocity dispersion measured within the light shaded area.\label{fig:VelocityProfile}}
\end{center}
\end{figure*}

In Figure \ref{fig:VelocityProfile} we plot the stellar rotation velocity and velocity dispersion along the intermediate axis of the moment-of-inertia tensor of three groups $G1-G3$. Projecting along this axis results in the \emph{largest} line-of-sight rotation velocity and we confirm that the stellar angular momentum is well aligned with the minor axis of the stellar component of the galaxy. The figure demonstrates that the central galaxies are supported by velocity dispersion. We note that the central galaxies in the groups $G2$ and $G3$ ($v/\sigma\approx{}0.7-0.9$) show a significant rotation component,  while the central galaxy in $G1$ is practically non-rotating. Interestingly, the last major mergers that the  central galaxies in the groups $G1$ and $G3$ undergo are between gas-poor, velocity dispersion supported stellar systems. We explain the relatively high rotational support of the remnant in the group $G3$ by the special orbital properties of merging galaxies. Indeed, a kinematic re-analysis of the binary merger simulations that have been presented in \cite{2008ApJ...684.1062F} shows that a gas-poor major merger on an eccentric orbit, e.g. with an apo-to-pericenter ratio of 6:1, typical for orbits of DM-halos in bound environment such as clusters and massive groups \citep{1998MNRAS.300..146G}, can also lead to $v/\sigma$ of order unity. Note that typical binary merger experiments assume parabolic orbits which are expected for mergers between field galaxies in isolated halos \citep{2006A&A...445..403K}, and which convert a lower amount of angular momentum into the spin of the merger remnant \citep{2006ApJ...650..791C}. 

The high masses of the simulated galaxies ($\sim{}4\times{}10^{11}M_\odot$ within a sphere of 20 kpc radius) and the scarcity of cold gas ($\lesssim{}10^9$ $M_\odot$ of gas colder than $3.2\times{}10^4$ K within a sphere of 20 kpc radius) are also consistent with the expectation of being early-type galaxies. We summarize the properties of the central galaxies at $z=0$ in tables \ref{tab:PropertiesZ0} and \ref{tab:PropertiesZ0Phot}. We conclude that based on visual appearance, lack of cold gas and rotational support, color, surface-mass profile and bulge-to-disk ratio, the central group galaxies resemble early-type galaxies. 

\begin{table*}
\begin{center}
\footnotesize
\caption[Structural and kinematic properties of the central group galaxies at $z=0$]{Structural and kinematic properties of the central group galaxies at $z=0$.\label{tab:PropertiesZ0}}
\renewcommand{\arraystretch}{1.25}
\begin{tabular}{ccccccccccc}
\tableline
\tableline
              &  $M_\mathrm{vir}$  & $R_\mathrm{vir}$ & $M_\mathrm{*}$               & $R_\mathrm{eff}$ & $R^\mathrm{deVau}_\mathrm{eff}$ & $R^\mathrm{Sersic}_\mathrm{eff}$  &     & $\sigma_\mathrm{eff}$ &    \\  
Group  & ($10^{12} M_\odot$) &  (kpc)   &    ($10^{10} M_\odot$)     &            (kpc)             &          (kpc)                                               &          (kpc)                                             &   $n^\mathrm{Sersic}$  & (km/s) & $v_\mathrm{rot}/\sigma_\mathrm{eff}$ \\
\\ \tableline
G1  & 11.0 & 447 & $39.6\pm{}3.6$  & $4.0\pm{}0.5$    &2.75$\pm$0.1  &2.50$\pm$0.1    &    5    & 299$\pm$20 & 0.20$\pm$0.05  \\
G2  & 12.0 & 459  &$44.5\pm{}3.6$  & $3.7\pm{}0.5$     &3.81$\pm$0.2   &3.81$\pm$0.2   &    4    & 326$\pm$20 & 0.79$\pm$0.05    \\
G3  & 15.6 & 502 &$42.9\pm{}3.6$  & $3.2\pm{}0.4$     &2.57$\pm$0.1   &1.57$\pm$0.1   &     $\gtrsim{}8$   & 323$\pm$20 & 0.86$\pm$0.06  \\
    \end{tabular}
\tablecomments{The first column lists the group name. Further columns denote: the virial mass and virial radius of the group, the stellar mass $M_\mathrm{*}$ within 20 physical kpc, the three-dimensional radius that contains half of $M_\mathrm{*}$, the half-mass radius from a deVaucouleurs-fit, the half-mass radius from a Sersic fit, the Sersic index, the stellar velocity dispersion within the effective radius, the ratio of stellar rotation velocity and velocity dispersion. Errors in $M_\mathrm{*}$ and $R_\mathrm{eff}$ are derived from varying the applied star formation correction in the range 6-16 M$_\odot$/yr. Errors in $R^\mathrm{Sersic}_\mathrm{eff}$ and $R^\mathrm{deVau}_\mathrm{eff}$ are formal fit errors. Errors in the kinematic properties are derived from a bootstrapping error analysis of the line-of-sight velocity data. Masses and sizes derived without the minimal star formation correction scheme can be found in Table \ref{tab:PropertiesZ0Corr}. In Table \ref{tab:ComparisonZ0d1} we compare the results of the simulations $G2$ and $G2-HR$ at $z=0.1$.}
\end{center}
\end{table*}
\normalsize

\begin{table*}
\begin{center}
\footnotesize
\caption{Photometric properties of the central group galaxies at $z=0$.\label{tab:PropertiesZ0Phot}}
\renewcommand{\arraystretch}{1.25}
\begin{tabular}{cccccccc}
\tableline
\tableline
                    &$R^\mathrm{Petrosian}$ & $R^\mathrm{Petrosian}_\mathrm{eff}$ &                         &                       &                        &            \\ 
       Group &(kpc) & (kpc)                                                                                                    & $^{0.1}(g-r)$ & $^{0.1}(r-i)$ & $^{0.1}M_i$ & $M_r$\\
\\ \tableline
G1 &   $6.9\pm{}1.3$ &  $3.0\pm{}0.5$  & $0.87\pm{}0.02$ & $0.36\pm{}0.02$ & $-23.51\pm{}0.11$  & $-23.34\pm{} 0.11$ \\
G2 &   $6.7\pm{}1.8$ &  $2.6\pm{}0.6$ & $0.81\pm{}0.02$ & $0.34\pm{}0.02$ & $-23.87\pm{}0.10$  & $-23.72 \pm{} 0.10$ \\
 G3 &  $4.8\pm{}0.5$ &  $2.1\pm{}0.4$ & $0.86\pm{}0.02$  & $0.35\pm{}0.02$ &  $-23.54\pm{}0.11$ & $-23.38\pm{}0.11$  \\
\end{tabular}
\tablecomments{The first column lists the group name. Further columns denote: the Petrosian radius in the SDSS r-band, the radius containing half of the light within an annulus of 2 Petrosian radii, $^{0.1}(g-r)$ color, $^{0.1}(r-i)$ color, and $^{0.1}i$ band magnitude in the redshifted SDSS u,g,r,i,z system \citep{2003ApJ...594..186B} and the restframe SDSS r-band magnitude. Errors include variations due to different viewing angles and due to a star formation correction in the range 6-16 M$_\odot$/yr.}
\end{center}
\end{table*}
\normalsize

Our simulated central galaxies have larger masses and velocity dispersions and/or smaller sizes \citep{2003ApJ...590..619M}, when compared with galaxies observed in the nearby Universe (\citealt{2003MNRAS.343..978S}, \citealt{2008ApJ...688...48V}, Fig 3). The compactness of our objects is less extreme, however, than e.g. in \cite{2003ApJ...590..619M} (the effective densities being an order of magnitude lower), which we ascribe to our efficient supernova blastwave model and, partially, to the increased resolution. In order to follow the average mass-size relation of local early-type galaxies \citep{2003MNRAS.343..978S} or local central group galaxies \citep{2009arXiv0901.1150G} the simulated central galaxies would need to reduce their mass by a factor of 2-3. This is also the amount necessary to put the galaxies right onto the red-sequence in the  $^{0.1}(g-r)$ and $^{0.1}(r-i)$ vs. $^{0.1}M_i$ color-magnitude diagrams of nearby galaxies \citep{2003ApJ...594..186B}. Similar to what is obtained from simulations of central galaxies in galactic halos \citep{2009arXiv0903.1636N}  our objects appear to be too massive, by about a factor of 2-3, for their halo mass when compared with weak lensing observations \citep{2006MNRAS.368..715M}, or compared to SDSS groups \citep{2007ApJ...671..153Y, 2008ApJ...676..248Y, 2009arXiv0901.1150G}, although this comparison is hampered by the fact that masses are determined in observations and simulations in a very different manner and a systematic (but not statistical) uncertainty of about a factor of 2 is not unlikely. Finally, in Fig. \ref{fig:IntegratedMassFunction} we plot the integrated mass function of \cite{2004ApJ...600..681B}. If we assume a 1:1 relationship between central galaxy and their hosting dark matter halo we can associate the number densities of galaxies with the number densities of their parent dark matter halos. The central galaxy in a $\sim{}10^{13}$ $M_\odot$ dark matter halo \cite{2002MNRAS.336..112M} is then expected to have a stellar mass of $\sim{}2\times{}10^{11}$ $M_\odot$, i.e. a factor of 2 smaller than the masses of the central group galaxies in the simulations. Taken together these results indicate that the masses and sizes of the central group galaxies in high-resolution cosmological simulations are still somewhat biased towards larger masses and/or smaller sizes, reminiscent of the size problem in simulations of disk galaxies. This is not entirely unexpected. The size problem is usually associated with the loss of angular momentum due to inefficient resolution at \emph{high} redshift when the low-mass progenitor galaxies are resolved by only a small number of particles and with the particular feedback implementation (see \citealt{2008ASL.....1....7M} for a recent review). Hence, it is likely that our simulations are also affected to some extent by this problem. Nonetheless, it should be stressed that reducing the stellar mass by a factor 2-3 would bring our galaxies in good agreement with observations and we note that this factor is likely not much larger than the systematical uncertainties in the masses determined from observations. To summarize: Our simulations produce massive early-type galaxies at the group centers that compare reasonable well with observations although they are somewhat biased w.r.t. masses, sizes, colors and/or magnitudes. Possibly a combination of higher resolution and an increased star formation threshold compatible with the cold molecular phase and/or additional feedback mechanisms such as AGN feedback could resolve these remaining discrepancies (\citealt{2003ApJ...590L...1K}, \citealt{2008ApJ...680.1083R}, \citealt{2008PASJ...60..667S}, \citealt{Governato2009}).

%%%%%%%%%%%%%%%%%%%%%%%%%%%%%%%%%%%%%%%%%%%%%%%%%%%%%%
\subsection{The $z\sim{}2$ progenitors of $z=0$ central group galaxies}

The main progenitors at $z\gtrsim{}2$ of our studied central group galaxies have very compact ($<1$ kpc) stellar components and given our spatial resolution we cannot reliably estimate their structural or kinematic properties. However, global properties such as total mass or luminosity can be robustly measured. We therefore classify the main progenitors at high redshifts based on colors, SFR and cold gas content. In addition, we compare their halo masses with observational estimates and clustering measurements.

\begin{figure}
\begin{center}
\includegraphics[width=80mm]{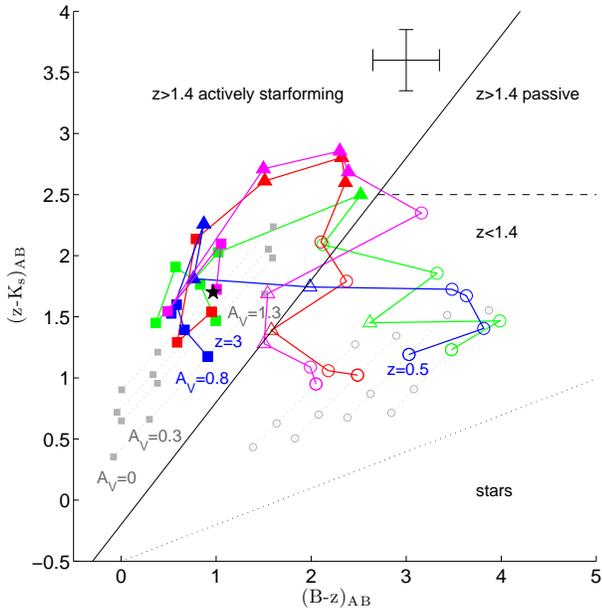}
\caption[The BzK colors of the main central galaxy progenitors between $z=3$ and $z=0.5$]{The BzK colors of the main progenitors of the central group galaxies within a projected radius of 8 kpc between $z=3$ and $z=0.5$. The different lines corresponds to $G1$ (green), $G2$ (red), $G3$ (blue) and $G2-HR$ (magenta). The gray symbols indicate the change of the colors due to different extinction corrections \citep{2000ApJ...533..682C} with $A_V$ ranging from 0 to 1.3. The color evolution is shown for the default value $A_V=0.8$. Filled (empty) symbols denote $z>1.4$ ($z<1.4$). Squares, triangles and circles indicate that the specific star formation rate within 20 kpc is $>0.5$ Gyr$^{-1}$, between 0.2 and 0.5 Gyr$^{-1}$ or below 0.2 Gyr$^{-1}$, respectively. The black star indicates the BzK colors of a typical $z\sim{}2$ star forming galaxy \citep{2006Natur.442..786G}. The error bar at the top shows the maximum changes that result from changing the projection direction and when the minimal star formation correction scheme is applied. Satellite galaxies that happen to lie along the line of sight are excluded because their presence can affect the overall colors.\label{fig:BzK}}
\end{center}
\end{figure}

Frequently used classification and selection schemes for high redshift galaxies are the BzK color-color classification \citep{2004ApJ...617..746D}, the BM/BX selection \citep{2004ApJ...607..226A, 2004ApJ...604..534S} and the selection as distant red galaxies (DRG) \citep{2003ApJ...587L..79F}. In Fig.~\ref{fig:BzK} we show the evolution of the colors of the main progenitors in the BzK plane between $z=3$ and $z=0.5$. At $z>2$ the galaxies fall into the star forming BzK (sBzK) regime consistent with their high star formation rates of $20-60 M_\odot$ yr$^{-1}$ and specific star formation rates of $>0.2$ Gyr$^{-1}$. The colors at $z\sim{}2$ ($B-z\sim{}1$, $z-K_s\sim{}2$) are typical of star forming galaxies observed at those redshifts, e.g. \cite{2009arXiv0903.1872F}. At $z\sim{}1.3$ the galaxies cross the line in the BzK plane that is typically used to select either $z<1.4$ galaxies or $z>1.4$ star forming galaxies. 
%Around this time the specific star formation rate drops below $0.2$ Gyr$^{-1}$. 
None of our galaxies classifies at any time as a proper passively evolving BzK (pBzK) galaxy, although the central galaxies in the groups $G1$ and $G2$ come temporarily close to the pBzK area at $z\sim{}1.5$. \cite{2008ApJ...682..896K} have shown that massive $K$-bright galaxies observed at $z=2-3$ divide into (i) red, post-burst galaxies that form a red sequence in a mass vs. $(U-B)_\mathrm{rest-frame,Vega}$ plane and (ii) blue star forming galaxies with small star formation timescales. The blue $(U-B)_\mathrm{rest-frame,Vega}$ colors of our objects at $z>2$ (-0.2 to 0) would put them into the second category. At $z\sim{}1.5-2$, however, the central galaxies of the groups $G1$ and $G2$ are redder ($>0.1$) and could be classified as post-burst galaxies. In Fig.~\ref{fig:DRG} and Fig.~\ref{fig:BMBX} we show how the progenitor galaxies would classify according to the DRG and BM/BX scheme, respectively. These selection schemes are somewhat more susceptible to dust extinction and we find that our classification outcome depends on the adopted extinction value. For $A_V\lesssim{}1$ (and in particular for the assumed value $A_V=0.8$, see section \ref{sect:Methodology}) our simulated progenitors classify as BX galaxies in the redshift range $z\sim{}2.7$ and $z\sim{}2$. The less massive progenitor galaxies (the progenitors of the central galaxies in the groups $G1$ and $G3$) qualify also as BM galaxies below $z=2$ down to $z\sim{}1.7$ ($G1$) and $z\sim{}1.5$ ($G3$). Only if the progenitors were strongly dust obscured ($A_V\gtrsim{}1.3$) they would qualify at $z>2.3$ as distant red galaxies  \citep{2003ApJ...587L..79F}. We summarize the photometric properties in Table \ref{tab:PropertiesZ25Phot}.

Masses and star formation rates of the central galaxies at $z\sim{}2$ are summarized in Table \ref{tab:PropertiesZ25}. Consistent with the blue colors are the significant star formation rates, see Fig. \ref{fig:SFR}, which are rapidly declining with time. At $z\sim{}2$ they still  amount to 20-60 $M_\odot$ yr$^{-1}$. The progenitor galaxies also host a significant reservoir of cold gas ($5\times{}10^{9}$ $M_\odot$ at $z\sim{}2-2.5$  within a 10 kpc radius) which is typically arranged in either a gas disk of $\sim{}3$ kpc radius (in $G2$ and $G3$) or it has a more irregular morphology ($G1$). When we look at the kinematic properties of this cold gas disk in our best resolved object ($G2-HR$) we measure a line-of-sight velocity dispersion of 200 km/s and a two times higher rotation velocity in the gas disk. 
We note that compared to e.g. the star forming galaxy reported in \cite{2006Natur.442..786G} the cold gas masses in our objects are significantly lower (by about an order of magnitude) and our disk rotation speed is larger (factor of 2) and reaches its maximum value at smaller radii. The latter is likely a consequence of the fact that the simulated galaxies are rather compact with stellar half-light radii below 1 kpc. We note that also about 1/3 of massive, star forming galaxies seen at $z\sim{}1.5-2.5$ are compact and have high velocity dispersions \citep{2009arXiv0903.1872F}. On the other hand, the size difference at $z\sim{}1.5$ between the simulations $G2$ and $G2-HR$ (see Fig. \ref{fig:MassSizevsTime}) indicates that our galaxies might suffer to some degree from artificially enhanced angular momentum loss, often seen in simulations of disk galaxies, despite the fact that our model invokes relatively energetic supernova feedback. Clearly, this issue needs further studies at higher resolution. Overall, we find that the colors and star formation rates of the simulated galaxies match the properties of optically/UV (BM/BX) selected $z\sim{}2$ star forming galaxies \citep{2006Natur.442..786G, 2008ApJ...687...59G} or that of sBzK galaxies \citep{2004ApJ...617..746D}, but not that of high redshift ``red-and-dead`` galaxies.

Measurements of the angular correlation functions allow to determine the typical halo masses in which sBzK galaxies of a given magnitude reside \citep{2006ApJ...638...72K,2007ApJ...660...72H, 2008ApJ...681.1099B, 2008MNRAS.391.1301H}. The parent halo mass increases rapidely with K-magnitude \cite{2007ApJ...660...72H}: by more then 2 orders of magnitude per 2.5 mag in K brightness. The brightest sBzK galaxies ($K_s<21$) thus populate halos comparable to that of pBzK selected galaxies. Typical halo masses of sBzK galaxies reported in the literature are: $2.8\times{}10^{11}  M_\odot$ for a $K_s<23.2$ sample \citep{2007ApJ...660...72H}, $6\times{}10^{11}  M_\odot$ for a $K_s<23$ sample \citep{2008MNRAS.391.1301H} and $\sim{}10^{13} M_\odot$ for a $K_s<22$ sample \citep{2006ApJ...638...72K,2007ApJ...660...72H,2008ApJ...681.1099B}. The halo masses ($0.5-1.8\times{}10^{12} M_\odot$) and $K_s$-band magnitudes ($21.5-22.4$) of the simulated central galaxies are consistent with their identification as sBzK galaxies of intermediate brightness and star formation intensity at $z\sim{}2$.

\begin{table*}
\begin{center}
\footnotesize
\caption{Photometric properties of the central galaxy progenitors at $z\sim{}2$.\label{tab:PropertiesZ25Phot}}
\renewcommand{\arraystretch}{1.25}
\begin{tabular}{ccccccccc}
\tableline
\tableline

Group & $z$ & $U_n$-G & $G-\mathcal{R}$ & $\mathcal{R}$ & $J_s - K_s$ & $B-z$  & $z-K_s$ & $K_s$ \\
\tableline

$G1$ & 2.4 & 0.57 & 0.25 & 24.3$_{-1.7}^{+1.1}$ & 1.20 & 0.58 & 1.91 & 22.2$_{-0.7}^{+0.4}$ \\
& 2.0 & 0.40 & 0.16 &  23.1$_{-1.6}^{+1.1}$ & 0.62 & 0.37 & 1.45 & 21.5$_{-0.6}^{+0.4}$  \\
& 1.5 & 0.86 & 0.98 & 24.5$_{-1.4}^{+0.9}$  & 0.96 & 2.52 & 2.50 & 20.8$_{-0.5}^{+0.3}$ \\ 
\tableline

$G2-HR$ & 2.4 & 0.57 & 0.22 & 23.2$_{-1.6}^{+1.1}$ & 1.02 & 0.49 & 1.55 & 21.5$_{-0.7}^{+0.4}$ \\
& 2.0 & 0.86 & 0.56 & 24.9$_{-1.7}^{+1.0}$ & 1.11 & 1.50  & 2.71 & 21.5$_{-0.7}^{+0.4}$ \\
& 1.5 & 0.52 & 0.87 & 24.6$_{-1.5}^{+0.9}$ & 1.10 & 2.43 & 2.69 & 20.7$_{-0.5}^{+0.3}$ \\
\tableline
$G3$ & 2.4 & 0.61 & 0.26 & 24.2$_{-1.7}^{+1.1}$ & 0.99 & 0.59 & 1.6 & 22.4$_{-0.6}^{+0.5}$  \\
& 2.0 & 0.51 & 0.25 & 23.6$_{-1.6}^{+1.1}$ & 0.62 & 0.53 & 1.53 & 21.9$_{-0.6}^{+0.4}$\\
& 1.5 & 0.17 & 0.22 & 23.0$_{-1.5}^{+0.9}$ & 0.77 &  0.76 & 1.81 & 20.7$_{-0.5}^{+0.3}$\\
\tableline
\end{tabular}
\tablecomments{The first columns list the group name and the redshift of the simulation. The next columns contain the colors and magnitudes (all in the AB system)  of the main progenitors of the $z=0$ central group galaxies in various filter bands used for selecting high-redshift galaxies. $U_\mathrm{n}$, $G$ and $\mathcal{R}$ filter bands are extracted from Fig. 1 of \cite{2003ApJ...592..728S} and compare well with the transmission curves of \citep{2004ApJ...607..226A}. $J_s$ is a slightly modified $J$ filter \citep{2003AJ....125.1107L}. The $J_s$ and $K_{s,2}$ transmission curves are downloaded from the ISAAC (VLT) website. We further use Bessel $B$, $z$ (Subaru telescope), $K_{s,1}$ (Kitt Peak 4-m telescope) filter bands. We use the $K_{s,1}$  filter in order to compute $z-K$ colors and the $K_{s,2}$ filter for calculating $J_s-K_s$ colors. Colors and $K_s$ magnitudes are affected by much less than 0.1 mag when switching between $K_{s,1}$ and $K_{s,2}$ filter bands. Colors and magnitudes are measured in an observed frame and within a projected radius of 8 kpc ($\sim{}1\arcsec$) roughly mimicking the observations of \cite{2003ApJ...587L..79F,2003AJ....125.1107L, 2003ApJ...592..728S, 2004ApJ...604..534S, 2007ApJ...660...72H}. An extinction law \citep{2000ApJ...533..682C} with $A_V=0.8$ is assumed. The quoted error in the magnitude results from variations of $A_V$ in the range $0-1.3$ which dominate the uncertainties in the magnitudes and colors. Since colors are strongly correlated with each other we do not quote corresponding uncertainties for them. Instead we refer to Fig. \ref{fig:BzK}, Fig. \ref{fig:DRG} and Fig. \ref{fig:BMBX} that demonstrate the impact of varying $A_V$ on apparent colors and magnitudes. At fixed $A_V$ variations in color and magnitude due to the chosen projection directions and the amount of SF correction in the minimal correction scheme (see Appendix \ref{sec:ResolutionTest}) typically amount to $\lesssim{}0.1$ mag.}
\end{center}
\end{table*}
\normalsize

\begin{table*}
\begin{center}
\footnotesize
\caption{Structural properties of the central galaxy progenitors at $z\sim{}2$.\label{tab:PropertiesZ25}}
\renewcommand{\arraystretch}{1.25}
\begin{tabular}{ccccccccc}
\tableline
\tableline
 & &                     $M_\mathrm{vir}$ & $R_\mathrm{vir}$ &                     $M_\mathrm{tot}$ & $M_\mathrm{star}$               & $M_\mathrm{gas}$    & $M_\mathrm{cgas}$     &   SFR           \\ 
Group & $z$ &   ($10^{12} M_\odot$) & (kpc)                 &                    ($10^{10}M_\odot$)  &    ($10^{10}M_\odot$)   & ($10^{10}M_\odot$)  &  ($10^{10}M_\odot$)  &  ($M_\odot$ yr$^{-1}$) \\
\tableline
$G1$& 2.4 & 0.69 & 53 & 16.7 & 4.9 & 1.2 & 0.39 & 38  \\
& 2.0 & 1.04 & 68 & 17.8 & 5.7 & 1.0 & 0.35 & 26  \\
& 1.5 & 1.38 & 89 & 23.2 & 9.9 & 0.6 & 0.19 & 16  \\ 
\tableline

$G2-HR$& 2.4 & 1.52 & 69 & 27.1 & 8.6 & 1.4 & 0.41 & 48  \\
& 2.0 & 1.82 & 82 & 28.2 & 9.5 & 0.8 & 0.18 & 22  \\
& 1.5 & 2.55 & 110 & 31.9 & 13.0 & 0.6 & 0.07 & 14  \\
\tableline
$G3$& 2.4 & 0.49 & 47 & 11.8 & 2.1 & 1.3 & 0.62 & 14   \\
& 2.0 & 0.61 & 57 & 16.9 & 6.2 & 1.6 & 0.71 & 60 \\
& 1.5 & 1.28 & 87 & 21.0 & 8.5 & 1.1 & 0.46 & 26 \\
\tableline
\end{tabular}
\tablecomments{The first columns list the group name, redshift, virial mass and physical virial radius of the parent halo of the main progenitors of the $z=0$ central group galaxies. The next columns show total mass, stellar mass, gas and cold gas ($T<3.4\times{}10^4$ K) mass within a sphere of 10 kpc radius ($\sim{}1\arcsec$) around the object's center. The last column contains the star formation rate within a spherical radius of 20 kpc around the progenitor galaxy, but excluding contributions from substructure. Cited stellar masses and SFR are corrected according to the minimal correction scheme which changes the stellar mass by $\lesssim{}10\%$. The mass subtracted from the stellar component is not added to the gas masses. The effective radius of the stellar component is not resolved, but smaller than $\sim{}1$ kpc. }
\end{center}
\end{table*}
\normalsize

%%%%%%%%%%%%%%%%%%%%%%%%%%%%%%%%%%%%%%%%%%%%%%%%%%%%%%
\subsection{Evolution of masses, sizes and densities}
\label{sect:MassSizeDens}

\subsubsection{Total masses and effective radii}

The left panel of Fig. \ref{fig:MassSizevsTime} presents the growth of stellar masses and effective radii of the central group galaxies as function of redshift. Mass and size evolution are closely linked and periods of slow/fast mass build-up correspond to periods of slow/fast size growth. The central galaxies in the groups $G1$ and $G3$ undergo at least two major galaxy mergers ($z\sim{}1$, $z\sim{}0.4$ in case of $G1$ and $z\sim{}0.8$, $z\sim{}0.1$ in the case of $G3$). Two of the 4 major mergers occur between rotation-dominated, disky, non gas-poor ($f_\mathrm{gas/stellar}$ of a few percent) galaxies, while the other two mergers take place later between an already gas-poor central galaxy and another gas-poor, non-rotation supported companion. The central galaxy in $G2$ ($G2-HR$) does not experience any major mergers below $z=4$.  

The overall behaviour of the mass-size evolution is shown in the right panel of Fig. \ref{fig:MassSizevsTime}. It roughly follows a relation $R\propto{}M^\alpha$. The large jumps with $\alpha<1$ are major mergers, consistent with estimates from binary merger simulations \citep{2006MNRAS.369.1081B}. Periods of minor merging and star formations show a continuos growth in mass and effective radius with $\alpha\gtrsim{}1$. We perform a robust linear regression (with a bisquare weighting scheme) in the $\log{}(M)-\log{}(R_\mathrm{eff})$ plane in order to determine the average value of $\alpha$ over the redshift range $z\sim{}0-1$, see Table  \ref{tab:RegressParams}. We now summarize the result of this analysis.

The two major merger below $z<0.5$ have $\alpha=0.25$ and  $\alpha=0.90$. In phases without major merging activities we identify three mechanisms that drive significant size growth at small or only moderate mass growth. These processes are (i) minor merging, (ii) non-central star formation and (iii) a redistribution of either pre-accreted or pre-formed stellar material. The latter process may originate in a physical mechanism, such as tidal heating due to orbiting satellites, and/or is caused by spurious numerical effects. To assess whether the latter is the case we compare the intermediate resolution $G2$ and its high resolution analogue  $G2-HR$ and find a resolution dependence of $\alpha$. More precisely, at higher resolution  the central galaxy seems to grow slower in size for a given mass ($\alpha\sim{}1$) compared to the corresponding simulation at intermediate resolution ($\alpha\sim{}2$). This behavior is partially explained by the fact that compared with $G2$ the central galaxy in $G2-HR$ is \emph{larger} at high redshift ($z>1$) but of similar size at $z\sim{}0$. It is clear that further work is necessary in order to tie down all resolution dependent effects that potentially contribute to this difference.

We conclude that major mergers alone typically result in a slow size growth $(\alpha<1)$. Group $G2$ is subject to both substantial minor merging and star formation, and experiences a faster size growth ($\alpha\sim{}1$ in the highest resolution simulation). Finally, groups $G1$ and $G3$ show an even faster size evolution at nearly constant mass in-between major mergers. At least parts of this evolution could be of spurious numerical origin. However, we do not exclude that, e.g., heating by tidal shocks from orbiting satellites could contribute to this redistribution of stellar mass and thus drive a very fast size evolution, see section \ref{sect:DriverSizeGrowth}.

\begin{table}
\begin{center}
\footnotesize
\caption[Linear regression parameters of the mass-size evolution of the central galaxies]{Linear regression parameters of mass-size evolution.\label{tab:RegressParams}}
\renewcommand{\arraystretch}{1.25}
\begin{tabular}{ccccc}
\tableline
\tableline
    & SF$_\textrm{corr}$                   &                                    &          \\
Group & ($M_\odot{}$ yr$^{-1}$)  & $z_\mathrm{max}$ & $\alpha$ \\ \tableline
\tableline
G1       &   0                                        & 0.54                           & 1.44$\pm{}$0.05 \\
G2       &   0                                        & 0.40                           & 2.06$\pm{}$0.03 \\
G3       &   0                                        & 0.69                           & 1.03$\pm{}$0.03 \\
G2-HR &  0                                        & 1.6                             & 1.03$\pm{}$0.02 \\
\tableline
G1       &   6                                        & 0.82                           & 1.56$\pm{}$0.04 \\
G2       &   6                                        & 0.54                           & 2.09$\pm{}$0.02 \\
G3       &   6                                        & 0.82                           & 0.97$\pm{}$0.04 \\
G2-HR &  6                                        & 1.6                             & 1.04$\pm{}$0.02 \\
\tableline
G1       &   11                                        & 1.00                         & 1.56$\pm{}$0.04 \\
G2       &   11                                        & 0.58                         & 2.20$\pm{}$0.03 \\
G3       &   11                                        & 0.82                         & 0.95$\pm{}$0.06 \\
G2-HR &  11                                        & 1.6                           & 0.98$\pm{}$0.03 \\
\tableline
G1       &   16                                        & 1.00                         & 1.48$\pm{}$0.05 \\
G2       &   16                                        & 0.64                         & 2.24$\pm{}$0.04 \\
G3       &   16                                        & 0.79                         & 0.88$\pm{}$0.09 \\
G2-HR &  16                                        & 1.6                           & 0.95$\pm{}$0.04 \\
\tableline
\end{tabular}
\tablecomments{The columns denote: (1) the name of the simulated galaxy group, (2) the amount of central star formation subtracted according to the minimal correction scheme (a value of 0 means no corrections, while we use SF$_\textrm{corr}$=11 is the default value), (3) the highest redshift at which the effective radius is resolved (i.e. $R_\mathrm{eff}\geq{}2\times{}\epsilon_\mathrm{bar}$), (4) the best fit exponent of a robust linear regression to the mass-size relation from $z_\mathrm{max}$ to $z\sim{}0$ and its error (one standard deviation).}
\end{center}
\end{table}
\normalsize

\begin{figure*}
\begin{center}
\begin{tabular}{cc}
\includegraphics[width=80mm]{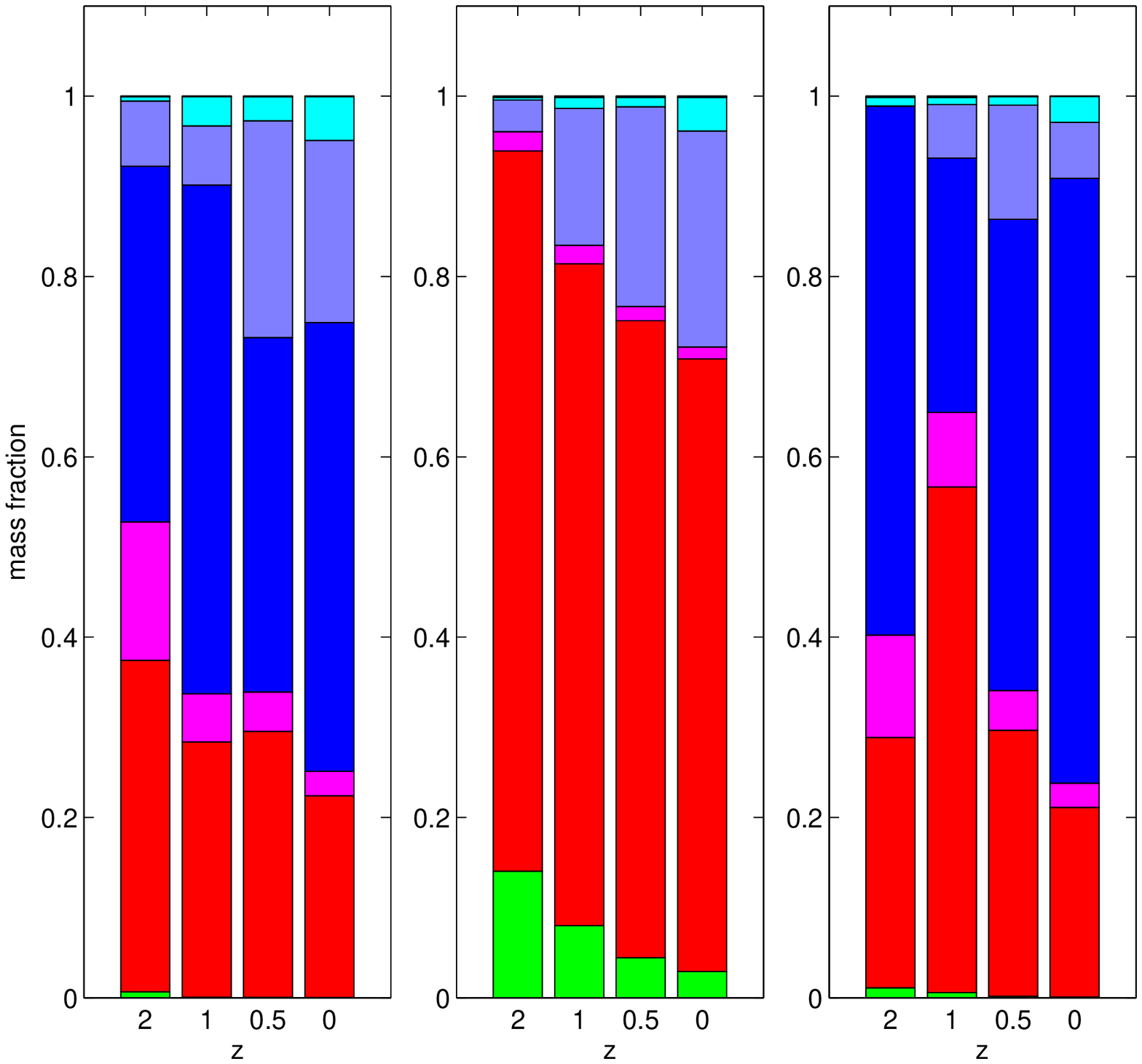} &
\includegraphics[width=80mm]{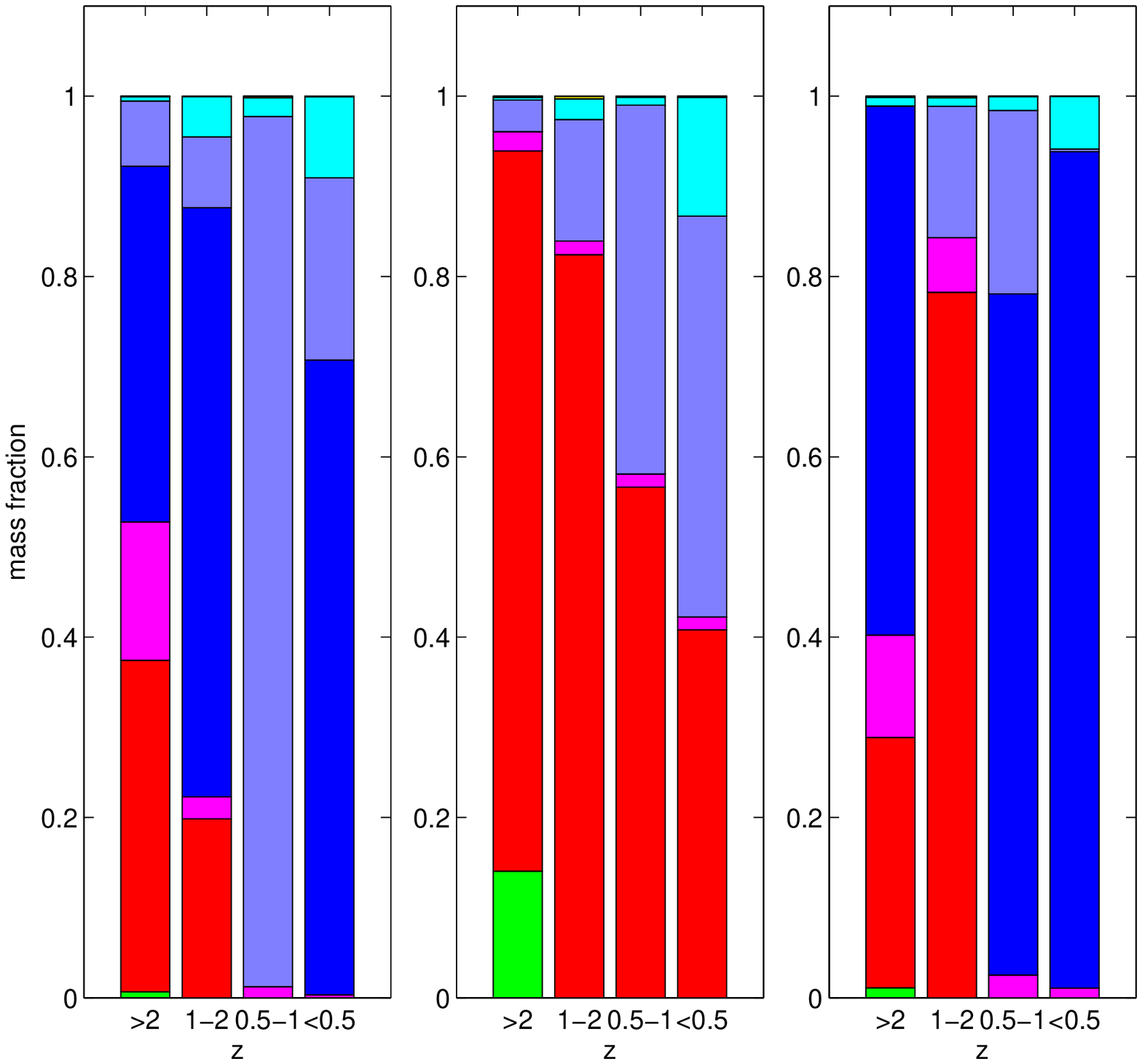} 
\end{tabular}
\caption[The origin of the stellar mass contained within 20 kpc of the central galaxies]{The origin of the stellar mass contained within 20 kpc of the central galaxies. The stellar mass fractions are normalized to 1. (Left panel) Stellar particles that are found in the central galaxy at a given specified redshift. (Right panel) Stellar particles that have been accreted or formed in the specified redshift interval. In each panel the sub-diagrams refer to $G1$ (left), $G2$ (middle) and $G3$ (right). The different colors correspond to a different origin of the stellar mass: formed before $z=4$ (green), in situ star formation (red), star formation in a merging satellite (magenta), formed in a satellite and accreted in a major merger (dark blue) or minor merger (light blue), formed in a satellite and accreted after being stripped from the satellite (cyan). The fraction of accreted stars formed in unresolved substructures is negligible. \label{fig:AccOrigin}}
\end{center}
\end{figure*}

\begin{figure*}
\begin{center}
\begin{tabular}{cc}
\includegraphics[width=80mm]{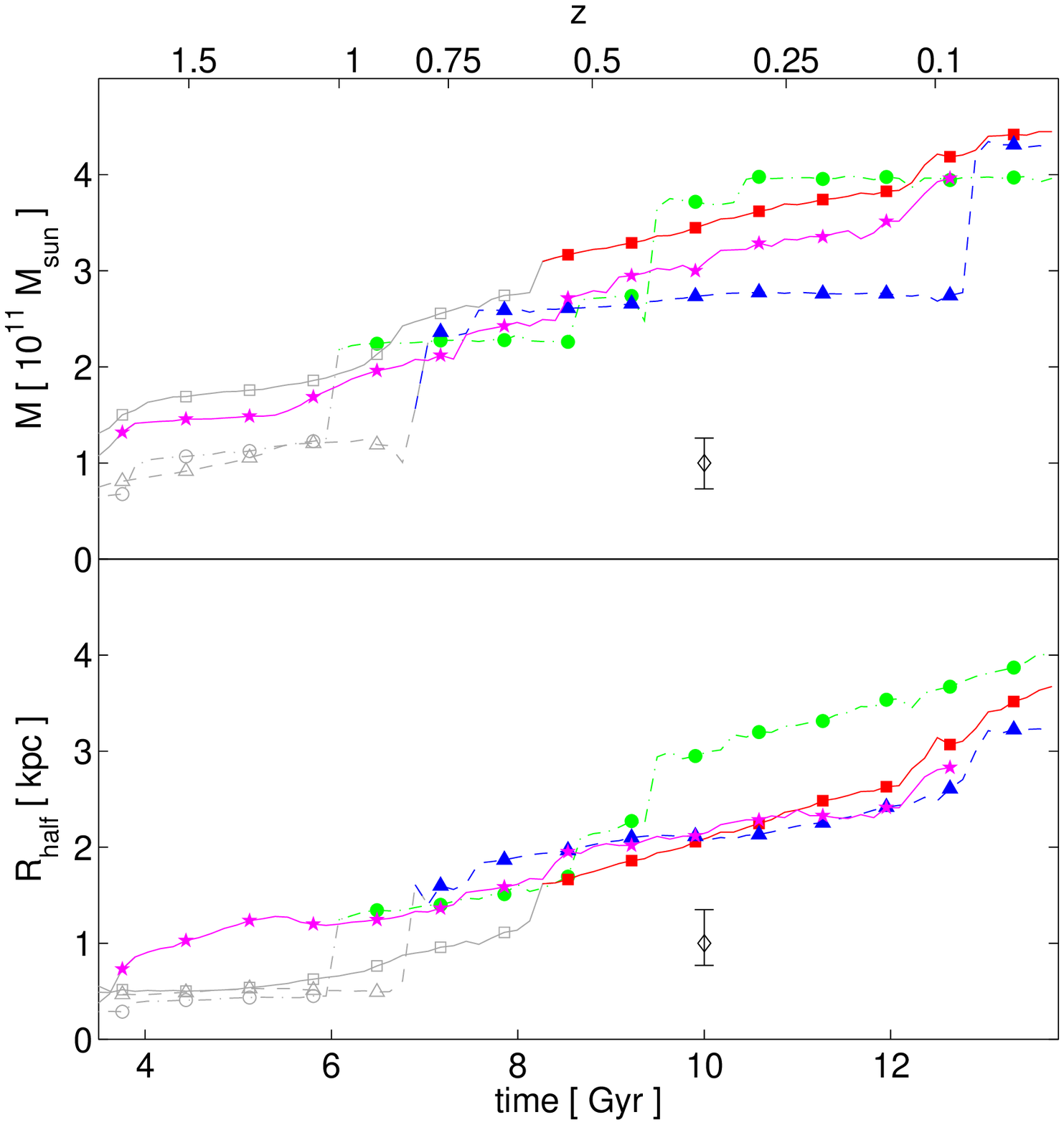} & \includegraphics[width=80mm]{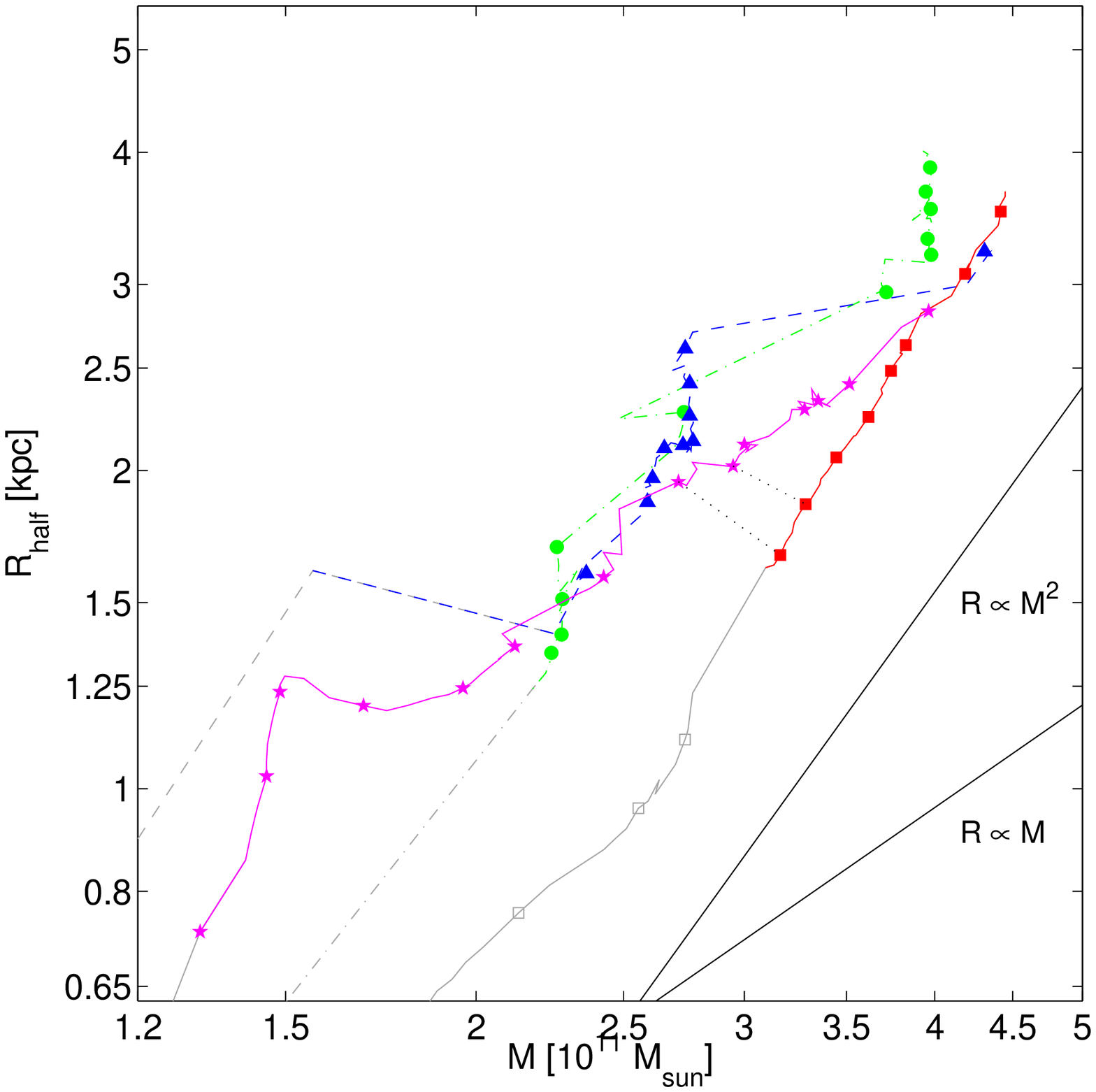} \\
\end{tabular}
\caption[Masses and half-mass radii of the central galaxies]{Masses and half-mass radii of the central galaxies in the simulations $G1-G3$ and $G2-HR$. The left plot shows the stellar mass within 20 kpc (top panel) and the half-mass radii (bottom panel) as function of age of the universe (bottom axis) or redshift (top axis). The results for simulations $G1$ (green dot-dashed), $G2$ (red solid), $G2-HR$ (magenta solid) and $G3$ (blue dashed) are indicated. Grey lines are measurements of masses and sizes that are not properly resolved. The point with error bars indicates the typical effect at $t\sim{}10$ Gyr of varying the residual SFR between 6, 11 (our fiducial value, see Appendix \ref{sec:ResolutionTest}) and 16 $M_\odot\,\mathrm{yr}^{-1}$. The right plot depicts the mass-radius evolution of the different galaxies. The symbols correspond to the specific times in left plot.\label{fig:MassSizevsTime}}
\end{center}
\end{figure*}

In Fig. \ref{fig:AccOrigin} we decompose the stellar mass found in central galaxies at given redshift into mass accreted by major/minor merging, mass produced by in situ star formation in the central galaxy, mass produced in the satellite while it merges with the central galaxy, mass accreted after being stripped from a satellite and mass accreted that has been formed in an unresolved substructure outside the central object. As the left panel shows only a small amount of stellar mass is accreted smoothly, e.g. either after being stripped from a satellite or after being formed in an unresolved substructure. Most of the stellar mass is accreted by major or minor mergers or is produced by in situ star formation within the central galaxy. Group $G2$ which does not experience a single major galaxy merger below $z=4$ forms $70-90\%$ of its stellar mass in situ, while both $G1$ and $G3$ accrete about $70\%$ of their final mass by merging. Minor merging contributes at the $5-25\%$ level to the mass build-up of the central group galaxies. The right panel shows that the contribution of in situ star formation is overall reduced compared to the effect of merging for $z<1$, indicating that the star formation rate drops faster than the merger rate, as predicted by semi-analytic modeling \citep{2008MNRAS.384....2G}. 
It is clear that a larger sample of galaxies with representative merger and gas accretion histories will be required in order to pin down in a statistical manner the contributions from the various processes (major merging, minor merging or star formation) to the mass build-up.

\subsubsection{Central and effective densities}

\begin{figure}
\begin{center}
\begin{tabular}{c}
\includegraphics[width=80mm]{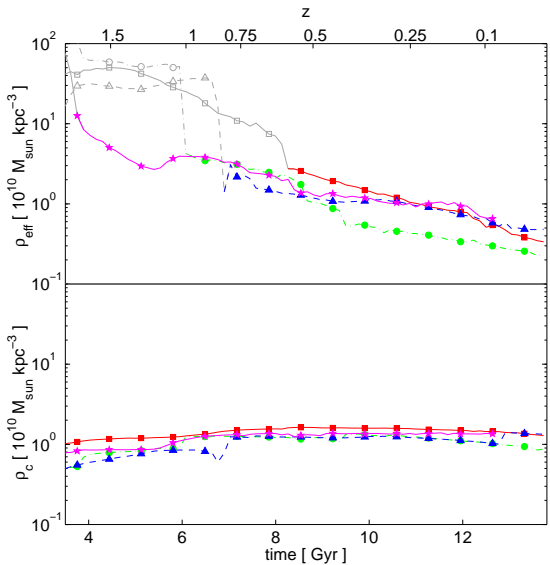}
\end{tabular}
\caption[Evolution of the stellar densities of the central galaxies]{Evolution of the stellar densities in the simulations $G1-G3$ and $G2-HR$. (Top panel) the effective density $\rho_\mathrm{eff}=M(<r_\mathrm{eff})/V(<r_\mathrm{eff})$ as function of time. The density evolution is determined by both the change of the effective radius and the total stellar mass of the central galaxy. (Bottom panel) The central density $\rho_\mathrm{c}=M(<2\,\mathrm{kpc})/V(<2\,\mathrm{kpc})$ remains roughly constant indicating that the mass within the central 2 kpc of the central galaxy does not change much with time. Varying the residual SFR correction by $\pm{}5\,M_\odot\,\mathrm{yr}^{-1}$ results to first order in an overall shift of the plots by $\sim{}0.3$ dex, but does not change the general behaviour. Symbols are as in Fig. \ref{fig:MassSizevsTime}.\label{fig:RhovsTime}}
\end{center}
\end{figure}

In Fig. \ref{fig:RhovsTime} we plot the evolution of the effective and central density in the central galaxies as a function of time. The effective density is defined as the mass within the effective radius divided by the spherical volume within the effective radius. This density is affected by both mass and size changes and decreases by 1-2 orders of magnitude between $z=1.5$ and $z=0$. The density within the inner 2 physical kpc, on the other hand, stays roughly constant over the last 9 Gyr of cosmic evolution. Major mergers mildly increase this central density while mass losses by stellar winds tend to decrease it gradually. This constancy is not an artifact of our minimal star formation correction scheme since it remains even in the case of no correction. We infer that the central density of massive galaxies today should correspond closely to their central density at $z\sim{}1.5$, while the effective density is strongly evolving.

\begin{figure*}
\begin{center}
\begin{tabular}{ccc}
\includegraphics[width=55mm]{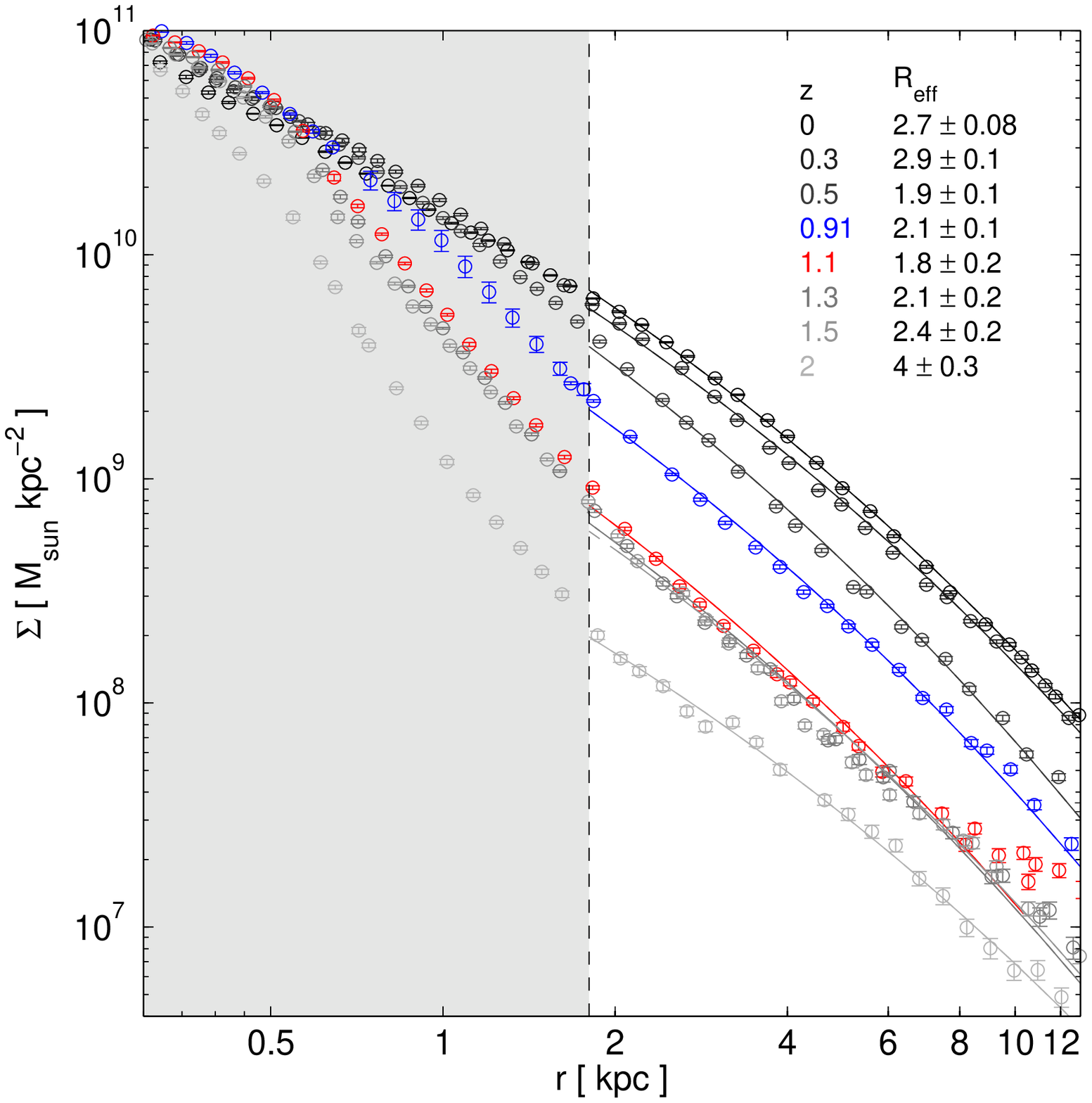} &
\includegraphics[width=55mm]{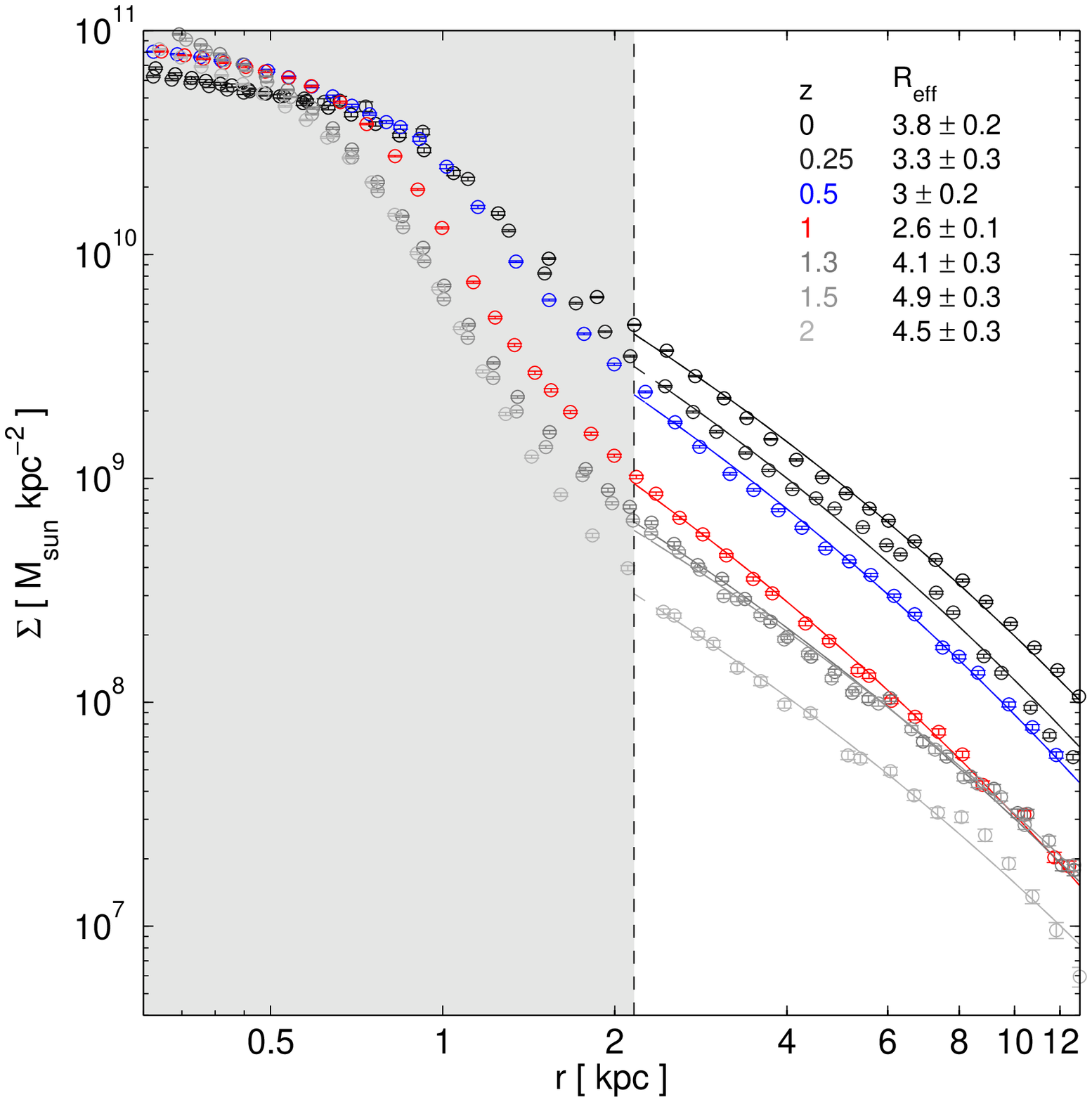} & 
\includegraphics[width=55mm]{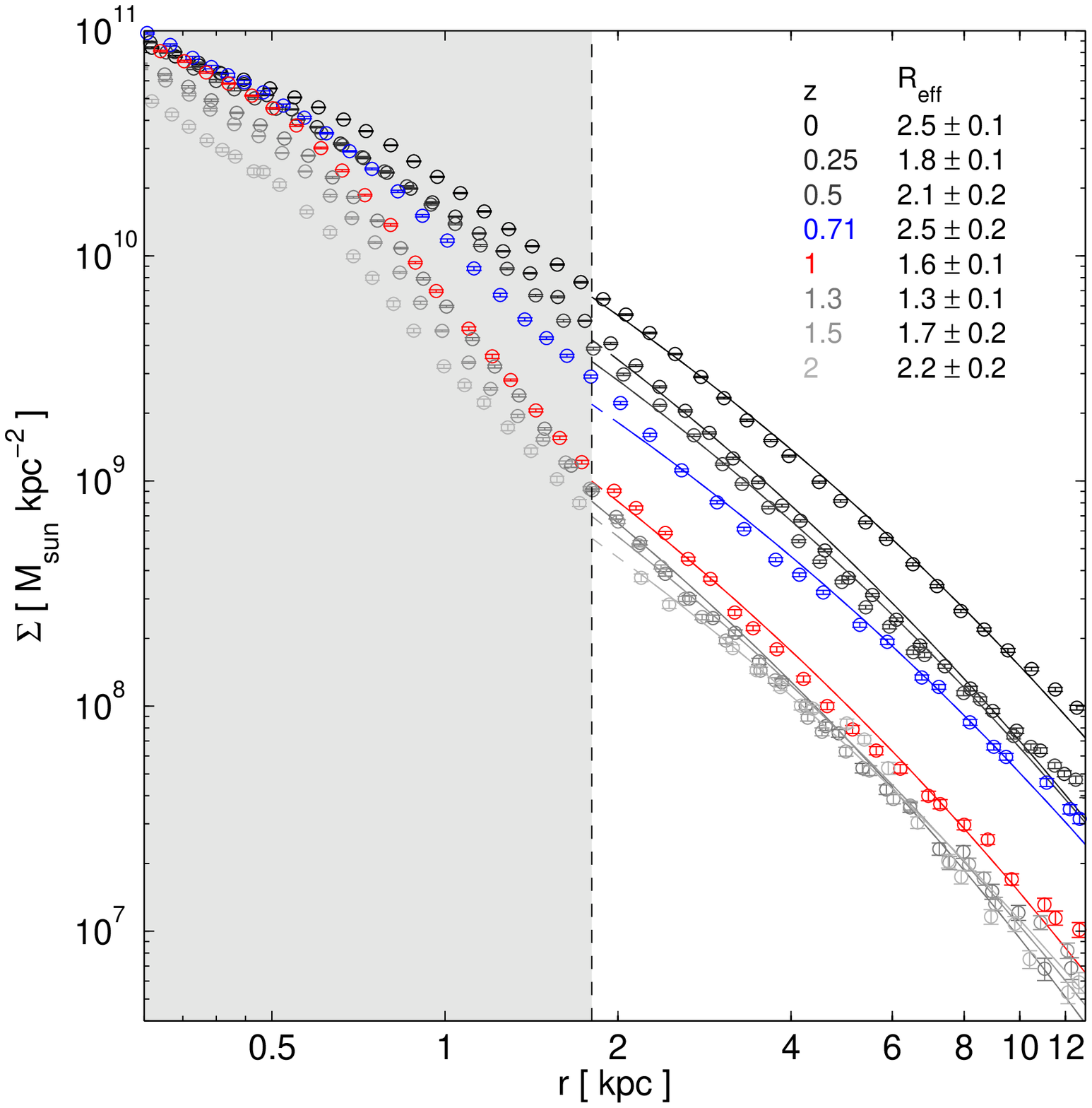}
\\
\end{tabular}
\caption[Evolution of the surface mass density profiles of the central galaxies]{Evolution of the surface mass density profiles of the central galaxies in the groups $G1$, $G2$, $G3$ (from left to right). A deVaucouleurs profile is fit to the data between 3$\times{}\epsilon_\mathrm{bar}$ and up to 20 kpc. The effective radii that are derived from the fit are denoted in the top right corner of each panel. The profile within 3$\times\epsilon_\mathrm{bar}$ is  affected by numerical resolution and this region is shaded gray to caution the reader. A particular time span of significant evolution in the surface mass density is shown for each of the central galaxies (blue and red symbols). \label{fig:SBevolution}}
\end{center}
\end{figure*}

Between $z\sim{}0-1.5$ the mass in the central 2 kpc stays roughly constant, but the galaxies increase their total stellar mass by a factor of 3-4. Therefore, they need to accrete or form mass \emph{outside} this central region. To explore this issue we plot in Fig. \ref{fig:SBevolution} the time evolution of the surface profile of the stellar mass outside 2 kpc.  The increase in the mass surface density of the envelope is not smooth, but undergoes phases of fast and slow growth. For example the envelope of $G2$ is almost non-evolving between $z=1.5$ and $z=1$, but increases rapidely before and after this period. A similar behaviour can also be observed at different times for $G1$ and $G3$. The growing stellar envelope can roughly be fitted with a deVaucouleurs profile, although deviations at both low radii, e.g. caused by the central ``bulge'' or the formation of a stellar disk, and at large radii, e.g. due to tidal debris from satellites, are visible. 

We want to caution the reader that despite the constancy of the stellar mass $M_\mathrm{cen}$ within a small radius $r_\mathrm{cen}$ (``constant central density'') the profile within $r_\mathrm{cen}$ could evolve. For example, one could imagine that violent relaxation processes during the initial collapse stage and/or by subsequent merger events establish a mass surface profile of deVaucouleurs form. In this case one can, for each given effective radius $r_\mathrm{eff}$, adjust the $r=0$ surface mass density $\Sigma(0)$ such that the mass within $r_\mathrm{cen}$ is $M_\mathrm{cen}$. At fixed $M_\mathrm{cen}$ a larger $r_\mathrm{eff}$ implies a smaller $\Sigma(0)$, a larger $\Sigma$ at large radii, and an increase in total mass. Plugging in $r_\mathrm{cen}=2$ kpc and $M_\mathrm{cen}\sim{}10^{11} M_\odot$, the model predicts that a change in effective radius from 1 kpc to 5 kpc goes along with an increase in total mass from $\sim{}1.5\times{}10^{11} M_\odot$ to $\sim{}2.8\times{}10^{11} M_\odot$ and an increase in surface mass density at $r=10$ kpc from $2\times{}10^7 M_\odot$ kpc$^{-2}$ to $2\times{}10^8 M_\odot$ kpc$^{-2}$. However, our simulations show that the fitted effective radii of the envelopes do not strongly change between $z\sim{}2$ and $z=0$. We thus conclude that at $z\sim{}2$ the profiles of the progenitor galaxies should differ substantially from a single deVaucouleurs law.

\subsubsection{The driver of the size growth}
\label{sect:DriverSizeGrowth}

\begin{figure}
\begin{center}
\begin{tabular}{c}
\includegraphics[width=85mm]{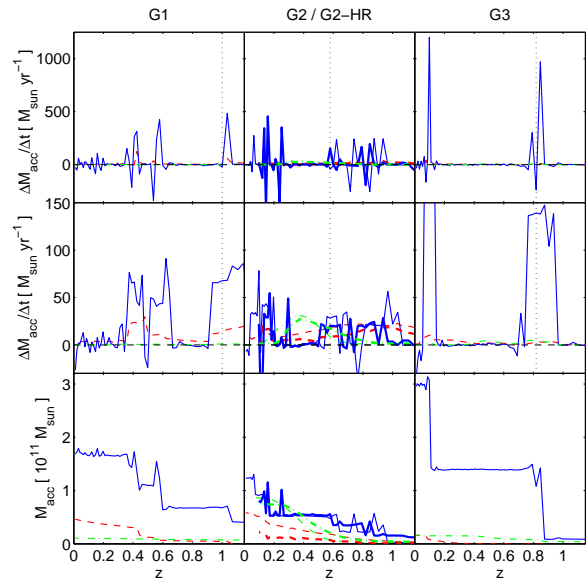}
\end{tabular}
\caption[Mass transfer into a shell outside $R_\mathrm{eff}$]{Mass transfer into a shell outside $R_\mathrm{eff}$.  Different rows correspond to the different groups: $G1$ (left column), $G2$ (middle column, thin lines), $G2-HR$ (middle column, thick lines), $G3$ (right column). (Top row) the net fluxes in the region between $R_\mathrm{eff}$ and 20 kpc averaged over 13 Myr; (Middle row) as the top row but averaged over 100 Myr; (Bottom row) cumulative net fluxes into the region between $R_\mathrm{eff}(z=0)\sim{}3.5$ kpc and 20 kpc. The different lines correspond to the net fluxes of accreted stars (blue solid line), of stars formed in situ outside the effective radius (green dash-dotted line) and of stars formed in situ within the effective radius but redistributed outwards (red dashed line), respectively.\label{fig:StellarTransport}}
\end{center}
\end{figure}

We now address how the stellar envelope is built and thus which processes drive the size growth. To this end we measure the accreted stellar mass, the stellar mass redistributed through the effective radius and the mass formed by in situ star formation as function of time. The top row in Fig. \ref{fig:StellarTransport} shows that mass accretion rates in mergers can be extremely high ($\sim{}1000 M_\odot{} \mathrm{yr}^{-1}$) for a very short period of time ($\sim{}10$ Myr). This number is consistent with what one would expect for groups with velocity dispersions of a few hundred km/s, satellite sizes of a few kpc and masses of order of $10^{10}M_\odot$. The middle row indicates that (i) the accretion flux is due to resolved, singular merger events and not due to a smooth accretion of halo stars, (ii) that the accretion by merging is predominantly driving the mass evolution and (iii) in periods in which no mergers occur either in situ star formation ($z\sim{}0.2-0.5$ in $G2$) or a redistribution of stellar mass ($z\sim{}0.7-0.8$ in $G1$, $z\sim{}0.1-0.2$ in $G3$) can drive the mass deposition in the region $R_\mathrm{eff}$-20 kpc and hence the size evolution. The bottom row demonstrates that, when accumulated over the history of the central galaxy, accretion of stars by merging is the dominant contributor of stellar mass outside $R_\mathrm{eff}(z=0)\sim{}3.5$ kpc, and hence the dominant mechanism that determines the sizes of group central galaxies by $z=0$. Averaged over the three groups merging contributes\footnote{The mass fractions are averaged over the simulations $G1$-$G3$. Quoted as uncertainties are the differences between the average mass fraction and the largest and smallest mass fraction among the three groups.}  to $70^{+20}_{-15}\%$ of the mass outside $R_\mathrm{eff}(z=0)$. In addition, non-central star formation ($14^{+18}_{-9}\%$) and a redistribution of preformed stellar mass ($16^{+6}_{-11}\%$) are also of significance, especially in phases without merging activity.
For the latter, secular, process we observe a correlation between the reshuffling of mass and the orbiting of satellites in/through the central regions of the respective group, indicating that satellites may be directly involved in heating the stellar component via tidal shocks or by dynamical friction. In addition, we also see that the reshuffling is substantial for the central galaxy of group $G2$ when it forms a large stellar disk. In this case satellites may induce stellar bars or spiral waves and indirectly lead to a redistribution of the angular momentum of the stellar component \citep{2002MNRAS.336..785S, 2008ApJ...675L..65R, 2008ApJ...684L..79R, 2008ApJ...688..254K, 2009MNRAS.398..591S}, see column 3 of Fig.~\ref{fig:EvolutionFaceOn}. On the other hand, the flux of redistributed stellar mass into the region between 3.5 and 20 kpc (bottom row of Fig. \ref{fig:StellarTransport}) is smaller by a factor of about 2 in $G2-HR$ compared to $G2$. This indicates that numerical resolution may, at least partially, affect the reshuffling of stellar matter, e.g. by changing the bar strength.

\subsection{Beyond mass: The role of assembly history}

The halo masses and radii of the three studied groups $G1-G3$ below $z=1$ differ by less than a factor of 2. At $z=0.7$ the virial masses are (by selection) almost identical, while at $z=0$ they range from $1.1$ to $1.6\times{} 10^{13} M_\odot$, see Fig \ref{fig:AccretionHistory}. Nonetheless, we observe strong differences in morphology of the central galaxies at $z<1$, see Fig.~\ref{fig:EvolutionFaceOn} and Fig.~\ref{fig:EvolutionEdgeOn}. In this section we will discuss the connection between the morphology of the central galaxies and their mass assembly histories and cooling properties.

\subsubsection{Star formation histories and the cold gas  reservoir}
\begin{figure*}
\begin{center}
\begin{tabular}{ccc}
\includegraphics[width=55mm]{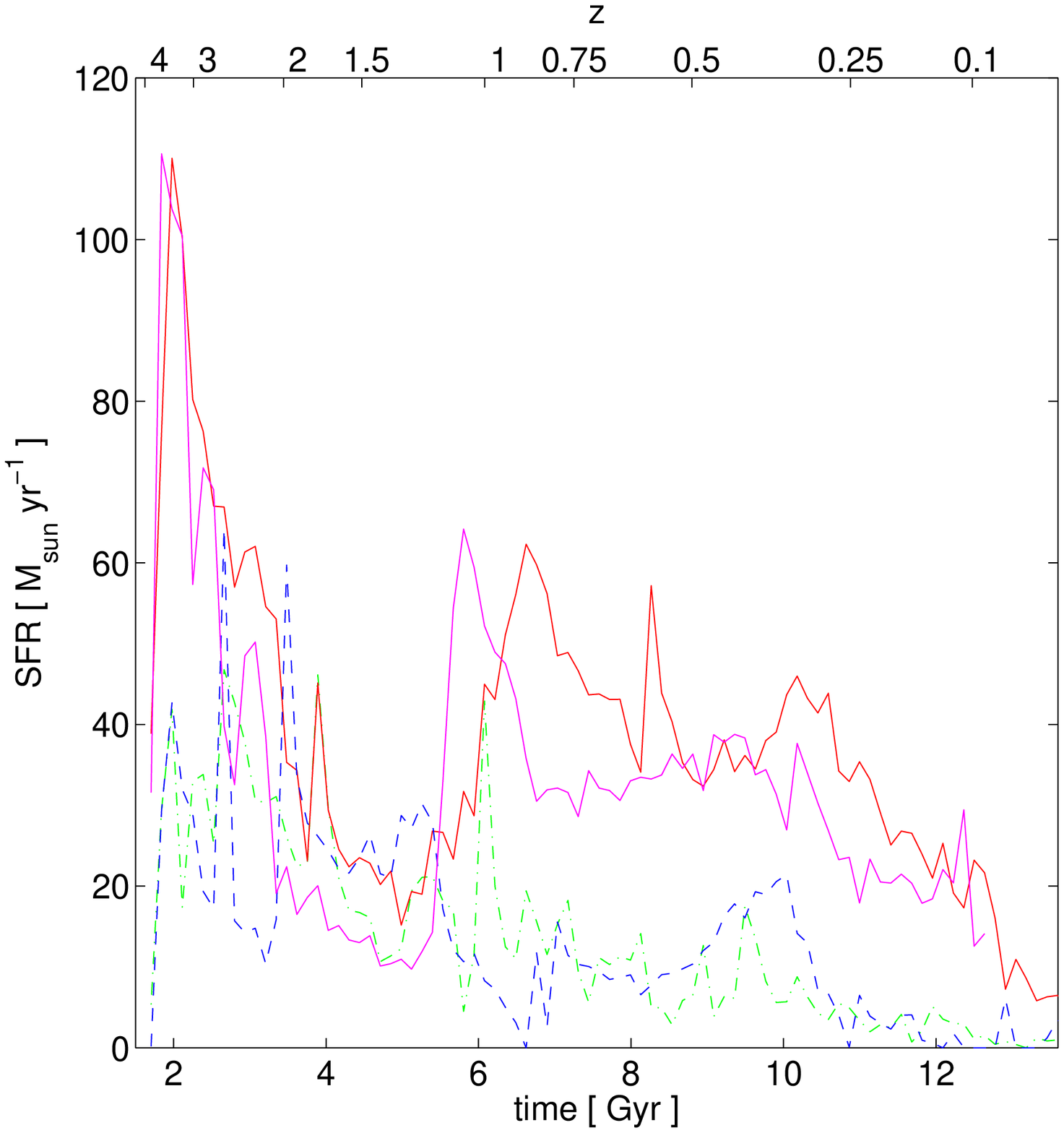} &
\includegraphics[width=55mm]{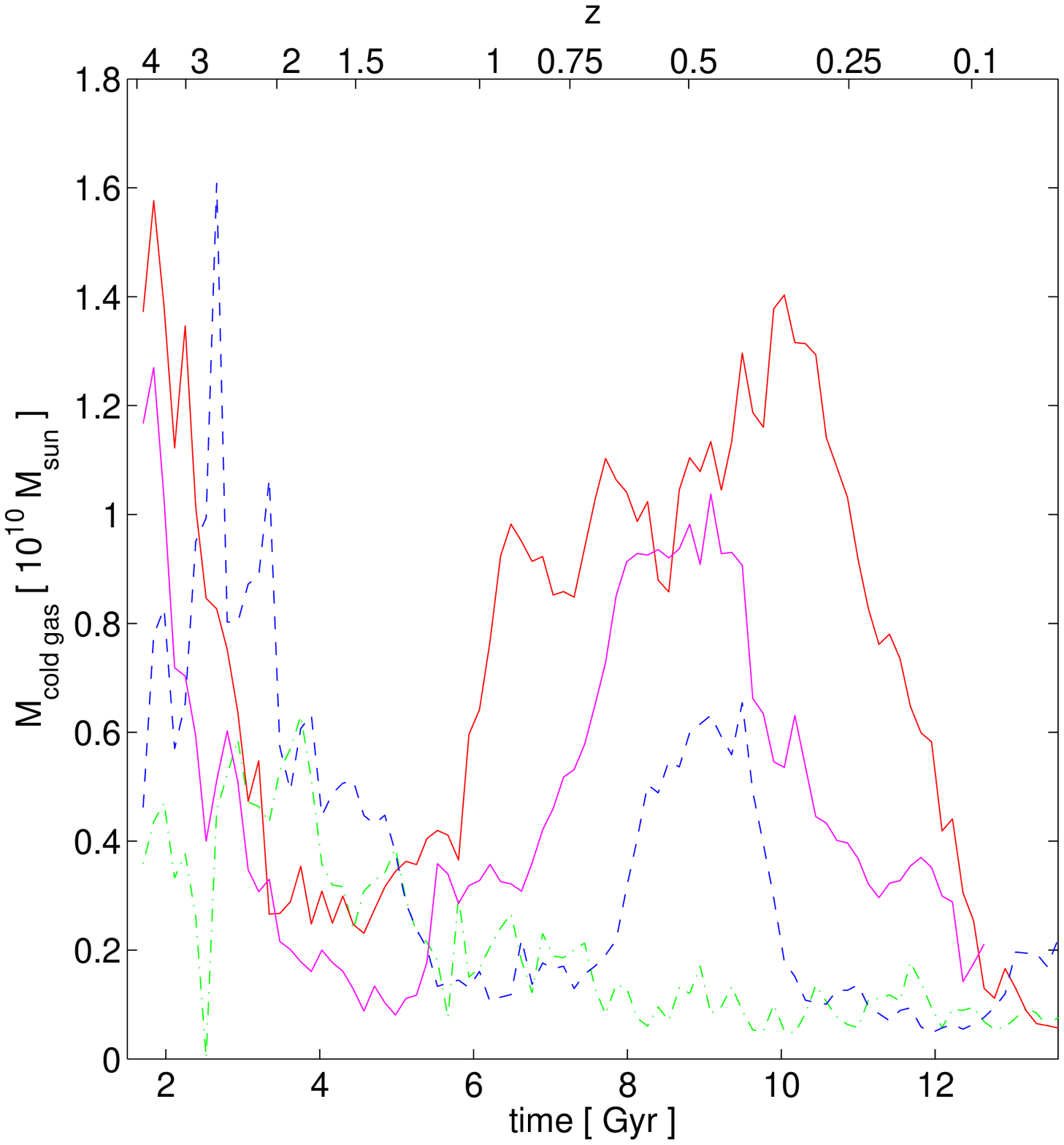} &
\includegraphics[width=55mm]{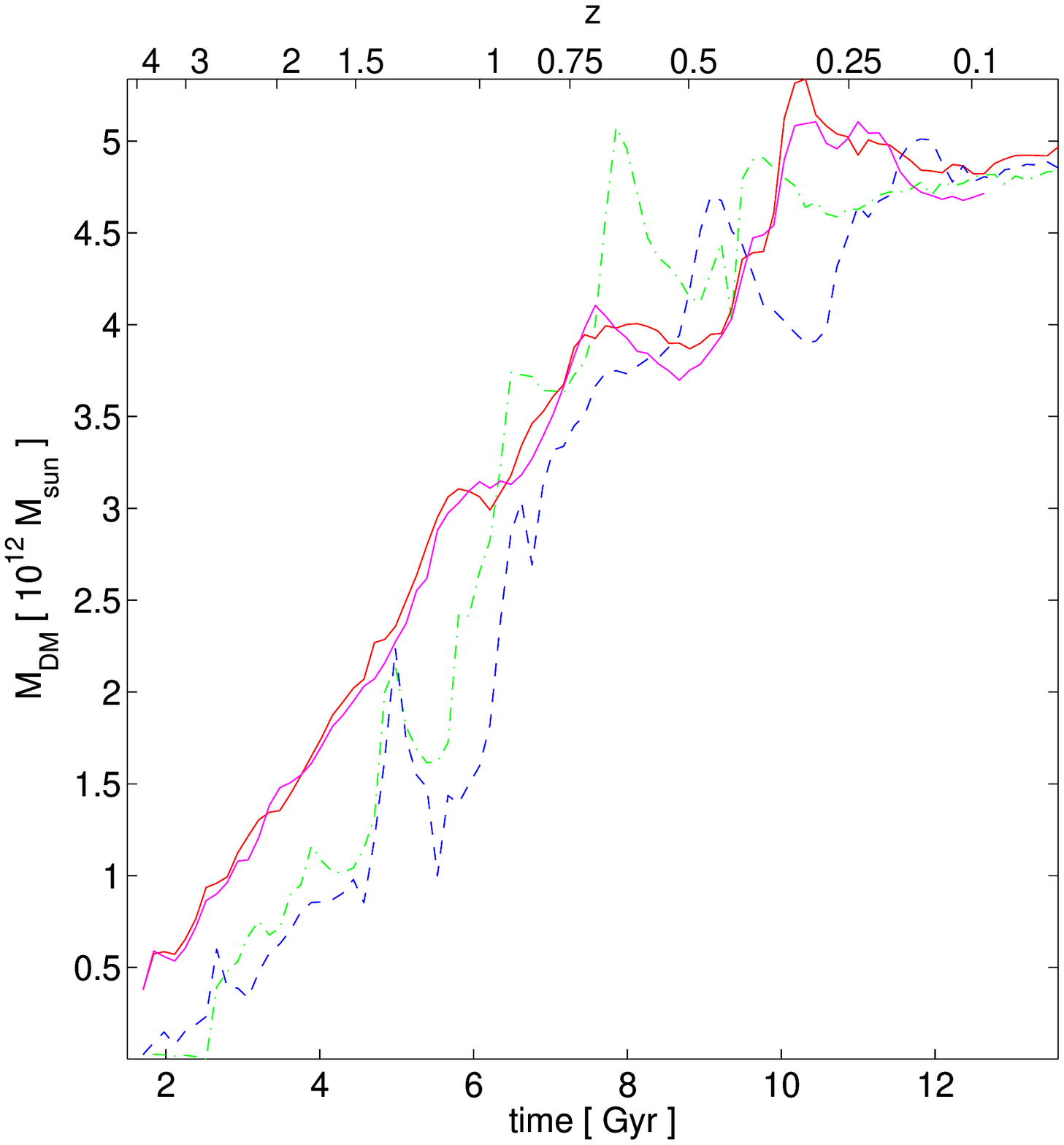}
\end{tabular}
\caption[Evolution of the SFR, cold gas mass and dark matter mass]{(Left panel) The star formation rate within 20 kpc around the central galaxy of the groups $G1$ (green dot-dashed line), $G2$ (red solid line), $G2-HR$ (magenta solid line) and $G3$ (blue dashed line) a function of cosmic time and redshift. The star formation is corrected for an artifical central star formation rate of $\sim{}11 M_\odot$ yr$^{-1}$, see appendix \ref{sec:ResolutionTest}. (Middle panel) The mass of cold gas within the inner 20 kpc of the central group galaxies. (Right panel) The dark matter mass within a fixed physical radius of 200 kpc around the main progenitor.\label{fig:SFR}} 
\end{center}
\end{figure*}

In order to understand the differences between the three central group galaxies we begin by investigating their star formation histories. The star formation histories peak at $z\sim{}4$, when the universe was about 2 Gyr old (left panel of Fig \ref{fig:SFR}). Later several short bursts of star formation occur of which some are connected to merging events. The star formation history of group $G2$ differs significantly from the groups $G1$ and $G3$. The overall star formation rate in the former group is generally larger, especially at high redshift ($z\sim{}3-4$) and below $z=1$. At high redshifts the star formation timescale ($M_*$/$SFR$) of the three groups at a given instant of time is similar. It ranges from $\sim{}0.9$ Gyr at $z=3$ (i.e. small compared to the age of the Universe of $2.2$ Gyr at $z=3$) to $2-5$ Gyr at $z=2$ (which is similar or even longer than the age of the Universe of $3.4$ Gyr at $z=2$) - indicating that the importance of star formation reduces with time and becomes subdominant around $z=2$. Despite their similar star formation timescales the central galaxies reach a stellar mass of $10^{11} M_\odot$ at different times: $G2$ at $z\sim{}2.4$, while $G1$ and $G3$ cross this boundary much later at around $z\sim{}1.5$. When we compare the star formation history with the available cold gas mass within 20 kpc we generally see a co-evolution (compare left and middle panel of Fig \ref{fig:SFR}). The groups cross the predicted virial mass threshold of $M_\mathrm{shock}\sim{}7\times{}10^{11}$ for the generation of a stable shock near the virial radius (\citealt{2003MNRAS.345..349B}) at $z\sim{}2.3$ (G1), $z\sim{}2.9$ (G2) and $z\sim{}2.0$ (G3). Consistent with the transition from a cold to a hot gas accretion mode (\citealt{2006MNRAS.368....2D}, \citealt{2005MNRAS.363....2K, 2009MNRAS.395..160K}, \citealt{2008MNRAS.390.1326O}) we find that the cold gas mass within the central 20 kpc of the groups drops roughly at these predicted times, as seen in the middle panel of Fig \ref{fig:SFR}. The time delays of 0.5-1 Gyr between the decrease in available cold gas and the reduction in the star formation rates compare well with the gas consumption timescales $M_\mathrm{cold gas}/$SFR $\sim{}0.2-0.4$ Gyr and indicate that the star formation rates mainly decrease due to the lack of cold gas in the central galaxies. 
Between $z\sim{}1$ and $z\sim{}0.2$ group $G2$ contains about $1\times{}10^{10}M_\odot$ of cold gas within 20 kpc. This exceeds the cold gas reservoir of $G1$ and $G3$ within that time-span by an order of magnitude and explains why the central galaxy in $G2$ experiences a large star formation activity at those intermediate redshifts of about 40-50 $M_\odot$ yr$^{-1}$. By $z\sim{}0.5$ it resembles a galaxy with prominent gaseous and stellar disks with spiral features and with a red bulge-like component. A bulge-to-disk decomposition with GALFIT results in B/D$\sim{}1-1.7$, depending on projection and whether mass density or I-band images are fitted. At the same time the rotational support of the galaxy smoothly increases from $v/\sigma=0.3$ at $z\sim{}1.1$ to $v/\sigma=1.1$ at $z\sim{}0.5$. On the other hand the central galaxies in $G1$ and $G3$ are relatively quiet (SFR $\lesssim{}10 M_\odot$ yr$^{-1}$) during that periode and their star formation is restricted to the central $\sim{}1$ kpc. Below $z\sim{}0.2$ the cold gas in $G2$ depletes quickly and its star formation rate drops substantially. Simultaneously, minor mergers and tidal interactions with smaller satellites in the group heat and damage the stellar disk and convert the spiral galaxy into an $S0$ over a timescale of 1-2 Gyr. Hence, the simulations indicate that, firstly, the cold gas fractions drop and thus the formation of new stars is suppressed, before subsequently major galaxy mergers ($G1$, $G3$) or minor mergers and tidal interactions with satellites ($G2$) heat the stellar system and transform it towards a more early-type galaxy. 

\subsubsection{The origin of the cold gas}
The substantial amount of cold gas of $G2$ below $z=1$ originates from three different sources: (i) cooling from the hot gas halo, (ii) cold gas brought in by mergers or in small gas clumps (but not by cold streams), and (iii) cooling from a warm, primordial gas phase. The first process contributes the most. More than $70\%$ of the cold gas at $z=0.5$ has cooled out of the hot halo below $z=1$. The cooling time of the hot gas within the central 10 (20) kpc of group $G2$ is $\sim100$ $(300)$ Myr below $z=1$, see Fig. \ref{fig:Cooling}. Therefore cooling can proceed quickly unless heating mechanisms keep the gas hot, such as adiabatic compression due to a smooth accretion of intergalactic gas onto the halo, shock heating of the ambient gas due to supersonic collisions of infalling gas lumps (e.g. \citealt{2003ApJ...593..599R}), heating by dynamical friction of supersonic galaxy motions \citep{1999ApJ...513..252O, 2004MNRAS.354..169E}, clumpy accretion \citep{2008MNRAS.383..119D}, heating by a background UV field \citep{1996ApJ...461...20H, 2006MNRAS.371..401H}, or heating due to feedback mechanisms such as radio mode AGN \citep{2006MNRAS.365...11C}. The latter is not included in our simulations. Roughly $15\%$ of the cold gas reservoir cools from a warm ($\sim{}10^5$ K), zero-metallicity (primordial) gas that is accreted from the surroundings of the group halo, possibly from the left-overs of cold filaments \citep{2009arXiv0905.2186K} or from underdense, unprocessed gas. The accretion of metal enriched, cold ($T<3.2\times{}10^4$ K) gas contributes to $\sim{}10\%$. At $z\sim{}0.4$ group $G2$ undergoes a merger with another, smaller group of virial mass of $\sim{}5\times{}10^{11}$ which has not yet crossed the threshold mass above which its cold accretion is terminated. It carries a cold gas mass of a few $10^{10}$ $M_\odot$. However, most of the cold gas in that subgroup is shock heated or consumed by star formation during the group-group merger, so that it barely contributes to the cold gas reservoir of the main halo. 

\subsubsection{The cooling time and assembly history}
We now address why the cooling in the group $G2$ and in the groups $G1$ and $G3$ behaves so differently. We believe that the mass assembly history, in particular the frequency of major halo-halo mergers, impacts the gas cooling in at least two different ways:

First, at fixed halo mass the absence of late halo-halo mergers implies an earlier formation time and therefore, on average, a more concentrated mass profile of the halo \citep{1997ApJ...490..493N, 2002ApJ...568...52W}. The more concentrated halo has a higher central density and may be more easily prone to cooling instabilities. In dark-matter-only re-simulations of the groups $G2$ and $G3$ 
the concentration\footnote{The concentration $c$ is defined as the ratio between virial radius (for the calculation of $c$ we use the definition of \cite{2002ApJ...568...52W}) and scale radius of the best fitting NFW-profile \citep{1997ApJ...490..493N}.} of group $G2$ amounts to about twice the concentration of group $G3$, namely $c=5.9$ vs $c=2.8$, see Fig. \ref{fig:DensProfiles}. It is straightforward to estimate that for an NFW profile a doubling in concentration increases the enclosed mass at small radii by a factor 2-2.8 (for $c\sim{}2-20$). However, this is clearly an oversimplification given the non-linear properties of the gas cooling, the adiabatic contraction of the dark matter and the fact that baryons and dark matter might decouple in the central region. In the right panel of Fig. \ref{fig:Cooling} we plot density weighted temperature, density, entropy and cooling time of the gas within 20 kpc for our three groups. At $z\sim{}0.7$ the central gas density of group $G2$ excels the central densities in the groups $G1$ and $G3$ by a factor $\sim{}4$, the central temperature of $G2$ is smaller by 10-20\%, and consequently its cooling time is lower (300 Myr vs. $\gtrsim$ 1 Gyr).

The second impact of halo-halo mergers it that they contribute to the heating of the intra-group gas. For example, infalling satellites and gas clumps can convert their potential and kinetic energies into thermal energy of the ambient medium \citep{2009ApJ...697L..38J}. This heating mechanism is boosted in a halo-halo mergers that bring in a large number of satellites. But also other channels such as large scale shock heating of the gas may be important or the decrease of the gas density by outward motions of the gas (increase in kinetic energy). In the right panel of Fig. \ref{fig:SFR} we plot the dark matter mass within 200 kpc around the central galaxy. Compared with the virial mass as function of time (Fig. \ref{fig:AccretionHistory}) this plot allows to identify both the first pericenter of the merging halos and the final coalescence, leading to a characteristic ``U'' shape. Together with Fig. \ref{fig:Cooling} (right panel) this plot shows that the halo-halo mergers that occur between $z=1.4$ and $z=1$ in both $G1$ and $G3$ lead to an increase in temperature and decrease in density at the centers of the groups. Similarly, the merger that occurs between $z\sim{}0.5$ and $z=0.25$ in group $G3$ terminates a short cooling event at $z\sim{}0.5$, and the merger that starts at $z\sim{}0.3$ in group $G2$ terminates the cooling below $z\sim{}0.25$.

The mass accretion history is therefore an important factor in order to explain the diversity that group halos of the same mass show in their cooling properties or the related properties (cold gas fractions, star formation activity, morphology, etc.) of their central group galaxies. On the one hand, halo-halo mergers inject energy into the intra-group gas and prevent or even terminate hot accretion episodes. On the other hand, groups with late halo-halo mergers, with a higher substructure fraction or with later formation times have on average less concentrated dark matter halos \citep{1997ApJ...490..493N, 2002ApJ...568...52W, 2004MNRAS.355..819G}, which increases their cooling times. We thus anticipate for groups of a given mass on average an anti-correlation between the amount of substructure (related to formation time) and the star formation rate of the central galaxy. On the same grounds we expect that groups which undergo mergers less likely host a disk or S0 galaxy at their centers compared to well virialized groups with a quiet merger history. We note that the assembly bias for the formation time and concentration is small at the group mass scale \citep{2007MNRAS.377L...5G, 2008MNRAS.389.1419L}. Therefore, the presented mechanism does not strongly differentiate between groups in underdense regions and groups near a cluster, respectively. We note that our analysis does not include the impact of AGN feedback on the galaxy group scale,  which is an important, yet open question for future work.

\section{Summary}
\label{sect:summaryCentralGalaxies}

% sims & results at z=0
We have simulated the evolution of galaxy groups with a final virial mass of $\sim{}10^{13}M_\odot$ in order to study the evolution of their central galaxies. We ran the Tree-SPH code GASOLINE \citep{2004NewA....9..137W} with the same parameters and at a resolution comparable to the one used in current state-of-the-art simulations that follow the formation and evolution of disk galaxies in a cosmological context \citep{2007MNRAS.374.1479G}. All our central group galaxies end up with roughly the same stellar mass of a few times $10^{11}$ $M_\odot$ at $z=0$ and their stellar profile can be well fitted with a Sersic index of $\gtrsim{}4$. Furthermore, our simulated central galaxies are kinematically hot systems supported by velocity dispersion, have colors of evolved stellar populations and an early-type morphology. However, the detailed morphology (E or S0) and the amount of rotation differs among the central galaxies. We trace such differences back to their different merger and gas accretion histories. Overall the basic properties of our simulated central galaxies match approximatively that of observed central group galaxies in the local Universe \citep{2008ApJ...676..248Y, 2009arXiv0901.1150G, Cibineletal2009a, Cibineletal2009b, Cibineletal2009c}. 

% progenitor properties
At redshift $\sim{}2$ the most massive progenitor galaxies reside in dark matter halos of galactic size ($\sim{}10^{12} M_\odot$). The parent halos are just massive enough to support a stable shock at the virial radius that reduces the impact of smooth cold accretion \citep{2006MNRAS.368....2D}. However, some cold streams are still able to reach the central objects and so at $z\sim{}2$ the galaxies experience still a significant ($>20$ $M_\odot$ yr$^{-1}$) star formation that is sustained by a substantial reservoir of cold gas replenished by cold and hot accretion modes. At this time the galaxies have gathered, within a compact ($\sim{}1$ kpc), blue (unless it is heavily dust obscured) ``proto-bulge'', an already significant fraction (10-25\%) of their final stellar mass. They can be observed as $K^{Vega}_s\sim{}20$ star forming BzK galaxies or (as long as $A_V\lesssim{}1$) as UV-optical selected galaxies \citep{2004ApJ...617..746D, 2004ApJ...607..226A, 2007A&A...465..393G}. Only in case they are strongly dust obscured ($A_V\gtrsim{}1.3$) they would qualify at $z>2.3$ as distant red galaxies  \citep{2003ApJ...587L..79F}.  The cold gas arranges in form of extended gas disk of a few kpc radius in agreement to observations of massive, star forming disks \citep{2006Natur.442..786G, 2008ApJ...687...59G, 2008ApJ...682..896K, 2009arXiv0903.1872F}. Over the next two Gyr the galaxies gradually reduce their amount of star formation and the overall colors redden due to the aging stellar population. The galaxies are still very compact - they concentrate $\sim{}10^{11}$ $M_\odot{}$ within 1 kpc at $z\sim{}1.5$ and might thus contribute to compact galaxies observed at those redshifts. In the remaining 9 Gyr of cosmic time the galaxies are subject to major and minor mergers, episodes of gas accretion and a secular redistribution of pre-formed stellar mass that drive the growth of a stellar envelope around a now rather passively evolving ``bulge''. 

% merger history
The merger histories of the three central galaxies and their progenitors differ substantially. One galaxy grows its stellar mass since $z\sim{}1.5$ mainly by minor merging and star formation. 
In the other two groups the central galaxies experience each two major mergers below redshift $z\sim{}1$. Two out of the 4 occurring major mergers involve dissipative disk-disk mergers at $z\sim{}0.8$ and $z\sim{}1$ and produce gas-poor, pressure supported remnants which subsequently undergo a ``dry merger'' with another gas-poor, pressure supported galaxy at $z=0.4$ and $z=0.1$, respectively.

% density evolution & Comparison with passive galaxies
Despite the diversity in their merger histories the evolution of the stellar component of the progenitors of the central group galaxies proceeds in a similar fashion. The densities within the inner 2 kpc remain roughly constant since $z\sim{}1.5$ and the galaxies subsequently grow in size by building a stellar envelope around the early formed stellar bulge. The growth of this envelope reduces the density within the half-mass radius by almost 2 orders of magnitude since $z\sim{}1.5$. It is interesting to compare our result with the proposed size evolution of compact, passively evolving, red-and-dead galaxies observed at $z\sim{}1-2$ (e.g. \citealt{2008AA...482...21C, 2008ApJ...677L...5V}).
In particular, minor merging has been advocated as a driver since it is expected to lead to a stronger size evolution per unit mass than major merging \citep{2009arXiv0903.1636N, 2009ApJ...697.1290B} and is potentially required to connect the sizes of compact, passive galaxies at high $z$ to the local early-type galaxy population \citep{2009ApJ...697.1290B}. 
Although minor mergers happen numerously in our simulations the size growth is overall slower than predicted from minor merging alone. The simple reason is that major mergers occur frequently (in our case in 2 our of 3 simulated galaxy groups) and then naturally dominate the mass evolution. Also, in one of our groups non-central star formation temporarily drives a late mass and size growth. We expect this latter channel to become dominant for lower mass galaxies (see e.g. \citealt{2008MNRAS.384....2G}). In addition, we expect central star formation to slow down the size growth. Our results imply that the size growth per unit mass of massive, yet star forming galaxies at $z\sim{}2$ is qualitatively similar, but likely less steep, compared to the mass-size evolution of massive, compact and passively evolving galaxies observed at $z\sim{}2$.

% environment
Below $z=1$ the main progenitors of the central group galaxies differ substantially in their morphologies and star formation activities although they sit in groups of similar virial mass (cf. \cite{2009MNRAS.396..696S} for galactic halos). In one of three cases the central galaxy accretes a substantial amount of gas between $z\sim{}0.3-0.9$ that cools out of the hot intra-group medium due to the short cooling time at the center of the group.  This leads to an inside-out growth of a stellar disk and the morphology approaches that of a massive spiral galaxy. The cooling episode is eventually stopped when a halo merger with an infalling subgroup occurs. The central galaxy is subsequently transformed into an early-type (S0) galaxy as cooling from the hot gas halo is suppressed, the cold gas reservoir is exhausted by star formation, and minor mergers and tidal interactions with satellites damage and heat the stellar disk below $z=0.2$. In contrast, the other two studied groups are subject to several group-group mergers which provide an energy injection mechanism that increases the temperature and decreases the density of the intra-group gas and thus counteracts a hot accretion mode. In addition, group mergers are followed by major galaxy mergers between the former central galaxies of the subgroups creating remnants of elliptical morphology. Hence, in the absence of additional heating mechanism the mass accretion history plays an important role in limiting the strength of the hot accretion mode and thus the kinematics and morphology of the central group galaxies. If the cooling is suppressed galaxy mergers and tidal interactions with satellites are able to drive a rejuvenated central galaxy towards an early-type appearance.

\section{Open Issues}
\label{sect:openq}

A missing ingredient in our simulations is feedback from active galactic nuclei typically associated with super-massive black holes which are found in the centers of most galaxies \citep{1995ARA&A..33..581K, 1998AJ....115.2285M, 2000ApJ...539L..13G, 2000ApJ...539L...9F}.  AGN feedback could potentially rival with environmental effects and assembly history in setting the final galaxy properties, although
recent work shows that this is not necessarily the case (\citealt{2008arXiv0803.4003C}). Indeed our results argue against a major role
of AGN feedback, at least for the properties of central galaxies outside the inner kpc. This is because even without AGN feedback our simulations reproduce reasonably well many properties of $z\sim{}0$ early-type galaxies (c.f. \citealt{2008MNRAS.387...13K}), such as the colors, morphologies, kinematics and structure \emph{outside the $\sim$ central softening length}, albeit with masses and/or sizes that are somewhat biased w.r.t. average values seen in local surveys \citep{2003MNRAS.343..978S}. The tendency towards compactness of our simulated galaxies is the counterpart of the mass concentration problem still seen in the simulations of disk galaxies, whereby, despite the fact that disk scale lengths and overall disk sizes are reasonable reproduced at the highest resolution achieved so far, central bulges are still too massive and compact, see \cite{2008ASL.....1....7M}. This likely reflects the need for even higher resolution in the early stages of galaxy assembly, which are responsible for setting the central density and inner mass distribution, as shown clearly in this paper, and/or the necessity of additional heating mechanisms in the very innermost regions. Therefore, the study of AGN feedback and its effect on massive galaxies in galaxy groups may be the key to obtain more realistic central densities, and thus effective radii, by partially suppressing cooling and star formation in the \emph{central kiloparsec}. Alternatively, the same effect may be achieved by a more realistic modeling of the multi-phase interstellar medium and of star formation/supernovae feedback once molecular gas densities are resolved (\citealt{2008ApJ...680.1083R}, \citealt{Governato2009}).

Our results suggest that halo-halo mergers are an efficient means to quench cooling. The precise mechanism (or the chain of mechanisms) by which the potential energy of the pre-merged halos is converted into kinetic or thermal energy of the gas still needs to be worked out, and we do not exclude that major stellar mergers are involved. However, we want to point out that a correlation but not a causal connection between shutdown of hot accretion and major galaxy merging may arise even if major mergers are not responsible for the suppression of the cooling - simply because after the halo-halo merger terminates the cooling, the former central galaxies will engage in a major merger due to the short dynamical friction time for massive galaxies.

Another question that remains open is  how the (already massive) galaxies at $z\sim{}2$ relate to higher redshift objects, such as $z\sim{}3$ Lyman-Break galaxies \citep{1996ApJ...462L..17S} or Lyman-Alpha emitters (e.g. \citealt{1998ApJ...502L..99H}) that are a potential progenitor population given their clustering properties (e.g. \citealt{2008ApJ...681.1099B}). We intend to address this question with future simulations that are specifically designed to resolve the $z\gtrsim{}3$ progenitors of central group galaxies.

\acknowledgments
{\small
R.F. acknowledges founding by the Swiss National Science Foundation. G.Y. thanks MEC (Spain) for financial support under research grants FPA2006-01105 and AYA2006-15492-C03. The simulations have been carried out at the Swiss National Computing Center (CSCS) in Manno, on the Brutus Beowulf cluster hosted by the Informatikdienste at the ETH Zurich, and
at the Barcelona Supercomputer Center (BSC). This research has made use of SAOImage DS9, developed by Smithsonian Astrophysical Observatory. Correspondence and requests for materials should be addressed to the first author (e-mail: feldmann@phys.ethz.ch).}

\bibliography{feldmann}

\begin{appendix}
\section{Star formation correction schemes}
\label{sect:SFCorr}

The raw star formation rates of the central galaxies are about $\sim{}11$ $M_\odot$ yr$^{-1}$ at $z=0$ and are therefore an order of magnitude higher than the average star formation rates of massive galaxies in the local Universe \citep{2007ApJ...661L..41Z}. This high star formation biases masses, sizes, colors and luminosities of the studied galaxies. The star formation excess can (i) arise from a shortcoming of the implemented model, e.g. a lack of a physical heating mechanism such as radio mode AGN feedback \citep{2006MNRAS.365...11C}, (ii) be caused by a too coarse resolution to resolve the relevant heating mechanism, such as clumpy accretion \citep{2008MNRAS.383..119D} or (iii) be of purely numerical origin. Unfortunately we do not know the origin of the excessive star formation in our simulations, but we can study its properties and try to correct for it in a posteriori fashion. Ultimately, however, the simulation model and/or resolution should be improved in order to reduce the high star formation rate in a fully self-consistent manner. In Table \ref{tab:PropertiesZ0Corr} we summarize the consequences of various star formation correction schemes on masses and sizes of the central galaxies.

\begin{figure}
\begin{center}
\begin{tabular}{cc}
\includegraphics[width=75mm]{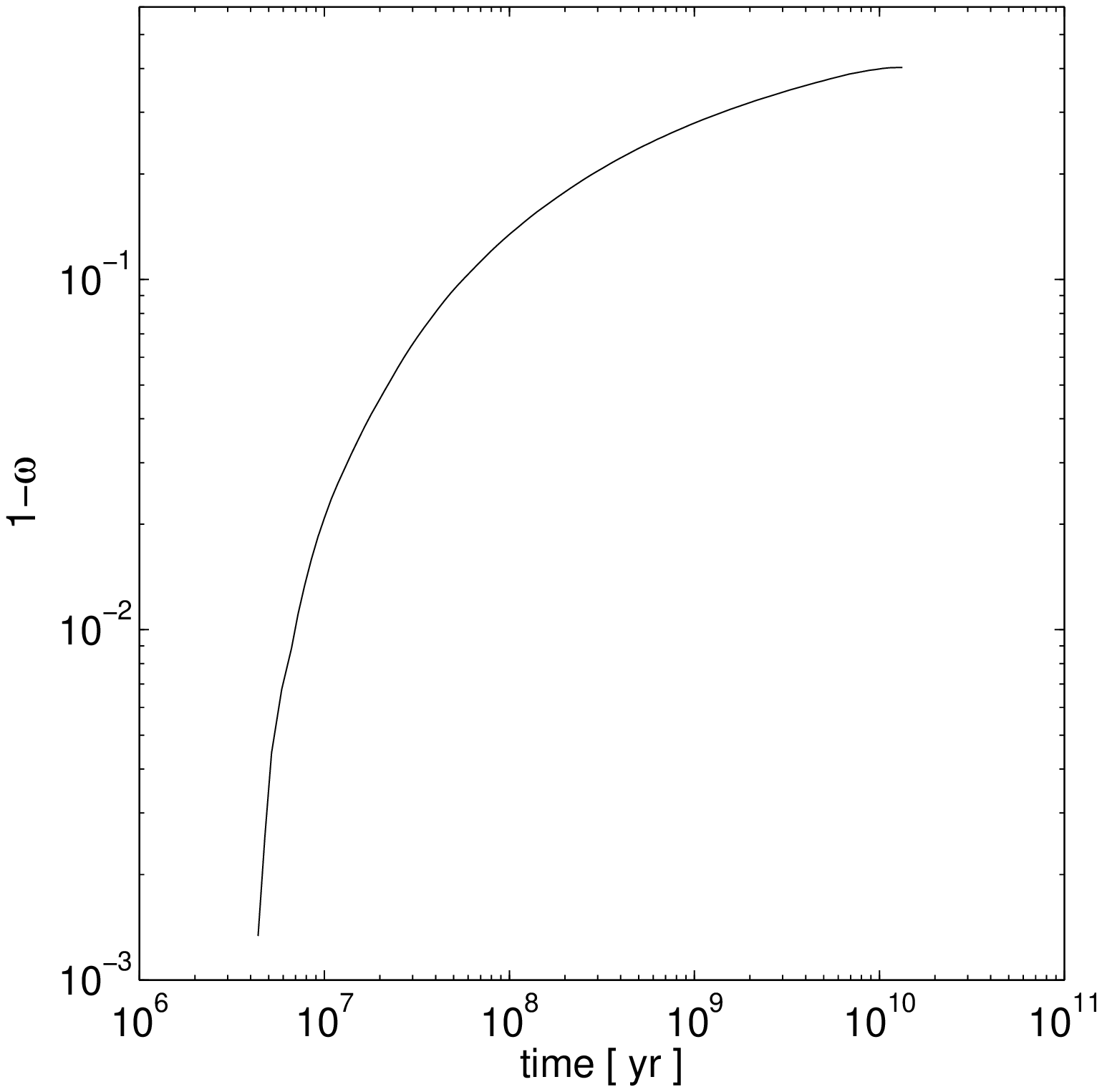} \\
\end{tabular}
\caption[Mass loss due to stellar winds of a star particle]{The mass fraction of a star particle, representing a simple stellar population with Scalo IMF, that is removed due to stellar winds vs. its age. The mass loss reaches $40\%$ after several Gyr.\label{fig:Omega}} 
\end{center}
\end{figure}

\begin{figure}
\begin{center}
\includegraphics[width=75mm]{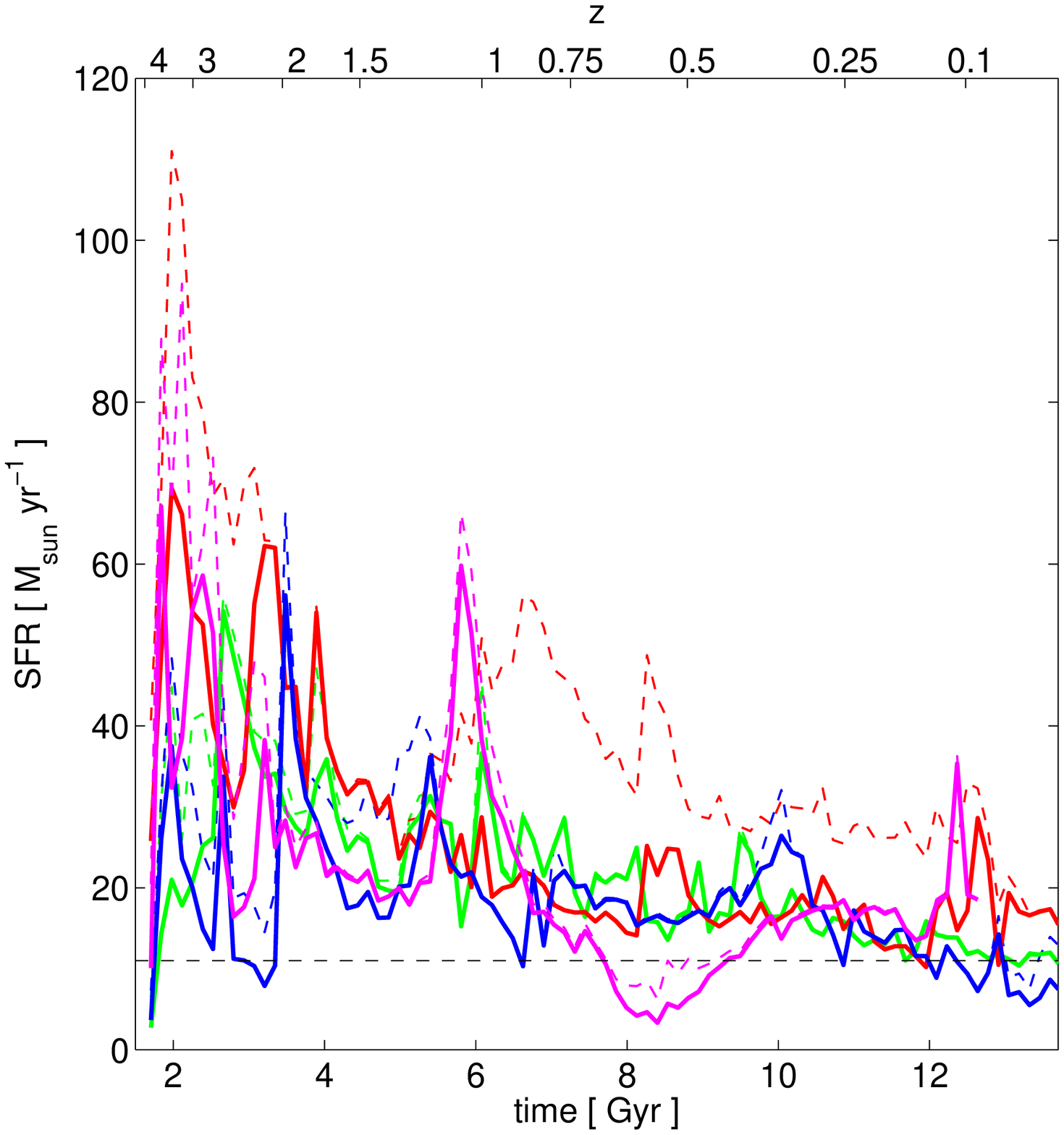}
\caption[The SFR within 1 and 2 baryonic softening lengths]{The star formation rate within 1 (solid lines) and 2 (dotted lines) baryonic softening lengths of the simulations $G1$ (green), $G2$ (red), $G2-HR$ (magenta) and $G3$ (blue). The horizontal dashed line indicates the estimate $\sim{}11\pm{}5 M_{\odot}\,\mathrm{yr}^{-1}$ of residual star formation rate in the central region.\label{fig:CentralSF}} 
\end{center}
\end{figure}

As Fig. \ref{fig:CentralSF} demonstrates a significant amount of star formation is occuring within the unresolved centers ($r<\epsilon_\mathrm{bar}$) of the galaxies. While within a radius of $2\times{}\epsilon_\mathrm{bar}$ the amount of SF varies substantially with time, group and resolution the SFR within $r<\epsilon_\mathrm{bar}$ depends only little on resolution and studied group. It still varies with time although it hardly drops below $11$ $M_\odot$ yr$^{-1}$. A natural way to correct for excessive central star formation is therefore to either (i) subtract a constant star formation rate of the order of $11\pm{}5$ $M_\odot$ yr$^{-1}$, or (ii) to subtract the complete star formation within $\epsilon_\mathrm{bar}$. In both approaches we can correct stellar masses by subtracting an excess-mass $M_\mathrm{exc}(t)$ from the central region that arises from central star formation since $z=4$ and takes mass losses by stellar winds  into account. 
\[
M_\mathrm{exc}(t) = \int_{t(z=4)}^{t}dt'\,SFR(t')\,\omega(t-t') 
\]
Here, $\omega(t)$ denotes the mass of a star particle at time $t$ after its formation (see Fig. \ref{fig:Omega}). If we assume a constant star formation rate of $11\pm{}5$ $M_\odot$ yr$^{-1}$ for $z\leq{}4$ we obtain an excess mass of $8.4\pm3.8\times{}10^{10} M_{\odot}$. Half-mass radii are computed after the excess mass is substracted from the inner region ($r<2\times{}\epsilon_\mathrm{bar}$).

We call the first approach the \emph{minimal star formation correction approach} since its effects are relatively weak: the masses, luminosities and sizes at $z=0$ change by less than $25\%$. The colors (Table \ref{tab:PropertiesZ0Phot}) are hardly affected by any star formation correction method because, as described in section \ref{sect:Methodology}, colors are effectively measured at a radius of $>2\times{}\epsilon_\mathrm{bar}$ by fixing the mass-to-light ratio of the central, unresolved region to its value at $2\times{}\epsilon_\mathrm{bar}$. 

\begin{figure}
\begin{center}
\includegraphics[width=75mm]{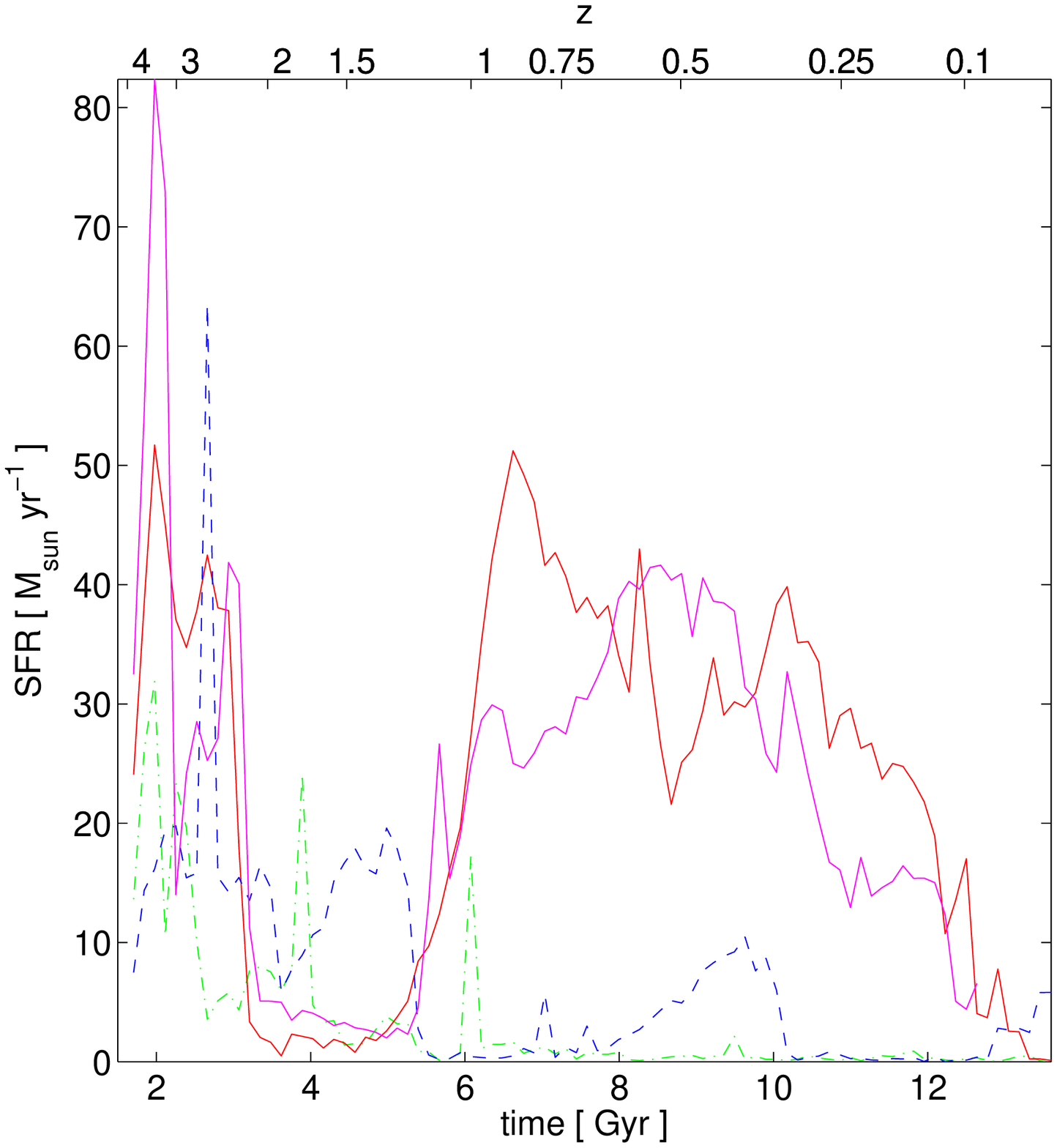}
\caption[The star formation rates of the central galaxies outside $\epsilon_\mathrm{bar}$]{The star formation rates of the central galaxies in the groups $G1$ (green), $G2$ (red), $G2-HR$ (magenta) and $G3$ (blue) within a spherical volume of 20 kpc around the central region after subtracting the star formation occurring within a baryonic softening length.\label{fig:SFRcorr1soft}} 
\end{center}
\end{figure}

\begin{figure}
\begin{center}
\includegraphics[width=75mm]{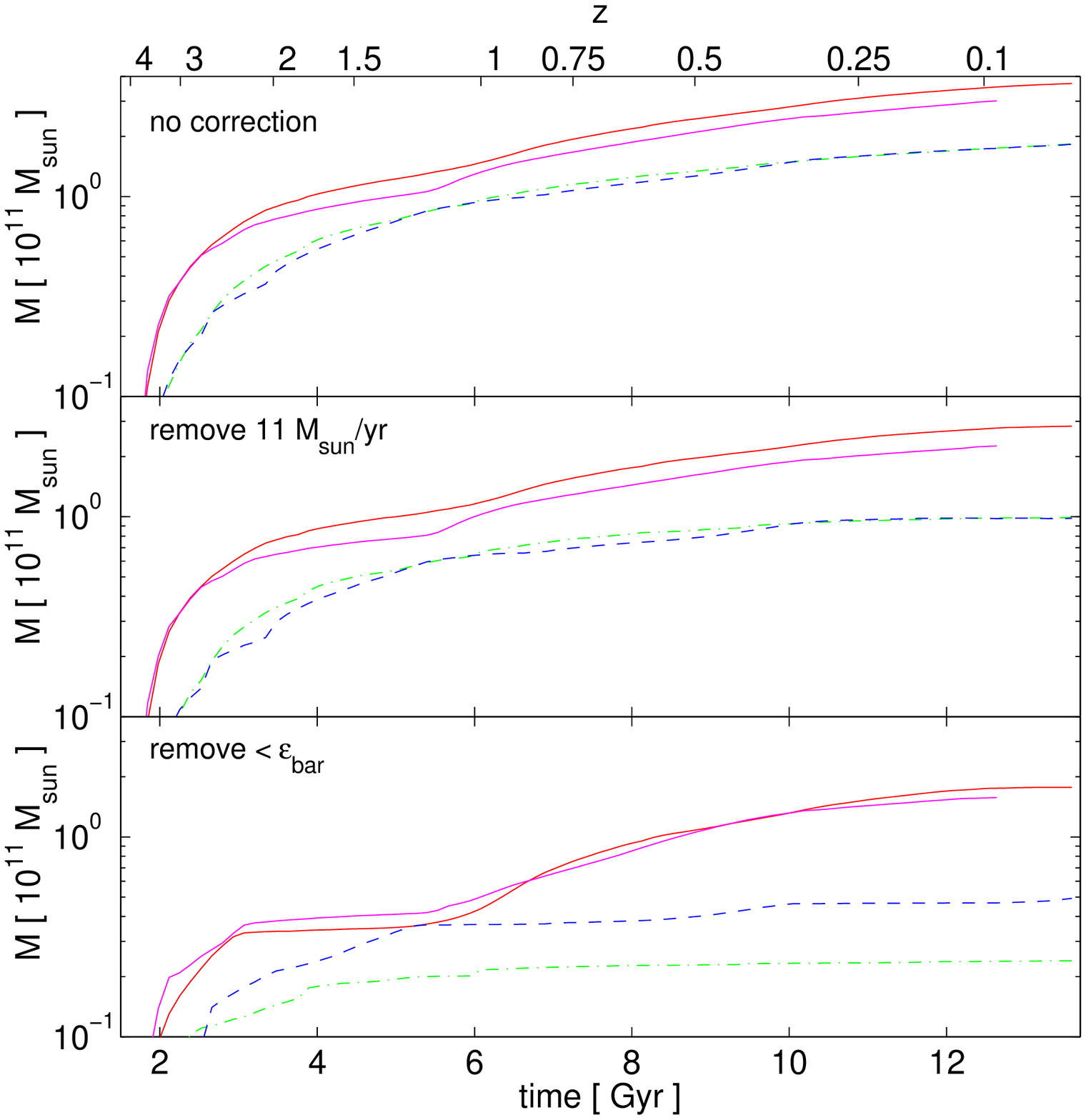}
\caption[The in situ formed stellar mass within a radius of 20 kpc]{The in situ formed stellar mass within a radius of 20 kpc around the center of the central galaxy. (Top panel) raw simulation output, (Middle panel) after correcting for an artifical star formation rate of $11$ $M_\odot$ yr$^{-1}$, (Bottom panel) after correcting for artificial star formation within the inner baryonic softening length. Mass losses by stellar winds are taken into account in the mass correction.\label{fig:inSituFormedMass}} 
\end{center}
\end{figure}

The second approach has a stronger impact on SFR, masses (and sizes), and luminosities. Comparing Fig. \ref{fig:SFRcorr1soft} and Fig. \ref{fig:SFR}, for example, demonstrates that the SFR is strongly affected by the second correction scheme. In particular, central galaxies are essentially evolving completely passive below $z=2$ except for short bursts associated with mergers and a few continuos, long lasting gas accretion episodes. The in situ formed stellar mass reduces by a factor of $\sim{}3$ compared to the no-correction scheme and by a factor of $\sim{}2$ compared to the minimal correction scheme, see Fig. \ref{fig:inSituFormedMass}. The correction is somewhat more important for the in situ formed stellar mass of $G1$ and $G3$ and less for $G2$. However, the central galaxies in $G1$ and $G3$ build their mass mostly by merging. We do not know how the amount of artificial star formation in less massive galaxies/halos and we can only speculate on how a change in the in situ formed mass is reflected on the total stellar mass. If we simply correct for the amount of in situ formed stellar mass within $\epsilon_\mathrm{bar}$ of the central galaxy, the stellar mass will reduce by only $30-35\%$. If we assume, however, that the in situ star formation in \emph{all} progenitors, not only the main progenitor, needs the same correction factor, the masses are reduced strongly by a factor $\sim{}3$. A complication of the second correction scheme is that it is not clear how galaxy sizes have to be corrected. 

We chose to employ the minimal correction approach partially for this reason, partially in order to avoid overcorrections and also because the results we focus on (e.g. the exponent in the mass-size relation) do not seem to strongly depend on which correction we apply. However, the absolute masses and sizes of the produced $z=0$ galaxies are in much better agreement with observations if a strong star formation correction is applied (c.f. the discussion in section \ref{sect:PropZ0}, Table \ref{tab:PropertiesZ0Corr}). We caution the reader that our correction scheme neglects the \emph{dynamical} effects of having an excess mass in the center and that of artificially enhanced gas inflow from larger radii. The excess mass changes the potential of the galaxy, decreases the dynamical time of orbiting matter, and the cooling time of gas in the hot diffuse medium near the center. The excess mass is thus likely to increase \emph{non-linearly} over time. In addition, artificially enhanced central gas cooling can lead to a stronger baryonic contraction of the stellar and dark matter and hence to a reduction of the half-mass radii of the central galaxies. Better resolution (especially at higher redshifts), a higher star formation threshold or an even more efficient feedback in the lower mass progenitor galaxies could obliterate the need for an explicit correction of the star formation and help to remedy any remaining discrepancies between simulated and observed galaxy properties. 

\begin{table*}
\begin{center}
\footnotesize
\caption[Masses and sizes of the central group galaxies at z=0 with and without star formation corrections]{Masses and Sizes of the central group galaxies at z=0 with and without star formation corrections.\label{tab:PropertiesZ0Corr}}
\renewcommand{\arraystretch}{1.25}
\begin{tabular}{crrr}
\tableline
\tableline
              &                               & $M_\mathrm{*}$               & $R_\mathrm{eff}$ \\
Group  & SF$_\textrm{corr}$ &   ($10^{11} M_\odot$)     &            (kpc)           \\
\\ \tableline
G1  & no     & $4.8$                 & $3.1$             \\
       & subtract  $11\pm{}5 M_{\odot}\,\mathrm{yr}^{-1}$    & $4.0\pm{}0.4$  & $4.0\pm{}0.5$ \\    
       & subtract $<\epsilon_\mathrm{bar}$ (central) & $3.2$ &  --- \\
       & subtract $<\epsilon_\mathrm{bar}$ (all) & $1.6$ &  --- \\ \tableline
G2  & no     & $5.3$                 & $2.7$  \\                 
       &  subtract  $11\pm{}5 M_{\odot}\,\mathrm{yr}^{-1}$   &$4.5\pm{}0.4$  & $3.7\pm{}0.5$     \\
       & subtract $<\epsilon_\mathrm{bar}$ (central) &                $3.4$     &  --- \\
       & subtract $<\epsilon_\mathrm{bar}$ (all) &                $1.8$     &  --- \\  \tableline
G3  & no     & $5.1$                 & $2.5$                   \\
       & subtract  $11\pm{}5 M_{\odot}\,\mathrm{yr}^{-1}$     &$4.3\pm{}0.3$  & $3.2\pm{}0.4$  \\    
       & subtract $<\epsilon_\mathrm{bar}$ (central) & $3.7$ & --- \\
       & subtract $<\epsilon_\mathrm{bar}$ (all) & $1.7$ & --- \\ \tableline
\end{tabular}
\tablecomments{The first column denotes the studied galaxy group. The second column states the correction method for artificial star formation that is applied: The top row (\emph{no}) refers to no correction. The next row refers to the minimal correction scheme. The third row shows the result of a scheme that removes in situ star formation of the central galaxy (and its most massive progenitor) within one baryonic softening length. The 4th row refers to a correction scheme in which the total stellar mass of the central galaxy is corrected with the same factor as its in situ formed stellar matter, see text. Columns 3 and 4 denote the stellar mass within 20 physical kpc and the three-dimensional radius that contains half of $M_\mathrm{tot}$.}
\end{center}
\end{table*}
\normalsize

\section{Metal Cooling}
\label{sect:MetalCooling}
High-temperature metal cooling is expected to modify the star formation rates and masses of galaxies by increasing the rate at which gas can cool. Our simulations of galaxy groups are actually at a rather sweet spot ($T_\mathrm{vir} \sim{}10^7$ K) in which the cooling rate is only mildly dependent on metallicity \citep{1993ApJS...88..253S}. In contrast, the cooling rate differs by more than an order of magnitude at the virial temperatures typical of galactic halos. Quantifying the impact of metal cooling is a non-trivial issue, because higher star formation leads to enhanced stellar feedback that will inject more energy into the ambient gas. Numerical simulations that study this non-linear interplay between gas cooling, accretion and feedback find that  the inclusion of high-temperature metal cooling increases the cosmic star formation rate by 0.25-0.3 dex at $z\sim{}1$ and 0.1-0.16 dex at $z\sim{}3$  \citep{2009MNRAS.393.1595C, 2009arXiv0909.5196S}, and is hardly significant at higher redshifts \citep{2003MNRAS.341.1253H, 2009arXiv0909.5196S}. While metal cooling increases the stellar masses of low and intermediate mass galaxies, it leaves the stellar masses of  the most massive ($\gtrsim{}10^{11} M_\odot$) galaxies at $z\sim{}1$  almost unchanged \citep{2009MNRAS.393.1595C}. This is partially due to the lower metal-sensitivity of the cooling function at high temperatures and also because more massive galaxies form on average their stars at earlier times when the impact of metal cooling is reduced. Metal cooling extends the cooling function to temperatures far below $10^4$ K. We do not expect that this affects the masses of the galaxies, but it might have an effect on central concentrations as gas is allowed to get colder and denser. However, the Jeans mass of the 300 K cold gas phase is small: $\sim{}5\times{}10^5$ $M_\odot$ at an unrealistically low density of 0.1 cm$^{-3}$ and $10^3-10^4$ $M_\odot$ at the typical densities ($\sim{}200$ cm$^{-3}$) of a giant molecular cloud. Even our high resolution simulation is far from resolving these Jeans masses and thus not able to study reliably the properties of such a cold gas phase. We note that, for consistency, our star formation and stellar feedback parameters are constrained by disk galaxy simulations with primordial gas composition \citep{2006MNRAS.373.1074S} and chosen such as to reproduce the observed Kennicutt-Schmidt law.

\section{Resolution tests}
\label{sec:ResolutionTest}

Throughout the paper we present the results of simulation $G2$ alongside with the results obtained from its higher resolution counterpart $G2-HR$. In this section we present additional considerations regarding the sizes and masses of the simulated central galaxies. 
 
\begin{figure}
\begin{center}
\includegraphics[width=75mm]{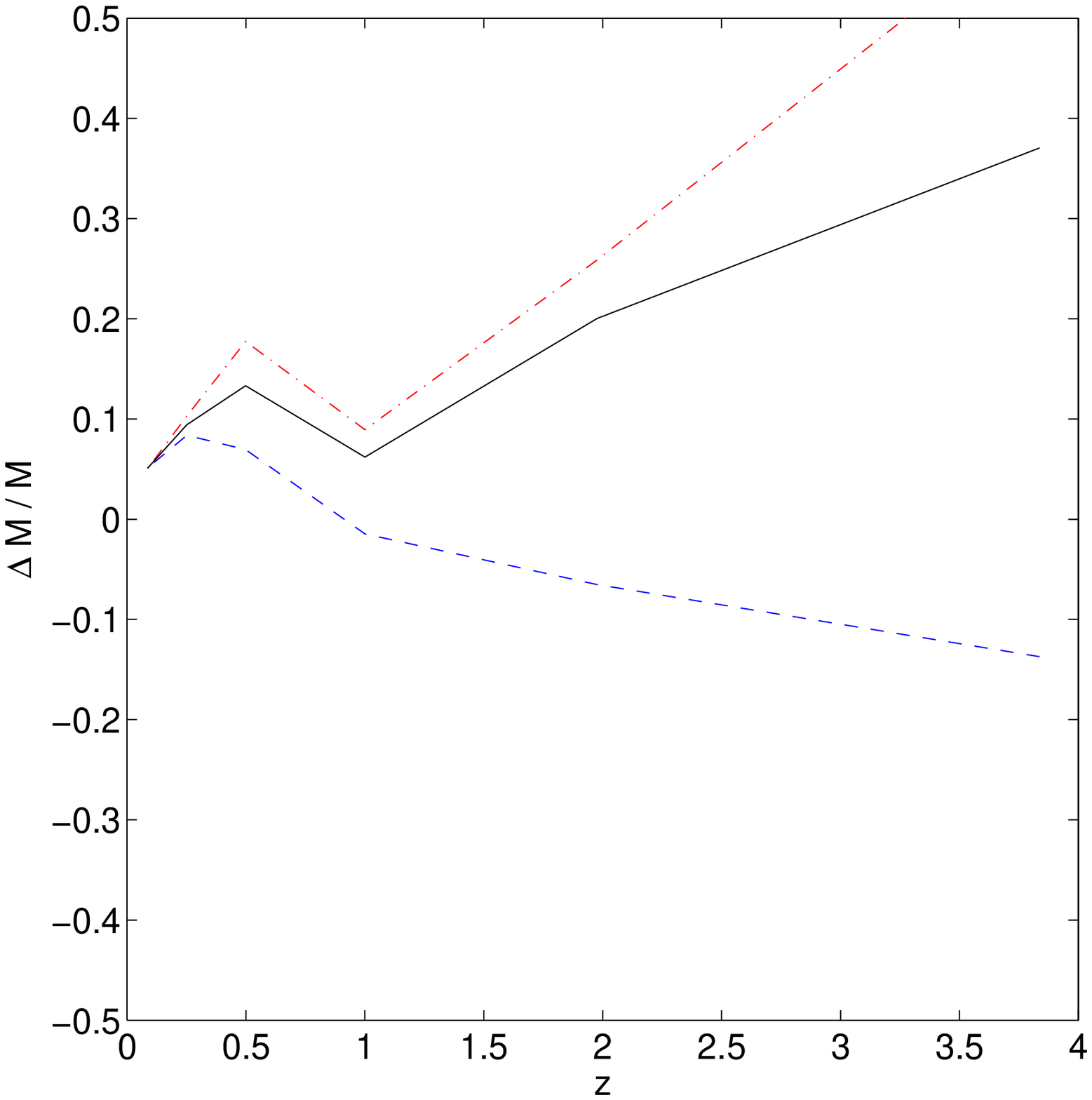}
\caption[Resolution comparison of the stellar masses]{Comparison of the stellar masses of the central galaxies between the intermediate resolution $G2$ and its high resolution counterpart $G2-HR$. The black solid line shows the mass difference of $G2$ and $G2-HR$ normalized to the mass of $G2-HR$ (all within 20 kpc). The red dot-dashed line shows the fractional difference of stellar mass within the inner 1.44 kpc, i.e. the unresolved region of $G2$, while the blue dashed curve shows the fractional mass difference between 1.44 kpc and 20 kpc. \label{fig:RegionOfTrust}
} 
\end{center}
\end{figure}

The measurement of masses and sizes is potentially affected by, e.g. overcooling, heating due to two-body relaxation, by the softening of the gravitational potential or by artificially enhanced mass deposition in the central region. In Fig. \ref{fig:RegionOfTrust} we therefore compare the stellar masses produced at intermediate ($G2$) and at high resolution ($G2-HR$). Below $z=2$ the stellar mass $M_\mathrm{*}$, the stellar mass within $r_\mathrm{min}\sim{}2\times{}\epsilon_\mathrm{bar}$ and the stellar mass outside $r_\mathrm{min}$  (and also the spherical and projected mass density profiles outside $3\times{}\epsilon_\mathrm{bar}$, not shown) converge to within a $\sim{}10-20$\% accuracy. Hence, we require that the half-mass radius is (at least) of the size $\sim{}2\times{}\epsilon_\mathrm{bar}$ for a reliable size measurement. In order to avoid that size determinations depend too strongly on the mass within $r_\mathrm{min}$ we also employ another safety criterion. We estimate the mass \emph{within} $r_\mathrm{min}$ by fixing the density \emph{at} $r_\mathrm{min}$ and extrapolate the density inwards with a power law of slope -1. This density profile corresponds to a flat surface density and we find that under these corrections the masses within $r_\mathrm{min}$ and half-mass radii change both by less than 30\% at $z=1$. However, at $z=1.5$, the mass is affected at the 50\% level and the half-mass radius by a factor of 2.  We therefore restrict the study of half-mass radii to $z\lesssim{}1.5$.
In Table \ref{tab:ComparisonZ0d1} we compare the structural and kinematic properties of the central galaxies in $G2$ and $G2-HR$ at $z=0.1$ (the lowest redshift reached in the simulation $G2-HR$) finding good agreement.

\begin{table*}
\begin{center}
\footnotesize
\caption[Resolution comparison of the structural and kinematic properties of the central galaxies at $z=0.1$]{Structural and kinematic properties of the central group galaxies at $z=0.1$.\label{tab:ComparisonZ0d1}}
%\caption[Resolution comparison of the structural and kinematic properties of the central galaxies at $z=0.1$]{Structural and kinematic properties of the central group galaxies at $z=0.1$.\label{tab:ComparisonZ0d1}}
\renewcommand{\arraystretch}{1.25}
\begin{tabular}{ccccccccccc}
\tableline
\tableline
              &  $M_\mathrm{vir}$  & $R_\mathrm{vir}$ & $M_\mathrm{*}$               & $R_\mathrm{eff}$ & $R^\mathrm{deVau}_\mathrm{eff}$ & $R^\mathrm{Sersic}_\mathrm{eff}$  &     & $\sigma_\mathrm{eff}$ &    \\  
Group  & ($10^{12} M_\odot$) &  (kpc)   &    ($10^{10} M_\odot$)     &            (kpc)             &          (kpc)                                               &          (kpc)                                             &   $n^\mathrm{Sersic}$  & (km/s) & $v_\mathrm{rot}/\sigma_\mathrm{eff}$ \\
\\ \tableline
G2         & 11.8  & 420  & $41.9\pm{}3.7$  & $3.1\pm{}0.5$   &3.9$\pm$0.3   & 4.4$\pm$0.3   &    3   & 332$\pm$20 & 0.87$\pm$0.05    \\
G2-HR  & 11.4  & 416  & $39.6\pm{}3.6$  & $2.9\pm{}0.4$   &3.2$\pm$0.2   & 2.7$\pm$0.2   &    5   & 348$\pm$20 & 0.92$\pm$0.05  \\
    \end{tabular}
\tablecomments{The columns are as in Table \ref{tab:PropertiesZ0}. The kinematical and structural properties of the central galaxy in the simulations $G2$ and $G2-HR$ are both inferred from radii that exclude the central 2-3 baryonic softening lengths in units of the simulation $G2$.}
\end{center}
\end{table*}
\normalsize

\section{Supplementary Material}

\label{sect:SupplementaryFigures}

%\begin{figure}
%\begin{center}
%\includegraphics[width=75mm]{fC1}
%\caption[Filter transmission curves]{The transmission curves of all filters used in this work. $U_\mathrm{n}$, $G$ and $\mathcal{R}$ filter bands are extracted from Fig. 1 of \cite{2003ApJ...592..728S} and compare well with the transmission curves of \citep{2004ApJ...607..226A}. $^{0.1}g$, $^{0.1}r$ and $^{0.1}i$ are $g$, $r$ and $i$ filters in the $z=0.1$ redshifted SDSS filter system of \cite{2003ApJ...592..819B}. $J_s$ is a slightly modified $J$ filter \citep{2003AJ....125.1107L}. The $J_s$ and $K_{s,2}$ transmission curves have been downloaded from the ISAAC (VLT) website. The other transmission curves correspond to standard filter bands: Bessel $U$, Bessel $B$, $r$ (SDSS), $z$ (Subaru telescope), $K_{s,1}$ (Kitt Peak 4-m telescope). We use $K_{s,1}$ in order to compute $z-K$ colors and $K_{s,2}$ for calculating $J_s-K_s$ colors. Colors and $K_s$ magnitudes are affected by much less than 0.1 mag when switching between $K_{s,1}$ and $K_{s,2}$ filter bands.\label{fig:Transmission}}
%\end{center}
%\end{figure}

\begin{figure}
\begin{center}
\includegraphics[width=75mm]{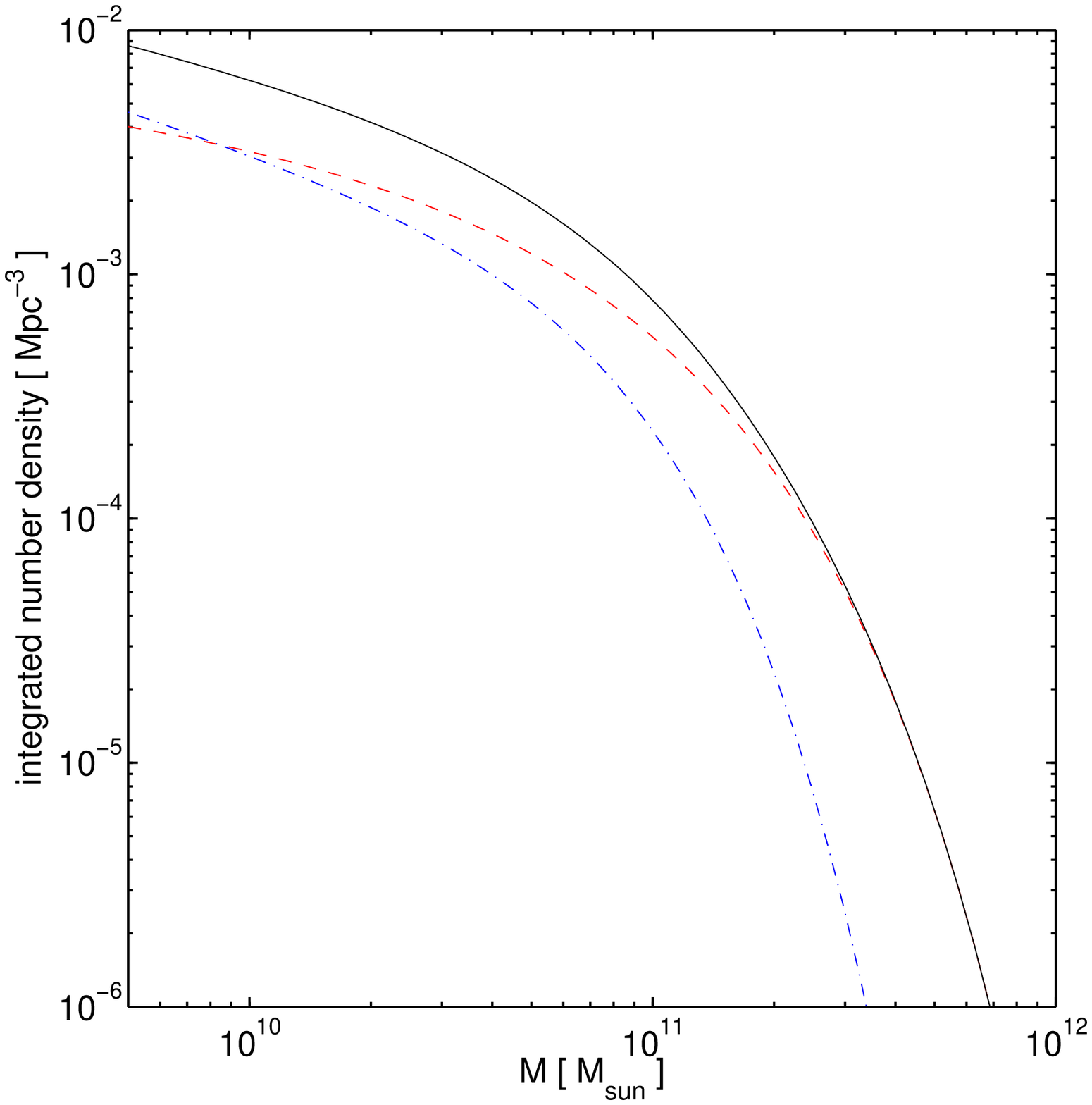}
\caption[Integrated stellar mass function]{The number density of galaxies exceeding a given stellar mass derived from the stellar mass function of \cite{2004ApJ...600..681B}. In addition to the integrated number density of all galaxies in the local universe (solid black line) the contributions from red (dashed red line) and blue (dot-dashed blue line) galaxies according to the color-magnitude split of \cite{2004ApJ...600..681B} are shown. The exponential cut-off at the high mass end causes the number density to drop by an order of magnitude when the stellar mass doubles from $2\times{}10^{11}$ $M_\odot$ ($n=1.4\times{}10^{-4}$) to  $4\times{}10^{11}$ $M_\odot$ ($n=1.8\times{}10^{-5}$).\label{fig:IntegratedMassFunction}} 
\end{center}
\end{figure}

\begin{figure}
\begin{center}
\includegraphics[width=75mm]{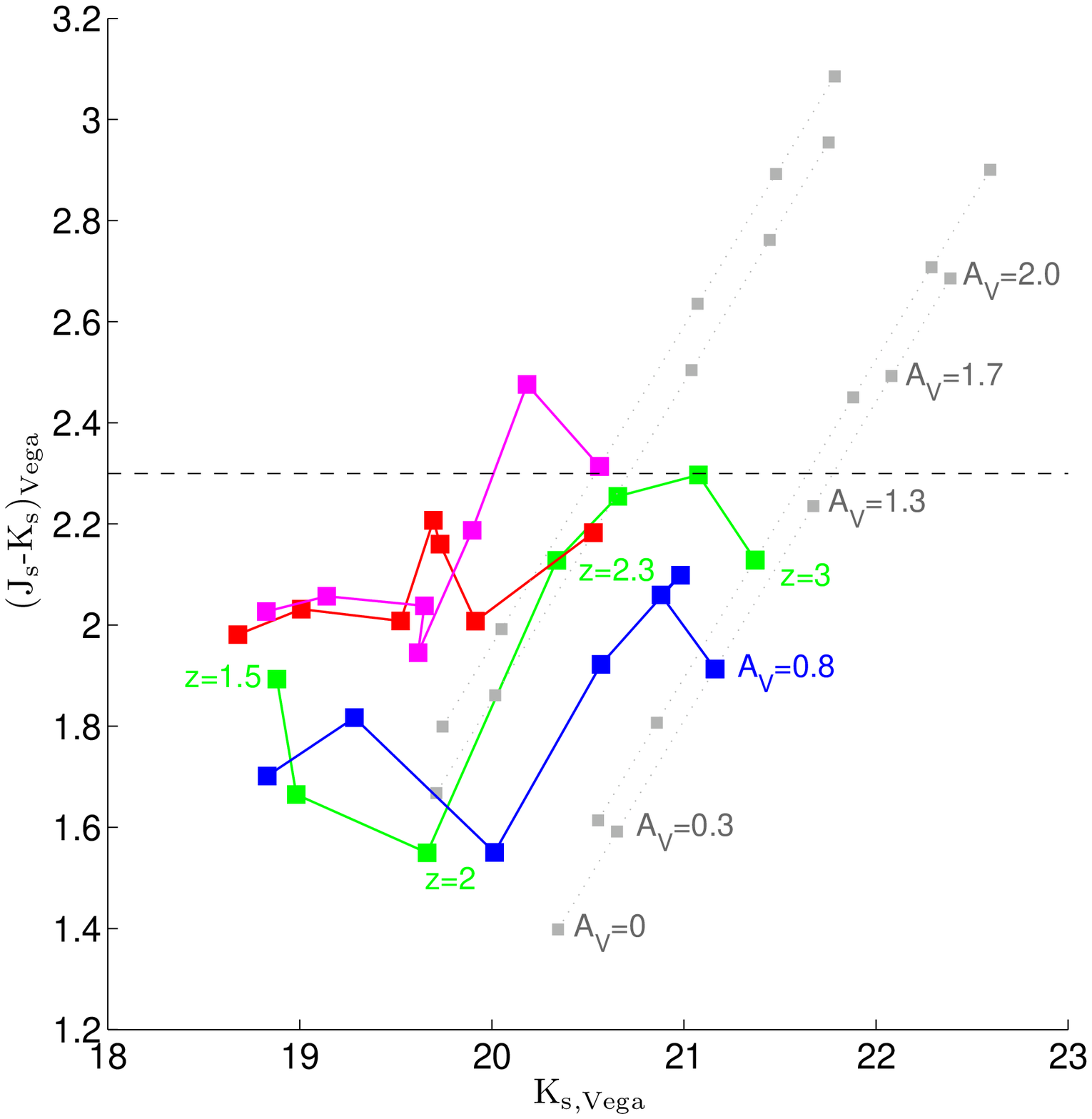}
\caption[Apparent $J_s-K_s$ color vs. $K_s$ magnitude of the main progenitors between $z=3$ and $z=1.5$]{Apparent $J_s-K_s$ color vs. $K_s$ magnitude of the main progenitors of the central group galaxies within a projected radius of 8 kpc between $z=3$ and $z=1.5$. The different lines corresponds to $G1$ (green), $G2$ (red), $G3$ (blue) and $G2-HR$ (magenta). The horizontal, dashed lines indicates the color cut that separates distant red galaxies (DRGs;  $J_s-K_s>2.3$) from non-DRG ($J_s-K_s<2.3$). Gray symbols indicate the change of the colors due to different extinction corrections \citep{2000ApJ...533..682C} with $A_V$ ranging from 0 to 2. The color evolution is shown for the default value $A_V=0.8$. Colors and magnitudes are normalized to the Vega system using the stellar template of \cite{1992IAUS..149..225K}.\label{fig:DRG}}
\end{center}
\end{figure}

\begin{figure}
\begin{center}
\includegraphics[width=75mm]{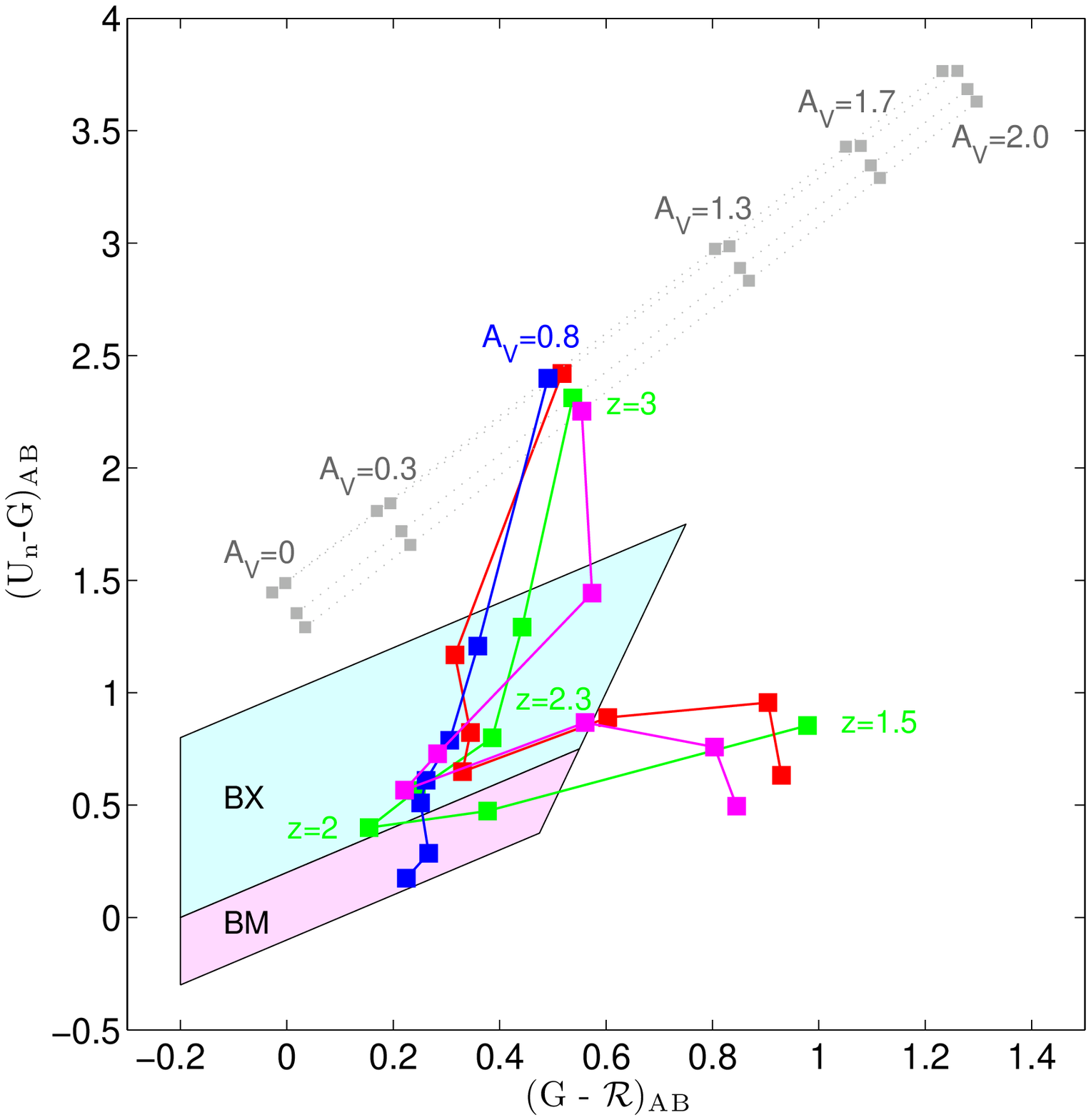}
\caption[Apparent $G-\mathcal{R}$ vs. $U_\mathrm{n}-G$ colors of the main progenitors between $z=3$ and $z=1.5$]{Apparent $G-\mathcal{R}$ vs. $U_\mathrm{n}-G$ colors of the main progenitors of the central group galaxies within a projected radius of 8 kpc between $z=3$ and $z=1.5$. The different lines corresponds to $G1$ (green), $G2$ (red), $G3$ (blue) and $G2-HR$ (magenta). The selection windows of the BM and BX techniques are indicated as colors boxes. Gray symbols indicate the change of the colors due to different extinction corrections \citep{2000ApJ...533..682C} with $A_V$ ranging from 0 to 2. The color evolution is shown for the default value $A_V=0.8$.\label{fig:BMBX}}
\end{center}
\end{figure}

\begin{figure}
\begin{center}
\includegraphics[width=75mm]{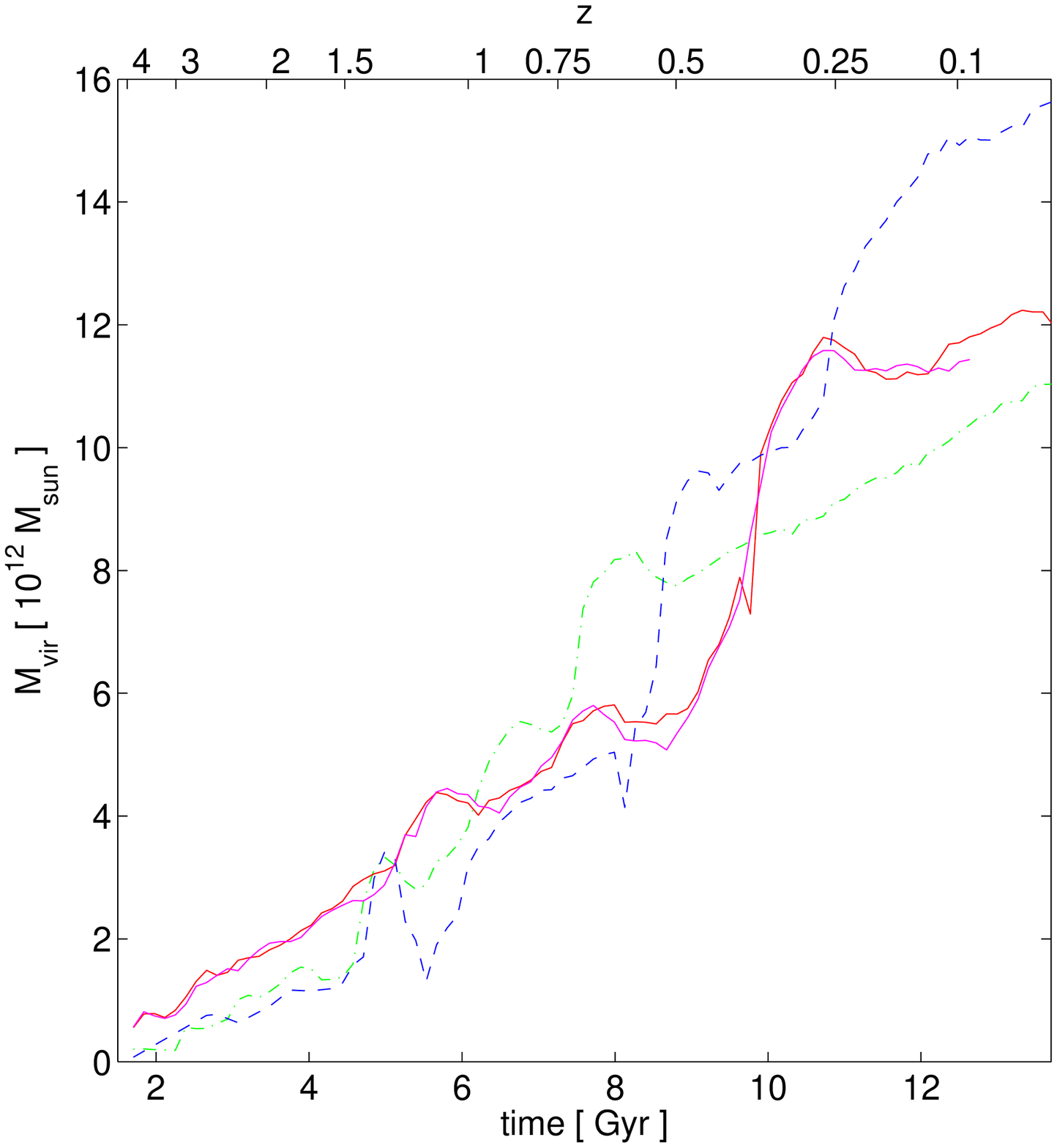}
\caption[The virial masses of the groups as function of time]{The virial mass of the groups $G1$ (green dot-dashed), $G2$ (red solid), $G3$ (blue dashed) and $G2-HR$ (magenta solid) as function of time (bottom axis) and redshift (top axis).\label{fig:AccretionHistory}}
\end{center}
\end{figure}

\begin{figure}
\begin{center}
\includegraphics[width=75mm]{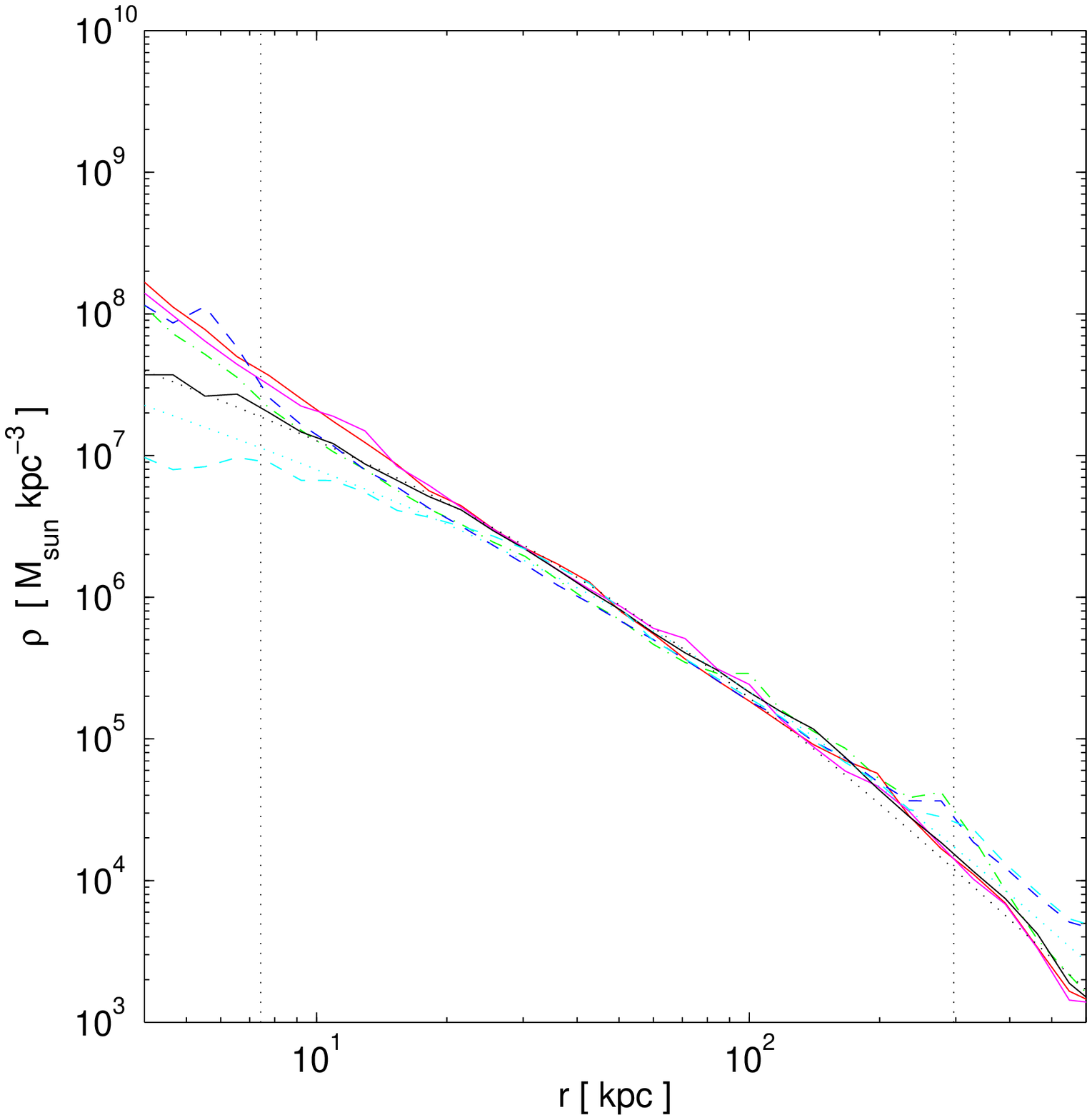}
\caption[The density profiles of the SPH vs. DM runs]{The density profiles of the SPH groups $G1$ (green dot-dashed), $G2$ (red solid), $G3$ (blue dashed) and $G2-HR$ (magenta solid) and the density profiles of corresponding dark-matter-only re-simulations: $G2-DM$ (black solid), $G3-DM$ (cyan dashed). The profiles of $G2-DM$ and $G3-DM$  are best fitted by NFW profiles (colored dotted lines) with $c=5.9$ ($G2$) and $c=2.8$ ($G3$). The fit ranges are indicated by vertical dotted lines. \label{fig:DensProfiles}}
\end{center}
\end{figure}

%\begin{figure}
%\begin{center}
%\includegraphics[width=75mm]{fC7}
%\caption[Enclosed mass according to a NFW profile]{The ratio of enclosed mass vs. radius for two NFW profiles with different concentrations $c_1$ and $c_2$ but equal virial radii \citep{1997ApJ...490..493N}: $(c_1,c_2)=(3,1.5)$ (red dot-dashed line), $(c_1,c_2)=(6,3)$ (black solid line), $(c_1,c_2)=(12,6)$ (blue dashed line),   $(c_1,c_2)=(24,12)$ (green dotted line). A doubling in concentration increases the enclosed mass at small radii by a factor 2-3.\label{fig:Concentration}}
%\end{center}
%\end{figure}

\begin{figure}
\begin{center}
\begin{tabular}{cc}
\includegraphics[width=75mm]{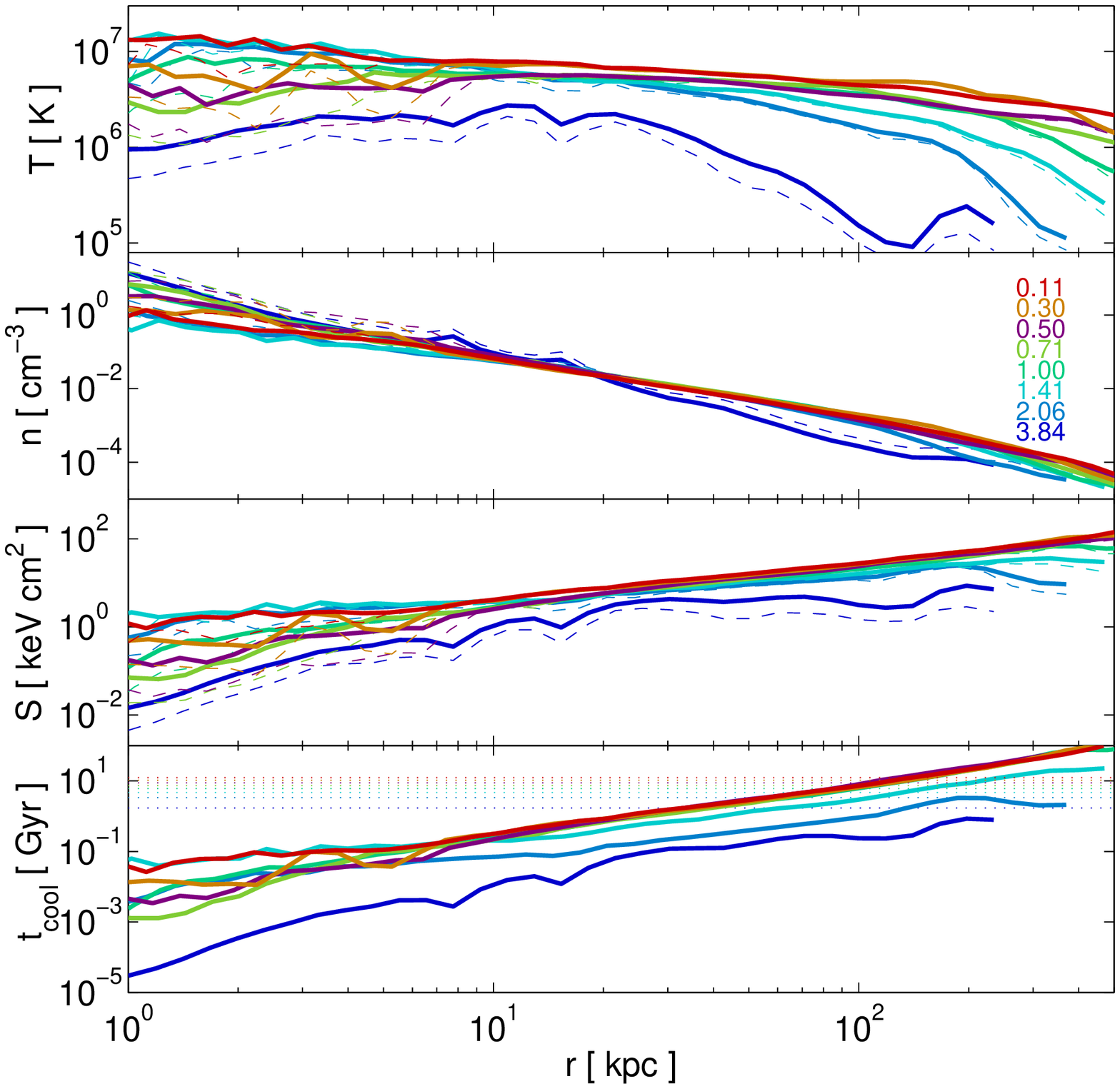} &
\includegraphics[width=75mm]{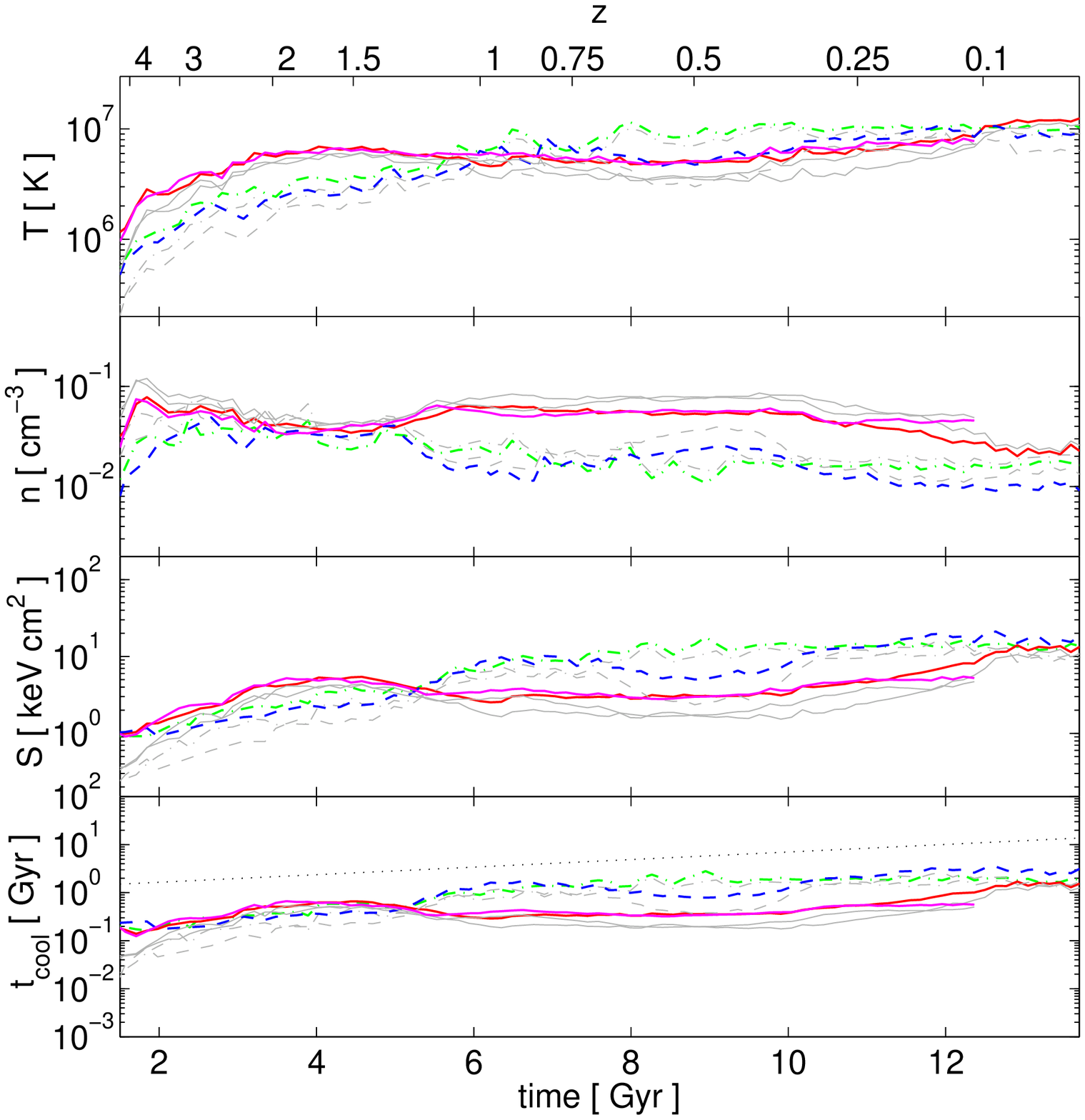}
\end{tabular}
\caption[Radial profiles and time evolution of temperature, density, entropy and cooling time]{(Left) Radial profile of density-weighted temperature, number density of charged particles, entropy and the cooling time of the intra-group medium in the group $G2-HR$ at different times. Temperature, density, entropy and cooling time are calculated from either all gas particles (thin dashed line) or from only the hot gas ($>3.2\times{}10^{4}$ K) particles (thick solid line) in spherical shells around the group center. (From top to bottom): (1) The density-averaged temperature initially raises quickly and by $z\sim{}2$ has reached an almost constant value of $\sim{}10^{7}$ in the central region, while it continues to raise in the outer region as the halo grows. The hot accretion below $z\sim{}1$ is clearly visible as a drop in temperature within the inner 10 kpc, (2) The number density of charged particles (electrons, protons and helium ions) $n=\rho/\mu$ under the assumption of complete ionization of a zero-metallicity gas ($\mu=0.59$ amu), (3) The entropy of the gas $S=T/n^{2/3}$, (4) The cooling time (eq. 14 of \citealt{2004ApJ...608...62S}) increases with increasing distance and becomes longer than the age of the Universe (dotted horizontal lines) at a radius of 100-200 kpc. (Right) Temperature, density, entropy and cooling time (defined as in the left panel) calculated from either all gas particles (light gray lines) or from only the hot gas ($>3.2\times{}10^{4}$ K) particles (colored lines) within a sphere of 20 kpc around the centers of the groups $G1$ (green dot-dashed), $G2$ (red solid), $G3$ (blue dashed) and $G2-HR$ (magenta solid) as function of time (bottom axis) and redshift (top axis). In all three groups the cooling time within 20 kpc is always lower than the age of the Universe (dotted line).\label{fig:Cooling}}
\end{center}
\end{figure}

\end{appendix}

\end{document}